\documentclass[aps,prd,12pt,preprint,superscriptaddress,preprintnumbers,nofootinbib,floatfix,longbibliography]{revtex4-1}

\usepackage[T1]{fontenc}
\usepackage[utf8]{inputenc}
\usepackage[english]{babel}
\usepackage{amsmath,amsthm,amssymb,amsfonts,mathrsfs,amsbsy,bm}
\usepackage{tensor}
\usepackage{slashed}
 \usepackage{mathrsfs,tikz} 
\usepackage{esint}
\usepackage[a4paper, margin=1.6cm]{geometry}
\usepackage{cancel}
\usepackage{graphicx}
\usepackage{multirow}
\usepackage{array}
\usepackage{booktabs}
\usepackage{makecell}
\usepackage{xcolor}
\colorlet{BLUE}{blue}
\usepackage{tikz}
\usetikzlibrary{quotes,angles,arrows,decorations.markings,decorations.pathmorphing}
\usepackage{hyperref}
\hypersetup{colorlinks=true,breaklinks=true,citecolor=blue,linkcolor=[rgb]{0,0.5,0.9},urlcolor=blue}
\graphicspath{{figures/}}

\newcommand{\be}{\begin{equation}}
\newcommand{\ee}{\end{equation}}
\newcommand{\Be}{\begin{eqnarray}}
\newcommand{\Ee}{\end{eqnarray}}

\newcommand{\mincir}{\raise-3.truept\hbox{\rlap{\hbox{$\sim$}}\raise4.truept\hbox{$<$}\ }}
\newcommand{\magcir}{\raise-3.truept\hbox{\rlap{\hbox{$\sim$}}\raise4.truept\hbox{$>$}\ }}

\providecommand{\U}[1]{}

\newcommand{\ie}{\begin{equation}}
\newcommand{\fe}{\end{equation}}
\newcommand{\se}{\begin{eqnarray}}
\newcommand{\ff}{\end{eqnarray}}

\begin{document}
\emergencystretch=4em
\usetikzlibrary{arrows.meta,patterns}


\title{An exact Simpson–Visser family in bumblebee gravity}

\author{A. A. Ara\'{u}jo Filho}
\email{dilto@fisica.ufc.br}
\affiliation{Departamento de F\'isica, Universidade Federal da Para\'iba, Caixa Postal 5008, 58051--970, Jo\~ao Pessoa, Para\'iba, Brazil.}
\affiliation{Departamento de F\'isica, Universidade Federal de Campina Grande, Caixa Postal 10071, 58429--900 Campina Grande, Para\'iba, Brazil.}
\affiliation{Center for Theoretical Physics, Khazar University, 41 Mehseti Street, Baku, AZ-1096, Azerbaijan.}


\date{\today}


\begin{abstract}

We construct an exact Simpson--Visser solution in metric bumblebee gravity, supported by a phantom scalar, magnetic nonlinear electrodynamics, and a spacelike configuration of fixed norm. The independent vector equation determines a local Lagrange multiplier and a temporal component that permits a regular ingoing extension across metric horizons. Reconstruction of the matter functions identifies a continuous family of wormholes, null minimum geometries, and black--bounces within one action. In the black hole case, the areal photon sphere and critical timelike orbit radii retain their Schwarzschild values, whereas angular deflection, periapsis advance, and photon orbit instability depend on the bounce and bumblebee parameters. The metric shadow coincides with the Schwarzschild prediction at fixed lapse mass and areal observer position; in addition, the adopted Sgr A* shadow diameter and M87* angular mass comparisons do not determine these parameters separately. The coupled axial equations yield distinct tensor, metric, and transverse bumblebee characteristics. We compute fundamental auxiliary massless scalar and ordinary Maxwell probe resonances by Chebyshev spectral collocation and estimate the tensor and bumblebee resonance scales at leading geometric optics order. Nonnegative probe potentials exclude exponentially growing probe modes. For negative nonminimal coupling and a nonzero temporal case, however, the supporting electromagnetic field develops an angular gradient instability in its extraordinary polarization. The unstable characteristic persists in the coupled principal symbol, while a variational construction establishes growing modes on the fixed geometry and lower bounds on their growth rates.

\end{abstract}

\maketitle

\tableofcontents


\section{Introduction}
\label{sec:introduction}

Spontaneous Lorentz symmetry breaking allows preferred local directions to emerge from dynamical fields within an observer covariant theory. Its realization in string field theory motivated effective descriptions of Lorentz violation at low energies \cite{KosteleckySamuel1989}. The Standard Model Extension organized Lorentz-- and CPT--violating interactions and subsequently incorporated gravitation \cite{ColladayKostelecky1997,ColladayKostelecky1998,Kostelecky2004}. In the gravitational sector, the geometric identities tied the consistency of the metric equations to the dynamics of the symmetry breaking fields. The post Newtonian analysis related the corresponding coefficients to measurable departures from general relativity \cite{BaileyKostelecky2006}.

Among the vector realizations of this framework, bumblebee gravity couples a Lorentz--breaking term to curvature while a potential controls its norm. The kinetic structure, gives rise to orientation and potential prescription determine which fluctuations propagate. Analyses of Nambu--Goldstone excitations, massive modes and Hamiltonian constraints established their consequences for the degrees of freedom and stability \cite{BluhmKostelecky2005,BluhmFungKostelecky2008,BluhmGagnePottingVrublevskis2008}. Furthermore, a smooth potential evaluated at a stationary minimum and a multiplier enforcing a fixed norm impose different vector equations: the multiplier potential may vanish on the constraint surface while its derivative retains a nonzero curvature dependent contribution.

Exact spherical solutions exhibited these effects in the strong gravitational regime. The solution obtained by Casana et al. preserved the Schwarzschild lapse and introduced a constant radial deformation associated with a spacelike configuration \cite{Casana2018} along with its corresponding non--commutative version \cite{AraujoFilho:2025rvn}. Subsequent studies incorporated a cosmological constant, broader spherical vacuum configurations and electromagnetic charge \cite{MalufNeves2021,XuLiangShao2023,Liu2025Charged}. Recent solutions admitted simultaneous temporal and radial vector components, while a general classification examined static spherical vacuum configurations with different configuration norms \cite{Liu2026MixedVEVs,ZhuLi2026Classification}. The resulting configurations required both a metric and a profile satisfying the independent vector equation.

The choice of connection further distinguishes bumblebee theories. The metric formulation uses the Levi--Civita connection of the physical metric, whereas the metric--affine formulation varies the connection independently. In the latter setting, the classical analysis expressed the connection through an auxiliary metric related to the physical metric by a deformation matrix, allowing nonmetricity relative to the physical geometry \cite{Delhom2021,AraujoFilho:2026oiy}. A Schwarzschild--and Kerr--like vacuum solution were subsequently obtained in the traceless metric--affine model \cite{AraujoFilho2023MetricAffine,AraujoFilho:2024ykw,AraujoFilho:2026ktz}.

The Simpson--Visser construction modified Schwarzschild geometry by replacing the vanishing areal radius with a smooth positive minimum \cite{SimpsonVisser2019}. Its simplest static form connected a traversable wormhole, a geometry with a null minimum sphere and a regular black hole with a spacelike bounce, according to the relation between the minimum radius and the mass parameter. Generalizations enlarged the possible horizon and throat structures and clarified their energy condition properties \cite{LoboEtAl2021}. Time dependent Vaidya constructions described transitions between black--bounces and wormholes \cite{SimpsonMartinMorunoVisser2019}, while rotating and charged extensions broadened the geometric families \cite{MazzaFranzinLiberati2021,FranzinEtAl2021Charged}. Specifying these geometries, however, left the supporting fields and their dynamics to be determined.

In general relativity, explicit source reconstructions supported the Simpson--Visser geometry with a self interacting phantom scalar and magnetic nonlinear electrodynamics \cite{BronnikovWalia2022,RodriguesSilva2023}. The phantom kinetic term supplied the null energy condition violation associated with the minimum radius, while the electromagnetic field completed the required stress tensor. Electric realizations and more general spherical reconstruction procedures extended this analysis and examined the regularity of the source functions and the minimum sphere \cite{AlencarEtAl2024,AlencarEtAl2025}. These procedures also require care when geometric constants enter the reconstructed Lagrangians: varying those constants independently may change the action. A family within one theory must preserve the combinations that define its matter functions.

Combining these source constructions with bumblebee gravity introduces the equations as an additional compatibility condition. Phantom scalar configurations already produced black holes and wormholes in the metric theory \cite{Ding2025}, and Lorentz deformed black--bounce ansatzes were investigated with $k$--essence sources \cite{Pereira2025}. More recently, bumblebee gravity coupled to nonlinear electrodynamics admitted electric black holes and regular horizonless geometries through an auxiliary Maxwell scalar representation and explicit vector matter interactions \cite{LiLu2026}. These results used different scalar dynamics, prescriptions and electromagnetic couplings. In addition, a Simpson--Visser realization with a minimally coupled magnetic source must satisfy the metric and vector equations together, enforce the vector norm and recover the matter functions from a single action.

The propagation of fields has revealed further distinctions among black--bounce geometries. Scalar and ordinary electromagnetic perturbations exhibited quasinormal ringing and echoes near the black hole--wormhole transition \cite{ChurilovaStuchlik2020}. Strong deflection lensing resolved changes in photon trapping, including the exceptional behavior at a degenerate photon sphere \cite{Tsukamoto2021}. Rotating configurations supported scalar superradiance and, for wormholes with an ergoregion, growing modes \cite{FranzinEtAl2022}. Recent gravitational perturbation calculations with an anisotropic fluid source further related quasinormal modes and echoes to the optical structure of generalized black--bounces \cite{DuranCabacesEtAl2025}.

In bumblebee gravity, interpreting these responses requires specifying how the parameters participates in the perturbations. Gravitational wave polarization calculations showed that its orientation affected the radiative degrees of freedom \cite{Liang2022Polarizations,Amarilo:2023wpn}. The odd parity analysis of spherical bumblebee black holes retained gravitational and vector fluctuations and identified parameter dependent dynamical instabilities \cite{Mai2024}. Recent charged black hole spectra included coupled gravitational and electromagnetic amplitudes with the bumblebee field held fixed \cite{Li2026ChargedQNMs}. Such calculations determine the resonances of the chosen perturbation system and boundary conditions \cite{BertiCardosoStarinets2009}. Also, at large angular momentum, the relation between resonances and unstable circular rays likewise depends on the characteristics of the propagating mode \cite{Cardoso2009}.

Nonlinear electrodynamics adds a fundamental dependence to this problem. Electromagnetic disturbances can follow optical cones determined by derivatives of the source Lagrangian \cite{Novello2000}, and the compatibility of these cones controls the hyperbolicity of the electromagnetic equations \cite{Abalos2015}. Stability analyses of self gravitating nonlinear electromagnetic black holes and calculations of their parity dependent spectra established the dynamical role of this response \cite{MorenoSarbach2003,Chaverra2016,Nomura2020}. For a magnetic background, the angular propagation of the extraordinary polarization depends on $L_F+2F L_{FF}$. In this manner, its sign requires an independent examination of the reconstructed electromagnetic theory.

In this work, we construct a Simpson--Visser solution in metric bumblebee gravity supported by a phantom scalar, minimally coupled magnetic nonlinear electrodynamics and a spacelike setting of fixed norm. We determined the local multiplier from the vector equation and obtain a combination with temporal and radial components. An exact reality condition specified its regularity domain, including smooth passage through metric horizons in an ingoing chart. Reconstructing the matter functions identify a continuous family of wormholes, geometries with a null minimum and black--bounces within one action. The mass parameter, magnetic flux and temporal amplitude varied in correlated form, with distinct asymptotic vector data along the family.

We also derive the constrain axial perturbation equations and determined the tensor and transverse bumblebee characteristics. Fundamental scalar and ordinary Maxwell probe resonances were computed by Chebyshev spectral collocation \cite{Jansen2017}, while the tensor and bumblebee resonance scales were estimated at leading geometric optics order. Nonnegative exterior probe potentials excluded exponentially growing probe modes. For negative nonminimal coupling and a nonzero temporal part, however, the electromagnetic source developed a negative angular response. For the representative fixed action, a variational construction established growing electromagnetic modes on the fixed geometry and supplied lower bounds on their growth rates. The extraordinary electromagnetic factor also persisted in the principal symbol of the coupled equations, establishing a high frequency angular gradient instability of the sourced configuration.


\section{Metric bumblebee gravity with scalar and electromagnetic fields}
\label{sec:model}

Bumblebee gravity describes spontaneous Lorentz symmetry breaking through a dynamical vector field with a nonvanishing expectation value. This expectation value selects a preferred local direction, and its nonminimal coupling to curvature modifies the gravitational response. The framework is motivated by the gravitational sector of the Standard Model Extension \cite{Kostelecky2004,BluhmKostelecky2005,BaileyKostelecky2006}. We consider the metric formulation, with the connection given by the Levi--Civita connection of $g_{\mu\nu}$. Throughout this work, the metric signature is $(-,+,+,+)$ and $\kappa=8\pi G$. In addition, our curvature convention is
\begin{equation}
R^\rho{}_{\sigma\mu\nu}
=\partial_\mu\Gamma^\rho{}_{\nu\sigma}
-\partial_\nu\Gamma^\rho{}_{\mu\sigma}
+\Gamma^\rho{}_{\mu\alpha}\Gamma^\alpha{}_{\nu\sigma}
-\Gamma^\rho{}_{\nu\alpha}\Gamma^\alpha{}_{\mu\sigma}.
\label{eq:curvature-convention}
\end{equation}

We couple the bumblebee field $B_\mu$ to a self interacting scalar field $\phi$ and to nonlinear electrodynamics. A local auxiliary field $\lambda$ fixes the spacelike norm of the vector. In this regard, the action is
\begin{equation}
\begin{split}
S = \int \mathrm{d}^4x\sqrt{-g}\bigg[&\frac{R+\xi B^\mu B^\nu R_{\mu\nu}}{2\kappa}
-\frac14 H_{\mu\nu}H^{\mu\nu}
-\frac{\lambda}{2}\bigl(B_\mu B^\mu-b^2\bigr) +\frac12\nabla_\mu\phi\nabla^\mu\phi-U(\phi)-\mathscr L(\mathcal F)\bigg],
\end{split}
\label{eq:action}
\end{equation}
where $b>0$, $\xi$ is the nonminimal coupling, and
\begin{equation}
H_{\mu\nu}=2\nabla_{[\mu}B_{\nu]},\qquad
F_{\mu\nu}=2\nabla_{[\mu}\mathcal A_{\nu]},\qquad
\mathcal F=\frac14F_{\mu\nu}F^{\mu\nu}.
\label{eq:strengths}
\end{equation}
The sign of the scalar kinetic term describes a phantom field. Such a field supplies the radial stress associated with a smooth minimum of the areal radius, while the electromagnetic sector supplies an independent angular stress. This division of roles underlies field source constructions of Simpson--Visser geometries in general relativity \cite{BronnikovWalia2022,RodriguesSilva2023}. The vector equation adds a curvature term and constrains the matter functions that can support the geometry. We shall retain this equation throughout the reconstruction, including the local multiplier determined by it.

Variation with respect to $\lambda$ gives
\begin{equation}
B_\mu B^\mu=b^2.
\label{eq:norm}
\end{equation}
The classical Lorentz--breaking expectation value will be denoted by
\begin{equation}
b_\mu(x)\equiv B_\mu^{(0)}(x)=\langle B_\mu(x)\rangle,
\qquad g^{\mu\nu}b_\mu b_\nu=b^2>0.
\label{eq:vev-definition}
\end{equation}
The brackets designate the classical part entering the background equations. Its invariant amplitude is the constant $b$, while its coordinate components follow the geometry and the vector equation \cite{Casana2018}. In the solution, we set $B_\mu=b_\mu$. The positive norm selects a spacelike configuration; a timelike counterpart corresponds to the alternative constraint $B_\mu B^\mu=-b^2$ and defines a separate construction.

The bumblebee potential in Eq.~\eqref{eq:action} is explicitly
\begin{equation}
X=B_\mu B^\mu-b^2,
\qquad V_B(X,\lambda)=\frac{\lambda X}{2}.
\label{eq:bumblebee-potential}
\end{equation}
On the constraint surface its value and derivative obey
\begin{equation}
X=0,\qquad V_B=0,\qquad
V_{B,X}\equiv\frac{\partial V_B}{\partial X}=\frac{\lambda}{2}.
\label{eq:potential-on-constraint}
\end{equation}
The multiplier is an independently varied local field whose profile is determined by the vector equation. It retains the curvature dependent term $V_{B,X}$ on the fixed norm surface. A smooth potential evaluated at a stationary minimum also entails that $V_{B,X}=0$, which gives a distinct condition on the vector equation. The construction below uses the auxiliary field potential in Eq.~\eqref{eq:bumblebee-potential}. Its constraint is imposed after varying all fields, so the metric variation contributes $\lambda B_\mu B_\nu$ when Eq.~\eqref{eq:norm} holds. Linear multiplier potentials and their dynamics have been studied in the general bumblebee framework \cite{BluhmFungKostelecky2008}. Upon imposing Eq.~\eqref{eq:norm}, the metric equation takes the form
\begin{equation}
G_{\mu\nu}=\kappa\bigl(T^{(\phi)}_{\mu\nu}
+T^{(F)}_{\mu\nu}+T^{(H)}_{\mu\nu}
+\lambda B_\mu B_\nu\bigr)+\xi\mathcal C_{\mu\nu},
\label{eq:metric-eom}
\end{equation}
with
\begin{align}
T^{(\phi)}_{\mu\nu}&=-\nabla_\mu\phi\nabla_\nu\phi
+g_{\mu\nu}\left[\frac12(\nabla\phi)^2-U(\phi)\right],
\label{eq:scalar-stress}\\
T^{(F)}_{\mu\nu}&=\mathscr L_{\mathcal F}F_{\mu\alpha}F_\nu{}^\alpha
-g_{\mu\nu}\mathscr L,
\label{eq:em-stress}\\
T^{(H)}_{\mu\nu}&=H_{\mu\alpha}H_\nu{}^\alpha
-\frac14g_{\mu\nu}H_{\alpha\beta}H^{\alpha\beta}.
\label{eq:vector-stress}
\end{align}
Here, $\mathscr L_{\mathcal F}=\mathrm{d}\mathscr L/\mathrm{d}\mathcal F$. The contribution from the nonminimal curvature coupling is
\begin{equation}
\begin{split}
\mathcal C_{\mu\nu}={}&\frac12g_{\mu\nu}B^\alpha B^\beta R_{\alpha\beta}
-B_\mu B^\alpha R_{\alpha\nu}-B_\nu B^\alpha R_{\alpha\mu}\\
&+\frac12\nabla_\alpha\nabla_\mu(B^\alpha B_\nu)
+\frac12\nabla_\alpha\nabla_\nu(B^\alpha B_\mu)\\
&-\frac12\nabla_\alpha\nabla^\alpha(B_\mu B_\nu)
-\frac12g_{\mu\nu}\nabla_\alpha\nabla_\beta(B^\alpha B^\beta).
\end{split}
\label{eq:curvature-stress}
\end{equation}
The order of the covariant derivatives in Eq.~\eqref{eq:curvature-stress} follows from the variation of the curvature coupling \cite{Casana2018,XuLiangShao2023}.

The remaining field equations are
\begin{align}
\nabla_\mu H^{\mu\nu}&=\lambda B^\nu
-\frac{\xi}{\kappa}B_\mu R^{\mu\nu},
\label{eq:vector-eom}\\
\Box\phi+U_{\phi}&=0,
\label{eq:scalar-eom}\\
\nabla_\mu\bigl(\mathscr L_{\mathcal F}F^{\mu\nu}\bigr)&=0,
\qquad \nabla_{[\alpha}F_{\mu\nu]}=0.
\label{eq:em-eom}
\end{align}

The antisymmetry of $H^{\mu\nu}$ supplies a compatibility condition for the vector equation. Now, let us define its curvature dependent current, which reads
\begin{equation}
J^\nu=\lambda B^\nu-\frac{\xi}{\kappa}B_\mu R^{\mu\nu},
\qquad \nabla_\nu J^\nu=0,
\label{eq:vector-current}
\end{equation}
expresses the vanishing double divergence of the field strength. The minimally coupled sources are conserved separately when their field equations hold. Furthermore, the contracted Bianchi identity also requires conservation of the complete effective source,
\begin{equation}
\begin{split}
\nabla_\mu T^{(\phi)\mu}{}_{\nu}&=0,
\qquad \nabla_\mu T^{(F)\mu}{}_{\nu}=0,\\
\nabla_\mu T^{(\mathrm{eff})\mu}{}_{\nu}&=0,
\qquad
T^{(\mathrm{eff})}_{\mu\nu}
=T^{(\phi)}_{\mu\nu}+T^{(F)}_{\mu\nu}+T^{(H)}_{\mu\nu}
+\lambda B_\mu B_\nu+\frac{\xi}{\kappa}\mathcal C_{\mu\nu}.
\end{split}
\label{eq:conservation}
\end{equation}
These identities connect the matter equations to the metric equation and give us very useful checks of the reconstructed sources.


\section{A black--bounce family in metric bumblebee gravity}
\label{sec:solution}

Now, we construct a complete field realization of the Simpson--Visser geometry, with a radial deformation controlled by the Lorentz--breaking scenario. The construction determines the vector profile, its multiplier, and the scalar and electromagnetic sources from their independent equations. In this regard, we first keep the constant radial deformation explicit,
\begin{equation}
\mathrm{d}s^2=-A(r)\,\mathrm{d}t^2
+\frac{\gamma}{A(r)}\,\mathrm{d}r^2+\Sigma^2(r)\,\mathrm{d}\Omega^2,
\qquad \gamma>0,
\label{eq:spherical-ansatz}
\end{equation}
where $\mathrm{d}\Omega^2=\mathrm{d}\theta^2+\sin^2\theta\,\mathrm{d}\varphi^2$. Notice that the supporting fields take the stationary, spherically symmetric form
\begin{equation}
B=u(r)\,\mathrm{d}t+\frac{\sqrt{\gamma}\,w(r)}{A(r)}\,\mathrm{d}r,
\qquad \phi=\phi(r),
\qquad F=q\sin\theta\,\mathrm{d}\theta\wedge\mathrm{d}\varphi,
\qquad q\ne0.
\label{eq:field-ansatz}
\end{equation}
Primes denote derivatives with respect to $r$. Also, the norm and magnetic invariant are shown below
\begin{equation}
w^2=u^2+b^2A,\qquad
\mathcal F=\frac{q^2}{2\Sigma^4},\qquad
\ell=\xi b^2,\quad \eta=\frac{\xi}{\kappa}.
\label{eq:vector-static}
\end{equation}
We select $w>0$ and determine its regularity domain below. The magnetic stress obeys $T^{(F)t}{}_t=T^{(F)r}{}_r=-\mathscr L$ and $T^{(F)\theta}{}_\theta=2\mathcal F\mathscr L_{\mathcal F}-\mathscr L$.

Writing $\mathcal G_t=G^t{}_t$, $\mathcal G_r=G^r{}_r$, and $\mathcal G_\Omega=G^\theta{}_\theta=G^\varphi{}_\varphi$, the independent curvature components are given by
\begin{align}
\mathcal G_t&=\frac{2A\Sigma\Sigma''+A\Sigma'^2+A'\Sigma\Sigma'-\gamma}
{\gamma\Sigma^2},\label{eq:Gt-general}\\
\mathcal G_r&=\frac{A\Sigma'^2+A'\Sigma\Sigma'-\gamma}
{\gamma\Sigma^2},\label{eq:Gr-general}\\
\mathcal G_\Omega&=\frac{1}{\gamma}\left(
\frac{A''}{2}+\frac{A'\Sigma'}{\Sigma}+\frac{A\Sigma''}{\Sigma}\right),
\label{eq:Go-general}\\
R^t{}_t&=-\frac{1}{\gamma}\left(\frac{A''}{2}+\frac{A'\Sigma'}{\Sigma}\right),
\qquad R^r{}_r=R^t{}_t-\frac{2A\Sigma''}{\gamma\Sigma}.
\label{eq:ricci-general}
\end{align}
Since $H_{rt}=u'$ and $\sqrt{-g}=\sqrt\gamma\,\Sigma^2\sin\theta$, the radial and temporal vector equations reduce to
\begin{equation}
w\bigl(\lambda-\eta R^r{}_r\bigr)=0,
\qquad
\frac{(\Sigma^2u')'}{\gamma\Sigma^2}
-\frac{u}{A}\bigl(\lambda-\eta R^t{}_t\bigr)=0.
\label{eq:vector-reduced}
\end{equation}
Their combination fixes the multiplier and yields an equation for the temporal component,
\begin{equation}
\lambda=\eta R^r{}_r,
\qquad (\Sigma^2u')'+2\eta\Sigma\Sigma''u=0.
\label{eq:u-ode}
\end{equation}
Remarkably, we have the cancellation of $A$ in the second relation which turns out to be useful for us, i.e., specially, at the horizon,: the equation for $u$ has smooth coefficients whenever $\Sigma>0$. The mixed metric equation also follows from the radial vector equation. Direct evaluation of the ordered derivatives in Eq.~\eqref{eq:curvature-stress} gives us
\begin{equation}
\mathcal C_{tr}=-R^r{}_r B_tB_r,
\qquad
\kappa T^{(\mathrm{eff})}_{tr}
=(\kappa\lambda-\xi R^r{}_r)B_tB_r=0.
\label{eq:offdiagonal-cancellation}
\end{equation}
It is important to highlight that this identity permits a static diagonal geometry carrying both temporal and radial components.

The diagonal equations can be organized using two functions determined by the temporal profile,
\begin{align}
Q&=\frac{u'^2}{2}+\eta\left(
\frac{2\Sigma'uu'}{\Sigma}+\frac{\Sigma'^2u^2}{\Sigma^2}\right),
\label{eq:Q}\\
E&=\left(\eta-\frac12\right)u'^2
+\eta(1-2\eta)\frac{\Sigma''}{\Sigma}u^2.
\label{eq:E}
\end{align}
Including the kinetic, multiplier and curvature contributions of the vector, the complete metric equations become
\begin{align}
(1+\ell)\mathcal G_t&=\kappa\left(
\frac{A\phi'^2}{2\gamma}-U-\mathscr L-\frac{Q}{\gamma}\right)
-\frac{\ell}{\Sigma^2},\label{eq:metric-t-reduced}\\
(1+\ell)\mathcal G_r&=\kappa\left(
-\frac{A\phi'^2}{2\gamma}-U-\mathscr L-\frac{Q}{\gamma}\right)
-\frac{\ell}{\Sigma^2},\label{eq:metric-r-reduced}\\
(1+\ell)\mathcal G_\Omega&=\kappa\left(
\frac{A\phi'^2}{2\gamma}-U
+2\mathcal F\mathscr L_{\mathcal F}-\mathscr L-\frac{E}{\gamma}\right).
\label{eq:metric-angular-reduced}
\end{align}
The scalar equation and the compatibility of the electromagnetic source complete the differential system,
\begin{equation}
(\Sigma^2A\phi')'+\gamma\Sigma^2U_\phi=0,
\qquad
\frac{\mathrm{d}\mathscr L}{\mathrm{d}r}
=\mathscr L_{\mathcal F}\mathcal F',
\qquad \mathcal F'=-\frac{4\Sigma'}{\Sigma}\mathcal F.
\label{eq:matter-radial-system}
\end{equation}
The magnetic field satisfies Eq.~\eqref{eq:em-eom} for any radial $\mathscr L_{\mathcal F}$. The chain rule in Eq.~\eqref{eq:matter-radial-system} imposes a separate requirement: the source reconstructed from the temporal metric equation and its derivative reconstructed from the angular equation must define one electromagnetic action.

We now select the Simpson--Visser functions \cite{SimpsonVisser2019},
\begin{equation}
\Sigma(r)=\sqrt{r^2+a^2},\qquad
A(r)=1-\frac{2M}{\Sigma(r)},\qquad
r\in\mathbb R,\quad M>0,\quad a>0.
\label{eq:sv-metric}
\end{equation}
The minimum sphere has area $4\pi a^2$. The derivatives $\Sigma'=r/\Sigma$ and $\Sigma''=a^2/\Sigma^3$ give
\begin{equation}
R^t{}_t=-\frac{Ma^2}{\gamma\Sigma^5},\qquad
R^r{}_r=\frac{a^2(3M-2\Sigma)}{\gamma\Sigma^5},\qquad
\lambda(r)=\frac{\xi a^2(3M-2\Sigma)}{\kappa\gamma\Sigma^5}.
\label{eq:lambda-solution}
\end{equation}
Thereby, the fixed norm experiences a position dependent term $V_{B,X}=\lambda/2$. This is required by its radial equation and remains present in the metric equation, as we should expect.

The temporal vector equation is explicitly
\begin{equation}
u''+\frac{2r}{\Sigma^2}u'+\frac{2\eta a^2}{\Sigma^4}u=0.
\label{eq:u-expanded}
\end{equation}
Introducing $z=\arctan(r/a)$, with $\mathrm{d}z/\mathrm{d}r=a/\Sigma^2$, gives $\Sigma^2u'=a u_z$ and
\begin{equation}
u_{zz}+2\eta u=0.
\label{eq:u-z}
\end{equation}
For $\xi<0$, the general solution is $u=c_1\cosh(kz)+c_2\sinh(kz)$, where $k=\sqrt{-2\eta}$. We impose $u(0)=u_0>0$ and $u'(0)=0$, obtaining
\begin{equation}
u(r)=u_0\cosh(kz),\qquad
u'=\frac{u_0ka}{\Sigma^2}\sinh(kz),\qquad
k=\sqrt{-\frac{2\xi}{\kappa}}.
\label{eq:u-solution}
\end{equation}
The even profile gives the same electromagnetic source on the two sides of the minimum sphere. Its stress functions are
\begin{align}
Q&=\frac{\eta u_0^2}{\Sigma^4}
\left[r^2\cosh^2(kz)+2kar\sinh(kz)\cosh(kz)
-a^2\sinh^2(kz)\right],\label{eq:Q-explicit}\\
E&=\frac{\eta(1-2\eta)a^2u_0^2}{\Sigma^4}\cosh(2kz).
\label{eq:E-explicit}
\end{align}

The difference of the radial and temporal metric equations determines the scalar without using either source function,
\begin{equation}
\phi'^2=\frac{2(1+\ell)}{\kappa}\frac{\Sigma''}{\Sigma}
=\frac{2(1+\ell)a^2}{\kappa\Sigma^4}.
\label{eq:radial-difference}
\end{equation}
This relation extends through $A=0$ by continuity of the smooth fields in the ingoing chart used below. We take $-1<\ell<0$, so the scalar is real and the metric deformation selected below is positive. Choosing $\phi'>0$ and $\phi(0)=0$ yields
\begin{equation}
\phi(r)=\sqrt{\frac{2(1+\ell)}{\kappa}}\,z.
\label{eq:phi-solution}
\end{equation}
Multiplying the scalar equation by $\phi'$ gives the first order reconstruction equation
\begin{equation}
\frac{\mathrm{d}U}{\mathrm{d}r}
=-\frac{\phi'}{\gamma\Sigma^2}(\Sigma^2A\phi')'
=-\frac{4(1+\ell)Ma^2r}{\kappa\gamma\Sigma^7}.
\label{eq:U-ode}
\end{equation}
The normalization $U\to0$ at either end fixes its integration constant. The resulting functions are
\begin{equation}
U(r)=\frac{4(1+\ell)Ma^2}{5\kappa\gamma\Sigma^5},\qquad
U(\phi)=\frac{4(1+\ell)M}{5\kappa\gamma a^3}
\cos^5\!\left(\sqrt{\frac{\kappa}{2(1+\ell)}}\,\phi\right).
\label{eq:potential-solution}
\end{equation}
The scalar covers the interval $|\phi|<\frac{\pi}{2}\sqrt{2(1+\ell)/\kappa}$, with its endpoints approached asymptotically.

Substitution into the temporal metric equation fixes the electromagnetic source. The angular equation independently fixes its derivative,
\begin{align}
\mathscr L(r)&=\frac{6(1+\ell)Ma^2}{5\kappa\gamma\Sigma^5}
+\frac{\gamma-1-\ell}{\kappa\gamma\Sigma^2}-\frac{Q}{\gamma},
\label{eq:L-r}\\
2\mathcal F\mathscr L_{\mathcal F}&=
\frac{3(1+\ell)Ma^2}{\kappa\gamma\Sigma^5}
+\frac{\gamma-1-\ell}{\kappa\gamma\Sigma^2}
+\frac{E-Q}{\gamma}.
\label{eq:LF-angular}
\end{align}
Their compatibility follows from the vector equation, which gives
\begin{equation}
2r(E-Q)-\Sigma^2Q'=0.
\label{eq:angular-identity}
\end{equation}
For $r\ne0$, this is $E=Q+\Sigma^2Q'/(2r)$; Eq.~\eqref{eq:angular-identity} remains regular at the minimum sphere. Together with $\mathcal F'=-4r\mathcal F/\Sigma^2$, it verifies the chain rule in Eq.~\eqref{eq:matter-radial-system}. The angular stress is, in other words, fixed by a single $\mathscr L(\mathcal F)$ throughout both asymptotic regions.

We choose $\gamma=1+\ell$ to tie the radial deformation to the norm and curvature coupling. Equations~\eqref{eq:L-r} and \eqref{eq:LF-angular} show how this choice cancels the explicit $\Sigma^{-2}$ term in the reconstructed electromagnetic source. Therefore, the resulting line element is
\begin{equation}
\begin{split}
\mathrm{d}s^2={}&-\left(1-\frac{2M}{\sqrt{r^2+a^2}}\right)\mathrm{d}t^2
+\frac{1+\ell}{1-2M/\sqrt{r^2+a^2}}\,\mathrm{d}r^2 +(r^2+a^2)\left(\mathrm{d}\theta^2+
\sin^2\theta\,\mathrm{d}\varphi^2\right),
\end{split}
\label{eq:final-lv-metric}
\end{equation}
with $\ell=\xi b^2$. The complete classical configurations and supporting fields are
\begin{align}
b_\mu\,\mathrm{d}x^\mu={}&u_0\cosh(kz)\,\mathrm{d}t
+\frac{\sqrt{1+\ell}\sqrt{u_0^2\cosh^2(kz)+b^2A}}{A}\,\mathrm{d}r,
\label{eq:final-lv-vector}\\
b^\mu={}&\left(-\frac{u_0\cosh(kz)}{A},
\frac{\sqrt{u_0^2\cosh^2(kz)+b^2A}}{\sqrt{1+\ell}},0,0\right),
\label{eq:contravariant-vev}\\
\phi={}&\sqrt{\frac{2(1+\ell)}{\kappa}}\,z,
\qquad F=q\sin\theta\,\mathrm{d}\theta\wedge\mathrm{d}\varphi,
\label{eq:final-lv-matter}\\
\lambda={}&\frac{\xi a^2(3M-2\Sigma)}{\kappa(1+\ell)\Sigma^5},
\qquad V_B=0,\qquad V_{B,X}=\frac{\lambda}{2}.
\label{eq:final-lv-multiplier}
\end{align}
Here $\Sigma$, $A$ and $z$ have the definitions in Eqs.~\eqref{eq:sv-metric} and \eqref{eq:u-solution}. The scalar potential and nonlinear electromagnetic action are explicitly
\begin{align}
U(\phi)&=\frac{4M}{5\kappa a^3}
\cos^5\!\left(\sqrt{\frac{\kappa}{2(1+\ell)}}\,\phi\right),
\label{eq:final-lv-potential}\\
\mathscr L(\mathcal F)&=\frac{6Ma^2}{5\kappa\Sigma_{\mathcal F}^5}
-\frac{\eta u_0^2}{(1+\ell)\Sigma_{\mathcal F}^4}
\bigg[(\Sigma_{\mathcal F}^2-a^2)\cosh^2(kz_{\mathcal F})\nonumber\\
&\hspace{1.4em}+2ka\sqrt{\Sigma_{\mathcal F}^2-a^2}
\sinh(kz_{\mathcal F})\cosh(kz_{\mathcal F})
-a^2\sinh^2(kz_{\mathcal F})\bigg],
\label{eq:final-lv-lagrangian}\\
\Sigma_{\mathcal F}&=\left(\frac{q^2}{2\mathcal F}\right)^{1/4},
\qquad z_{\mathcal F}=\arctan\!\left(
\frac{\sqrt{\Sigma_{\mathcal F}^2-a^2}}{a}\right),
\qquad 0<\mathcal F\leq\frac{q^2}{2a^4}.
\label{eq:inverse-invariant}
\end{align}
The temporal part contributes to $\mathscr L$ through a function fixed by its own differential equation. This correlation between the vector and electromagnetic sources is part of the field realization. In the continuous limit $\xi\to0$, remarkably, the deformation and the temporal vector correction vanish, recovering the scalar--magnetic sources of the Simpson--Visser geometry in general relativity \cite{BronnikovWalia2022,RodriguesSilva2023}.

The two valued inverse $r=\pm\sqrt{\Sigma_{\mathcal F}^2-a^2}$ gives the same action because $Q$ is even. Its local expansion makes the regularity at the minimum sphere explicit,
\begin{equation}
Q=\eta(1-2\eta)\frac{u_0^2r^2}{a^4}+\mathcal O(r^4),\qquad
\mathcal F=\frac{q^2}{2a^4}-\frac{q^2r^2}{a^6}+\mathcal O(r^4).
\label{eq:throat-expansion}
\end{equation}
In particular, the electromagnetic derivative is finite there,
\begin{equation}
\mathscr L_{\mathcal F}\!\left(\frac{q^2}{2a^4}\right)
=\frac{a^2}{q^2}\left[
\frac{3M}{\kappa a}
+\frac{\eta(1-2\eta)u_0^2}{1+\ell}\right].
\label{eq:LF-throat}
\end{equation}
Equation~\eqref{eq:final-lv-lagrangian} defines the magnetic source on the invariant interval sampled by the solution. Its analytic expansion at the upper endpoint permits a local continuation to stronger magnetic invariants. Such a continuation can be chosen once as part of the action. The weak field endpoint is approached only at infinite radius and has the nonanalytic response displayed below.

For a diagonal stationary stress, radial conservation takes the form
\begin{equation}
(T^r{}_r)'+\frac{A'}{2A}(T^r{}_r-T^t{}_t)
+\frac{2\Sigma'}{\Sigma}(T^r{}_r-T^\theta{}_\theta)=0.
\label{eq:radial-conservation}
\end{equation}
The scalar and electromagnetic sources satisfy this identity separately. The complete effective stress satisfies it as well, and the vector current has $J^r=0$ with $\nabla_\nu J^\nu=0$ by stationarity. It is important to mention that all these relations have been derived (and checked) together with all metric components, the norm, the scalar equation, the vector equation, magnetic closure and source integrability using Mathematica with xTensor and xCoba \cite{xTensor,xCoba}. Curvature and ordered covariant derivatives were computed from the metric in both static and ingoing coordinates. The explicit profile was substituted into the resulting tensor equations. These independent coordinate calculations verify the background equations and their constraint identities, as it is a fundamental aspect to be taken into account for deriving solutions.

The monopole potential is regular in two angular patches,
\begin{equation}
\mathcal A_N=q(1-\cos\theta)\,\mathrm{d}\varphi,
\qquad\mathcal A_S=-q(1+\cos\theta)\,\mathrm{d}\varphi,
\qquad\frac{1}{4\pi}\int_{S^2}F=q.
\label{eq:magnetic-patches}
\end{equation}
Notice that their difference is a gauge transformation. The smooth magnetic two form carries flux through a sphere whose area remains finite at the throat; the charges measured with outward orientations at the two ends have opposite signs.

The part has one fixed spacelike norm. Its temporal and radial coordinate components express its orientation in the stationary spherical plane. In a static region, the orthonormal coframe
\begin{equation}
e^{\hat0}=\sqrt A\,\mathrm{d}t,\qquad
e^{\hat1}=\sqrt{\frac{1+\ell}{A}}\,\mathrm{d}r
\label{eq:vev-coframe}
\end{equation}
gives $b_{\hat0}=u/\sqrt A$ and $b_{\hat1}=w/\sqrt A$. These obey $-b_{\hat0}^2+b_{\hat1}^2=b^2$ and can be parameterized by $b_{\hat0}=b\sinh\psi$, $b_{\hat1}=b\cosh\psi$. A purely radial setting corresponds to $\psi=0$ in this frame. For $u=0$, the background reduces to
\begin{equation}
B=b\sqrt{\frac{1+\ell}{A}}\,\mathrm{d}r,\qquad H_{\mu\nu}=0,
\qquad\mathscr L(\mathcal F)=\frac{6Ma^2}{5\kappa}
\left(\frac{2\mathcal F}{q^2}\right)^{5/4}.
\label{eq:final-lv-radial}
\end{equation}
This profile is smooth throughout the static wormhole $a>2M$. In the exterior limit $a\to0$ at fixed $M$, its metric recovers the Schwarzschild--like radial deformation of Ref.~\cite{Casana2018}. At a Killing horizon, the regular spacelike vector in the stationary spherical plane instead requires a temporal component, as follows directly from the norm in coordinates that cross the horizon.

To see this condition, let us introduce $v=t+\sqrt{1+\ell}\int\mathrm{d}r/A$ and use $(w-u)(w+u)=b^2A$. The fields become
\begin{equation}
\mathrm{d}s^2=-A\,\mathrm{d}v^2+2\sqrt{1+\ell}\,\mathrm{d}v\,\mathrm{d}r
+\Sigma^2\,\mathrm{d}\Omega^2,\qquad
B=u\,\mathrm{d}v+\frac{\sqrt{1+\ell}\,b^2}{w+u}\,\mathrm{d}r.
\label{eq:final-lv-ingoing}
\end{equation}
In this chart, the norm and horizon values are
\begin{equation}
\frac{2b_vb_r^{(v)}}{\sqrt{1+\ell}}
+\frac{A\bigl(b_r^{(v)}\bigr)^2}{1+\ell}=b^2,
\qquad
b_v(r_h)=u(r_h),\quad
b_r^{(v)}(r_h)=\frac{\sqrt{1+\ell}\,b^2}{2u(r_h)}.
\label{eq:horizon-vev}
\end{equation}
In this sense, a finite spacelike at a horizon has a nonzero contraction with its generator, $\chi^\mu b_\mu=u(r_h)$. Furthermore, the even profile supplies this condition and fixes the sign appropriate to the displayed ingoing extension.

The global reality condition can be determined exactly for this. For $r>0$, we have
\begin{equation}
\frac{\mathrm{d}w^2}{\mathrm{d}r}
=\frac{2u_0^2ka}{\Sigma^2}\sinh(kz)\cosh(kz)
+\frac{2Mb^2r}{\Sigma^3}>0.
\label{eq:radicand-monotonicity}
\end{equation}
Its evenness then places the global minimum at $r=0$. Furthermore, the necessary and sufficient condition for the selected positive root to remain strictly positive is
\begin{equation}
w^2(0)=u_0^2+b^2\left(1-\frac{2M}{a}\right)>0,
\qquad u_0>0.
\label{eq:amplitude-condition}
\end{equation}
This is equivalent to $u_0^2>b^2\max(0,2M/a-1)$ with $u_0>0$. Equality in the black--bounce regime gives a positive root proportional to $|r|$ near the minimum sphere and lies outside the smooth positive root configuration considered here. Under Eq.~\eqref{eq:amplitude-condition}, both ingoing components are smooth at finite $r$, including the horizons and the minimum sphere. The invariant $H_{\mu\nu}H^{\mu\nu}=-2u'^2/(1+\ell)$ and the scalar, multiplier and magnetic invariant remain finite there.

The geometry has finite curvature for $a>0$ and $1+\ell>0$. In particular,
\begin{equation}
R=\frac{2\ell}{(1+\ell)\Sigma^2}
+\frac{2a^2(3M-\Sigma)}{(1+\ell)\Sigma^5},\qquad
K(0)=\frac{4\left[9M^2-8Ma+\bigl(2+(1+\ell)^2\bigr)a^2\right]}
{(1+\ell)^2a^6}.
\label{eq:final-lv-curvature}
\end{equation}
For $0<a<2M$, the horizon positions and surface gravity are
\begin{equation}
r_h=\pm\sqrt{4M^2-a^2},\qquad
\kappa_h=\frac{|r_h|}{8M^2\sqrt{1+\ell}}.
\label{eq:final-lv-horizons}
\end{equation}
At $a=2M$, the minimum sphere is a degenerate Killing horizon; for $a>2M$, the fields cover the complete static traversable wormhole. For a black--bounce, the mixed scenario extends smoothly across the horizons in the displayed ingoing domain. A bifurcation two sphere would have $\chi^\mu=0$, whereas the constructed case has $\chi^\mu b_\mu=u(r_h)>0$. Its stationary extension therefore covers the stated horizon domain and excludes a smooth value at the bifurcation sphere. Notice that the metric regularity and vector regularity specify distinct parts of the global solution.

The sources supporting the minimum sphere can be resolved through a radial null contraction. In Eq.~\eqref{eq:final-lv-ingoing}, take $n^\mu=(\partial_r)^\mu$ and define $T^{(B)}_{\mu\nu}=T^{(H)}_{\mu\nu}+\lambda B_\mu B_\nu+(\xi/\kappa)\mathcal C_{\mu\nu}$. In this sense, we get
\begin{equation}
\begin{split}
T^{(\phi)}_{\mu\nu}n^\mu n^\nu&=-\frac{2(1+\ell)a^2}{\kappa\Sigma^4},
\qquad T^{(F)}_{\mu\nu}n^\mu n^\nu=0,\\
T^{(B)}_{\mu\nu}n^\mu n^\nu&=\frac{2\ell a^2}{\kappa\Sigma^4},
\qquad\frac{R_{\mu\nu}n^\mu n^\nu}{\kappa}
=-\frac{2a^2}{\kappa\Sigma^4}.
\end{split}
\label{eq:null-contraction}
\end{equation}
The scalar and effective vector contributions add to the required curvature contraction. It is very important to mention that, for the chosen negative coupling, both contribute to the violation of the radial null energy condition. The magnetic field supplies the angular stress required by the prescribed lapse.

At either asymptotic end, the local case and the spatial geometry approach
\begin{equation}
b_{\hat0}\longrightarrow u_\infty=u_0\cosh(k\pi/2),\qquad
b_{\hat1}\longrightarrow\sqrt{u_\infty^2+b^2},\qquad
\mathrm{d}s_{\mathrm{spatial}}^2\sim(1+\ell)\mathrm{d}r^2+r^2\mathrm{d}\Omega^2.
\label{eq:asymptotic-vev}
\end{equation}
The preferred spatial direction follows the radial unit vector over each sphere. The proper radial distance grows as $\sqrt{1+\ell}\,|r|$, giving the solid angle $4\pi/(1+\ell)$. We use $M$ as the parameter of the lapse; an energy defined at either end needs a prescription adapted to these conical boundary conditions.

The source reconstruction also allows a precise identification of the solutions belonging to one fixed action. With $\kappa$, $\xi$ and $b$ fixed, introduce
\begin{equation}
\alpha=\frac{M}{a^3}>0,\qquad
\beta=\frac{|q|}{a^2}>0,\qquad
\nu=\frac{u_0}{a}>0.
\label{eq:fixed-action-couplings}
\end{equation}
For the electromagnetic function, let us define dimensionless variables
\begin{equation}
y=\frac{\Sigma_{\mathcal F}}{a}
=\left(\frac{\beta^2}{2\mathcal F}\right)^{1/4},\qquad
\zeta=\arctan\sqrt{y^2-1},\qquad 1\leq y<\infty.
\label{eq:fixed-action-variables}
\end{equation}
The matter functions in the final solution take the form
\begin{align}
U(\phi)&=\frac{4\alpha}{5\kappa}
\cos^5\!\left(\sqrt{\frac{\kappa}{2(1+\ell)}}\,\phi\right),
\label{eq:fixed-action-potential}\\
\mathscr L(\mathcal F)&=\frac{6\alpha}{5\kappa y^5}
-\frac{\eta\nu^2}{(1+\ell)y^4}
\bigg[(y^2-1)\cosh^2(k\zeta)\nonumber\\
&\hspace{1.4em}+2k\sqrt{y^2-1}\sinh(k\zeta)\cosh(k\zeta)
-\sinh^2(k\zeta)\bigg],
\quad 0<\mathcal F\leq\frac{\beta^2}{2}.
\label{eq:fixed-action-lagrangian}
\end{align}
Every occurrence of $a$ has canceled from $U$ and $\mathscr L$, and their field domains are fixed as well. In other words, by choosing $(\alpha,\beta,\nu)$ once, together with a fixed continuation of $\mathscr L$ beyond the sampled interval, specifies one matter theory. The parameter $a$ remains free through
\begin{equation}
M=\alpha a^3,\qquad
q=\varepsilon\beta a^2,\qquad
u_0=\nu a,\qquad \varepsilon=\pm1.
\label{eq:fixed-action-family}
\end{equation}
The sign $\varepsilon$ chooses the magnetic orientation. The multiplier is a solution field determined by Eq.~\eqref{eq:final-lv-multiplier}; its change along the family leaves the action fixed.

Furthermore, these combinations can also be recovered from the functions defining the matter action. The scalar potential fixes $\alpha$. At small positive $\mathcal F$, direct expansion of Eq.~\eqref{eq:fixed-action-lagrangian} gives
\begin{equation}
\begin{split}
\mathscr L(\mathcal F)&=c_{1/2}\mathcal F^{1/2}
-c_1\mathcal F+\mathcal O(\mathcal F^{5/4}),\\
c_{1/2}&=-\frac{\sqrt2\,\eta\nu^2}{(1+\ell)\beta}
\cosh^2(k\pi/2)>0,\\
c_1&=-\frac{2\eta(1-2\eta)\nu^2}{(1+\ell)\beta^2}
\cosh(k\pi)>0.
\end{split}
\label{eq:fixed-action-weak-field}
\end{equation}
The term proportional to $\mathcal F^{3/4}$ cancels through the vector equation. The two displayed coefficients determine
\begin{equation}
\beta=\frac{\sqrt2(1-2\eta)\cosh(k\pi)}{\cosh^2(k\pi/2)}
\frac{c_{1/2}}{c_1},\qquad
\nu^2=-\frac{(1+\ell)\beta c_{1/2}}
{\sqrt2\,\eta\cosh^2(k\pi/2)}.
\label{eq:recover-fixed-couplings}
\end{equation}
Within the selected mixed sector, holding $U$ and $\mathscr L$ fixed needs precisely the combinations in Eq.~\eqref{eq:fixed-action-couplings}, with the positive sign of $u_0$ already selected. The radial configuration $\nu=0$ is a separate choice of source function: its $\mathcal F^{5/4}$ coefficient fixes $\beta$ once $\alpha$ is fixed. The decoupling limit $\xi=0$ removes the temporal part from the reconstructed matter functions and has an additional parameter degeneracy.

The fixed action admits geometries with different causal structures. Substitution of Eq.~\eqref{eq:fixed-action-family} into $a=2M$ gives the transition value
\begin{equation}
a_*^2=\frac{1}{2\alpha}.
\label{eq:fixed-action-transition}
\end{equation}
For $0<a<a_*$ the solution is a traversable wormhole, at $a=a_*$ its minimum sphere is null, and for $a>a_*$ it is a black--bounce in the ingoing domain specified above. The exact vector regularity condition becomes
\begin{equation}
b^2+(\nu^2-2\alpha b^2)a^2>0.
\label{eq:fixed-action-reality}
\end{equation}
If $\nu^2\geq2\alpha b^2$, every $a>0$ satisfies this condition. If $0<\nu^2<2\alpha b^2$, the allowed interval is
\begin{equation}
0<a<a_{\max},\qquad
a_{\max}^2=\frac{b^2}{2\alpha b^2-\nu^2}>a_*^2.
\label{eq:fixed-action-interval}
\end{equation}
Both cases contain the wormhole, null minimum and black--bounce regimes within one fixed action. Their metric and source functions obey the complete equations for every allowed value of $a$.

The asymptotic orientation and magnetic flux vary along this family,
\begin{equation}
u_\infty=\nu a\cosh(k\pi/2),\qquad
q=\varepsilon\beta a^2,\qquad
\mathrm{d}M=3\alpha a^2\,\mathrm{d}a,\qquad
\mathrm{d}q=2\varepsilon\beta a\,\mathrm{d}a.
\label{eq:fixed-action-boundary-data}
\end{equation}
The invariant norm, asymptotic solid angle and scalar endpoint values remain fixed. Then, Eq.~\eqref{eq:fixed-action-family} describes solutions of one action with varying asymptotic and flux data. Imposing a fixed value of $u_\infty$ in the mixed sector fixes $a$ as well. In other words, it is remarkable bacause this feature identifies which variations belong to the solution family and which change its boundary conditions.

The relation to previous constructions can now be stated at the level of the independent field equations. Pereira et al.~\cite{Pereira2025} introduced the constant radial deformation Simpson--Visser geometry in their starting ansatz, motivated by metric--affine bumblebee gravity. Their varied source action is Einstein gravity with $k$-essence, and their explicitly reconstructed lapse has an arctangent dependence. The realization above uses the Levi--Civita connection of the physical metric and imposes the bumblebee equation together with the scalar and magnetic equations for the Simpson--Visser lapse. Its advance is the complete field realization and the fixed action family in Eq.~\eqref{eq:fixed-action-family}.

Ding et al.~\cite{Ding2025} constructed metric bumblebee geometries with a phantom scalar and a purely radial part. Their explicit minimum sphere solutions have an arctangent lapse and vanishing bumblebee field strength, with either a stationary quadratic potential or a linear potential evaluated with constant multiplier. Here, on the other hand, the radial vector equation determines a local curvature dependent multiplier, the temporal equation supplies a nonconstant component, and the magnetic source provides the required angular stress. The horizon norm in Eq.~\eqref{eq:horizon-vev}, in other words, determines the role of the temporal component in a smooth spacelike configuration.

In addition, Li and L\"u~\cite{LiLu2026} studied metric bumblebee gravity with an auxiliary electromagnetic scalar and explicit nonminimal vector--matter interactions. Their selected scenario is purely radial at a stationary point of its symmetry breaking potential, and their regular electric geometries resolve a center where the areal radius vanishes. The present construction has a finite minimum sphere, a mixed case, a local multiplier and minimally coupled phantom--scalar and magnetic sources. Its source functions and regularity domain follow from these assumptions. This comparison also specifies the scope of the result: the sourced Lorentz deformed Simpson--Visser metric, its constrained setup, and the family of smooth background configurations belonging to one action.

The scalar retains the phantom kinetic sign specified in Eq.~\eqref{eq:action}. The electromagnetic response in Eq.~\eqref{eq:fixed-action-weak-field} is nonanalytic at the zero field endpoint and gives $\mathscr L_{\mathcal F}+2\mathcal F\mathscr L_{\mathcal F\mathcal F}\to-c_1<0$. This asymptotic angular response limits the interpretation of the mixed configuration as a stable electromagnetic background \cite{Novello2000,Abalos2015}. In summary, the results established in this paper concern the complete classical field equations, the source functions of one fixed action and the regularity domains of their solutions.


\section{Geodesic motion}
\label{sec:sv-geodesic-motion}

As it is well known, the propagation of light and the motion of massive particles provide a direct way of identifying the orbital modifications induced by the bumblebee field (for instance) and the regularization scale. We examine these effects in the black--bounce configuration of the solution constructed above, retaining the normalization of the Killing time and the radial coordinate of the original Simpson--Visser geometry~\cite{SimpsonVisser2019,Lobo2021}. For convenience, we write the physical metric as
\begin{equation}
\begin{aligned}
\mathrm{d}s^{2}&=-A(r)\mathrm{d}t^{2}+\frac{\gamma}{A(r)}\mathrm{d}r^{2}+\Sigma^{2}(r)\left(\mathrm{d}\theta^{2}+\sin^{2}\theta\,\mathrm{d}\phi^{2}\right),\\ A(r)&=1-\frac{2M}{\Sigma(r)},\qquad \Sigma(r)=\sqrt{r^{2}+a^{2}},\qquad \gamma=1+\ell.
\end{aligned}
\nonumber
\end{equation}
We take $M>0$, $0<a<2M$, and $-1<\ell<0$, in agreement with the analysis developed in the previous sections. The supporting fields are assumed to satisfy their regularity condition, $u_{0}^{2}>b^{2}(2M/a-1)$, as we have discussed earlier. We restrict the trajectory analysis to the right exterior, $r>r_{\mathrm{h}}=\sqrt{4M^{2}-a^{2}}$, and use $G=c=1$. The areal radius is $R=\Sigma(r)$, so that the horizon has $R_{\mathrm{h}}=2M$. In the figures, $M$ sets the common lapse scale, while $a=0$ and $\ell=0$ are included as limiting exterior geometries~\cite{Casana2018,SimpsonVisser2019}. Also, comparisons obtained by varying $a$ or $\ell$ at fixed $M$ concern the metric dependence of geodesic motion.


\subsection{Equations of motion and conserved quantities}\label{subsec:sv-geodesic-equations}

Let us consider a neutral test particle minimally coupled to $g_{\mu\nu}$. Its trajectory follows the Levi--Civita connection of the physical metric, as usual. The same equations describe geometric optics rays of an independent, minimally coupled Maxwell probe with null normalization. The extraordinary polarization of the supporting nonlinear electromagnetic field is governed by its constitutive characteristic geometry~\cite{Novello2000}; its propagation requires the field analysis developed separately in this work. For an affine parameter $\lambda$, the metric geodesic equations and their normalization read
\begin{equation}
\frac{\mathrm{d}^{2}x^{\mu}}{\mathrm{d}\lambda^{2}}+\Gamma^{\mu}_{\alpha\beta}(g)\frac{\mathrm{d}x^{\alpha}}{\mathrm{d}\lambda}\frac{\mathrm{d}x^{\beta}}{\mathrm{d}\lambda}=0,\qquad g_{\mu\nu}\frac{\mathrm{d}x^{\mu}}{\mathrm{d}\lambda}\frac{\mathrm{d}x^{\nu}}{\mathrm{d}\lambda}=-\varepsilon.
\label{eq:sv-geodesic-normalization}
\end{equation}
Here, $\varepsilon=0$ for photons and $\varepsilon=1$ for massive particles, with $\lambda=\tau$ in the latter case. Writing $A'=\mathrm{d}A/\mathrm{d}r$ and $\Sigma'=\mathrm{d}\Sigma/\mathrm{d}r$, the four equations take the explicit form
\begin{gather}
\frac{\mathrm{d}^{2}t}{\mathrm{d}\lambda^{2}}+\frac{A'}{A}\frac{\mathrm{d}t}{\mathrm{d}\lambda}\frac{\mathrm{d}r}{\mathrm{d}\lambda}=0,\label{eq:sv-time-geodesic}\\ \frac{\mathrm{d}^{2}r}{\mathrm{d}\lambda^{2}}+\frac{AA'}{2\gamma}\left(\frac{\mathrm{d}t}{\mathrm{d}\lambda}\right)^{2}-\frac{A'}{2A}\left(\frac{\mathrm{d}r}{\mathrm{d}\lambda}\right)^{2}-\frac{A\Sigma\Sigma'}{\gamma}\left[\left(\frac{\mathrm{d}\theta}{\mathrm{d}\lambda}\right)^{2}+\sin^{2}\theta\left(\frac{\mathrm{d}\phi}{\mathrm{d}\lambda}\right)^{2}\right]=0,\label{eq:sv-radial-geodesic}\\ \frac{\mathrm{d}^{2}\theta}{\mathrm{d}\lambda^{2}}+\frac{2\Sigma'}{\Sigma}\frac{\mathrm{d}r}{\mathrm{d}\lambda}\frac{\mathrm{d}\theta}{\mathrm{d}\lambda}-\sin\theta\cos\theta\left(\frac{\mathrm{d}\phi}{\mathrm{d}\lambda}\right)^{2}=0,\label{eq:sv-polar-geodesic}\\ \frac{\mathrm{d}^{2}\phi}{\mathrm{d}\lambda^{2}}+\frac{2\Sigma'}{\Sigma}\frac{\mathrm{d}r}{\mathrm{d}\lambda}\frac{\mathrm{d}\phi}{\mathrm{d}\lambda}+2\cot\theta\frac{\mathrm{d}\theta}{\mathrm{d}\lambda}\frac{\mathrm{d}\phi}{\mathrm{d}\lambda}=0.\label{eq:sv-azimuthal-geodesic}
\end{gather}

Spherical symmetry allows the orbital plane to be chosen as $\theta=\pi/2$. Initial data with $\mathrm{d}\theta/\mathrm{d}\lambda=0$ preserve this plane. The cyclic coordinates $t$ and $\phi$ supply the conserved energy and angular momentum,
\begin{equation}
\begin{aligned}
E&=A(r)\frac{\mathrm{d}t}{\mathrm{d}\lambda},\qquad L=\Sigma^{2}(r)\frac{\mathrm{d}\phi}{\mathrm{d}\lambda},\\ \gamma\left(\frac{\mathrm{d}r}{\mathrm{d}\lambda}\right)^{2}+V_{\varepsilon}(r)&=E^{2},\qquad V_{\varepsilon}(r)=A(r)\left(\varepsilon+\frac{L^{2}}{\Sigma^{2}(r)}\right).
\end{aligned}
\label{eq:sv-conserved-potential}
\end{equation}

For massive particles, $E$ and $L$ are measured per unit rest mass. We select future directed motion with $E>0$ and orient the nonradial trajectories so that $L>0$. For photons, $b_{\gamma}=L/E$ is invariant under a common affine rescaling and will denote the impact parameter. The potential determines the turning points at specified $(M,a,E,L)$, while $\gamma$ controls the radial rate. An equivalent system, convenient for integration through turning points and up to the horizon, is
\begin{equation}
\frac{\mathrm{d}r}{\mathrm{d}\lambda}=v_{r},\qquad \frac{\mathrm{d}v_{r}}{\mathrm{d}\lambda}=\frac{r}{\gamma}\left(-\frac{\varepsilon M}{\Sigma^{3}}+\frac{L^{2}}{\Sigma^{4}}-\frac{3ML^{2}}{\Sigma^{5}}\right),\qquad \frac{\mathrm{d}\phi}{\mathrm{d}\lambda}=\frac{L}{\Sigma^{2}}.
\label{eq:sv-numerical-system}
\end{equation}

The coefficients in Eq.~\eqref{eq:sv-numerical-system} remain finite at $r_{\mathrm{h}}$. The divergence of the static time equation there reflects the coordinate chart. For $L\ne0$, introducing the inverse areal radius $u_{\mathrm{orb}}=1/R$ gives a common orbital equation for null and timelike motion,
\begin{align}
\gamma\left(\frac{\mathrm{d}u_{\mathrm{orb}}}{\mathrm{d}\phi}\right)^{2}&=\left(1-a^{2}u_{\mathrm{orb}}^{2}\right)\left[\frac{E^{2}-\varepsilon}{L^{2}}+\frac{2M\varepsilon}{L^{2}}u_{\mathrm{orb}}-u_{\mathrm{orb}}^{2}+2Mu_{\mathrm{orb}}^{3}\right],\label{eq:sv-orbital-first-integral}\\ \gamma\frac{\mathrm{d}^{2}u_{\mathrm{orb}}}{\mathrm{d}\phi^{2}}+\left[1+\frac{a^{2}(E^{2}-\varepsilon)}{L^{2}}\right]u_{\mathrm{orb}}&=\frac{M\varepsilon}{L^{2}}+3M\left(1-\frac{\varepsilon a^{2}}{L^{2}}\right)u_{\mathrm{orb}}^{2}+2a^{2}u_{\mathrm{orb}}^{3}-5Ma^{2}u_{\mathrm{orb}}^{4}.\label{eq:sv-orbital-equation}
\end{align}
At fixed $a$ and fixed conserved quantities, the angular rescaling $\psi=\phi/\sqrt{\gamma}$ reduces this equation to that of the undeformed Simpson--Visser metric. For $a\ne0$, the remaining terms depend on the inverse radius and cannot be removed by a constant angular rescaling to Schwarzschild motion. Here, a important remark is worhty to be pointed out: since the physical azimuth retains its period $2\pi$, the factor $\sqrt{\gamma}$ remarkably changes the accumulated angle in a complete orbit.


\subsection{Null trajectories and capture}\label{subsec:sv-geodesic-null-trajectories}

For photons, the radial first integral becomes
\begin{equation}
\gamma\left(\frac{\mathrm{d}r}{\mathrm{d}\lambda}\right)^{2}=E^{2}\left[1-b_{\gamma}^{2}\frac{A(r)}{\Sigma^{2}(r)}\right],\qquad \frac{\mathrm{d}}{\mathrm{d}r}\left(\frac{A}{\Sigma^{2}}\right)=\frac{2r(3M-\Sigma)}{\Sigma^{5}}.
\label{eq:sv-null-radial}
\end{equation}

The exterior maximum occurs at $R_{\mathrm{ph}}=3M$, corresponding to $r_{\mathrm{ph}}=\sqrt{9M^{2}-a^{2}}$. This is the unstable circular null orbit associated with the photon sphere~\cite{Claudel2001,SimpsonVisser2019}. Its critical impact parameter is
\begin{equation}
b_{\mathrm{c}}=\frac{R_{\mathrm{ph}}}{\sqrt{A(r_{\mathrm{ph}})}}=3\sqrt{3}\,M.
\label{eq:sv-critical-impact}
\end{equation}

Thereby, an inward ray launched outside the photon sphere is captured for $b_{\gamma}<b_{\mathrm{c}}$, scattered for $b_{\gamma}>b_{\mathrm{c}}$, and approaches the unstable circular orbit at equality. Notice that the capture here denotes passage through the future exterior horizon. The areal photon sphere radius and the capture threshold are independent of $a$ and $\ell$ at fixed $M$, whereas the coordinate location of the photon sphere changes with $a$. Furthermore, a scattered ray has an outer turning point satisfying
\begin{equation}
b_{\gamma}^{2}=\frac{R_{\min}^{3}}{R_{\min}-2M},\qquad R_{\min}>3M,\qquad r_{\min}=\sqrt{R_{\min}^{2}-a^{2}}.
\label{eq:sv-null-turning}
\end{equation}

To display the trajectories, we first prescribe a common initial coordinate position and spatial coordinate velocities. The future directed time component follows from null normalization. With $\Sigma_{0}=\Sigma(r_{0})$, $A_{0}=A(r_{0})$, $v_{r0}=(\mathrm{d}r/\mathrm{d}\lambda)_{0}<0$, and $\omega_{0}=(\mathrm{d}\phi/\mathrm{d}\lambda)_{0}>0$, the initial integrals are
\begin{equation}
E=\sqrt{\gamma v_{r0}^{2}+A_{0}\Sigma_{0}^{2}\omega_{0}^{2}},\qquad L=\Sigma_{0}^{2}\omega_{0},\qquad \left.\frac{\mathrm{d}t}{\mathrm{d}\lambda}\right|_{0}=\frac{E}{A_{0}}.
\label{eq:sv-null-launch}
\end{equation}

We use $r_{0}=7.7M$, $\phi_{0}=3\pi/2$, $M\omega_{0}=0.022$, $a=1.2M$, and $v_{r0}\in\{-0.18,-0.21,-0.25,-0.31\}$, with $\ell\in\{-0.9,-0.6,-0.3,-0.1,0\}$. A common set of coordinate velocities produces different $E$ and $b_{\gamma}$ when $\ell$ changes. It also changes the local launch direction. Indeed, the angle $\beta_{0}$ between the ray and the inward radial direction, measured in a static orthonormal frame, satisfies
\begin{equation}
\sin\beta_{0}=\frac{b_{\gamma}\sqrt{A_{0}}}{\Sigma_{0}},\qquad \cos\beta_{0}=\frac{\sqrt{\gamma}\,|v_{r0}|}{E}.
\label{eq:sv-local-launch-angle}
\end{equation}

Table~\ref{tab:sv-null-impact} identifies the resulting capture and scattering regimes. Decreasing $\ell$ lowers $E$ at these fixed coordinate launch velocities and increases $b_{\gamma}$. The two smaller inward velocities give scattered rays in every displayed case. Also, the other two families cross the capture threshold, illustrating how the initial null normalization controls the change in the trajectory class.

\begin{table}[tbp]
\centering
\caption{Impact parameters $b_{\gamma}/M$ generated by Eq.~\eqref{eq:sv-null-launch} for $a/M=1.2$, $r_{0}/M=7.7$, and $M\omega_{0}=0.022$. The superscripts $\mathrm{C}$ and $\mathrm{S}$ denote capture and scattering. The common threshold is $b_{\mathrm{c}}/M=3\sqrt{3}\simeq5.1962$.}
\label{tab:sv-null-impact}
\begin{tabular}{cccccc}
\toprule
$|v_{r0}|$ & $\ell=-0.9$ & $\ell=-0.6$ & $\ell=-0.3$ & $\ell=-0.1$ & $\ell=0$ \\
\midrule
$0.18$ & $8.4349^{\mathrm{S}}$ & $7.1610^{\mathrm{S}}$ & $6.3314^{\mathrm{S}}$ & $5.9156^{\mathrm{S}}$ & $5.7362^{\mathrm{S}}$ \\
$0.21$ & $8.2448^{\mathrm{S}}$ & $6.7233^{\mathrm{S}}$ & $5.8189^{\mathrm{S}}$ & $5.3858^{\mathrm{S}}$ & $5.2026^{\mathrm{S}}$ \\
$0.25$ & $7.9703^{\mathrm{S}}$ & $6.1727^{\mathrm{S}}$ & $5.2164^{\mathrm{S}}$ & $4.7808^{\mathrm{C}}$ & $4.6003^{\mathrm{C}}$ \\
$0.31$ & $7.5327^{\mathrm{S}}$ & $5.4413^{\mathrm{S}}$ & $4.4755^{\mathrm{C}}$ & $4.0591^{\mathrm{C}}$ & $3.8902^{\mathrm{C}}$ \\
\bottomrule
\end{tabular}
\end{table}

\begin{figure}[tbp]
\centering
\includegraphics[width=0.58\linewidth]{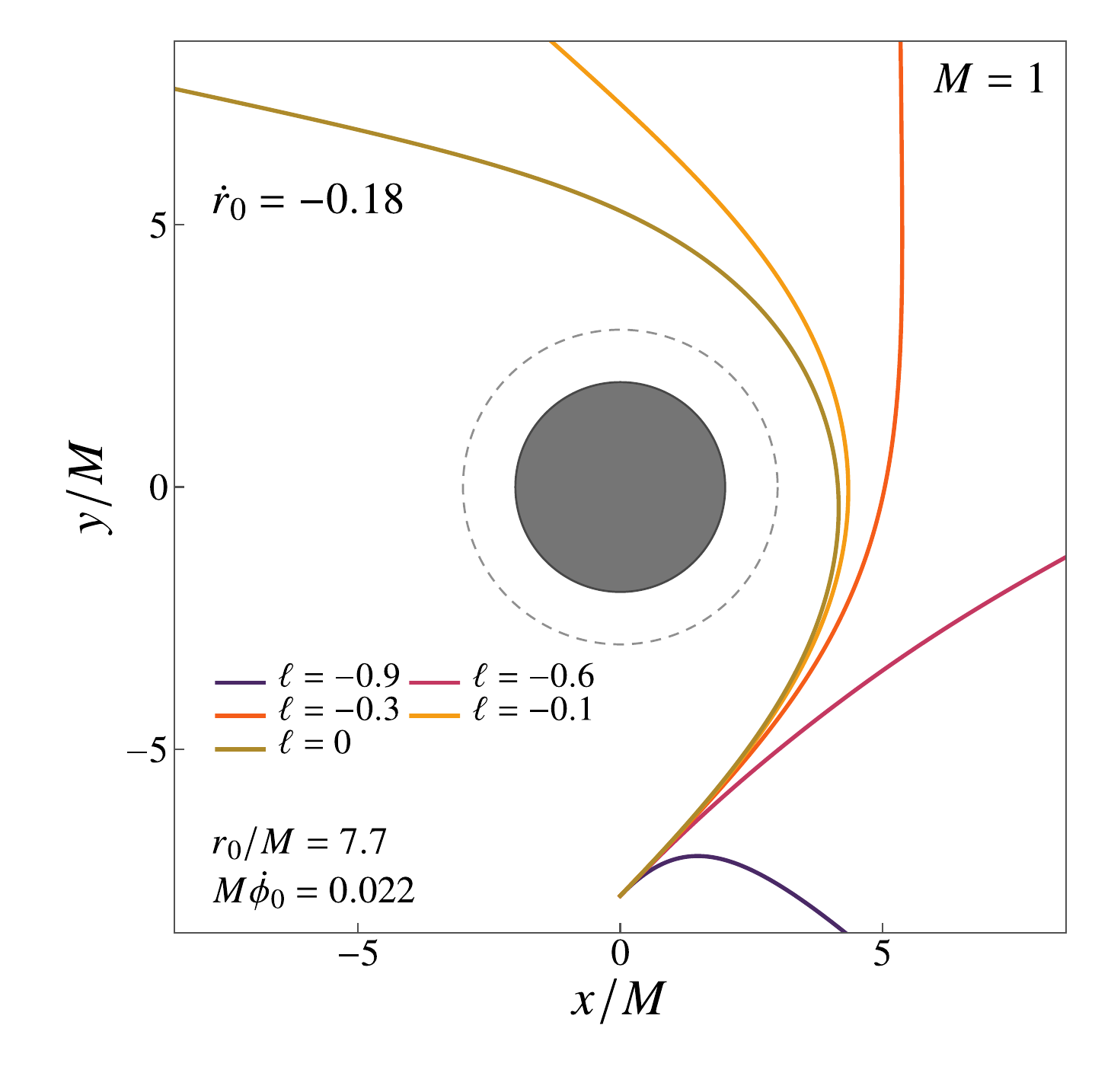}
\caption{Null trajectories with $v_{r0}=-0.18$, $r_{0}=7.7M$, $\phi_{0}=3\pi/2$, $M\omega_{0}=0.022$, and $a=1.2M$. All five values of $\ell$ give scattering. The gray disk and dashed circle have areal radii $2M$ and $3M$, respectively. The view is restricted to $|x|,|y|\leq8.5M$. The impact parameters vary because the coordinate launch velocities are held fixed.}
\label{fig:sv-null018}
\end{figure}

\begin{figure}[tbp]
\centering
\includegraphics[width=0.58\linewidth]{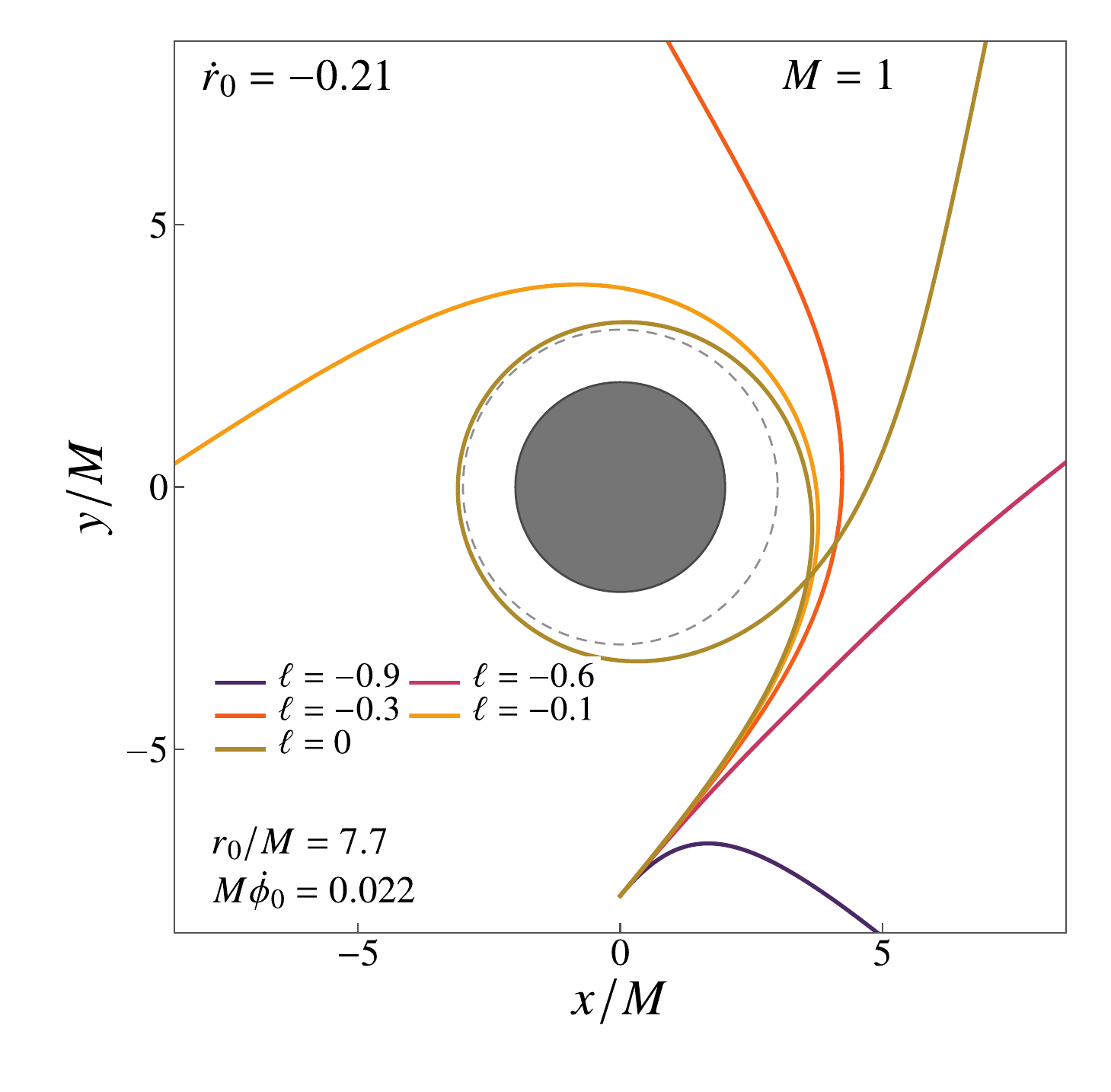}
\caption{Null trajectories with $v_{r0}=-0.21$ and the remaining launch data of Fig.~\ref{fig:sv-null018}. The plotting window is the same as in Fig.~\ref{fig:sv-null018}, and every ray scatters. The $\ell=0$ trajectory has $b_{\gamma}/M\simeq5.2026$, close to the critical value, and accumulates a larger angle near the photon sphere.}
\label{fig:sv-null021}
\end{figure}

\begin{figure}[tbp]
\centering
\includegraphics[width=0.58\linewidth]{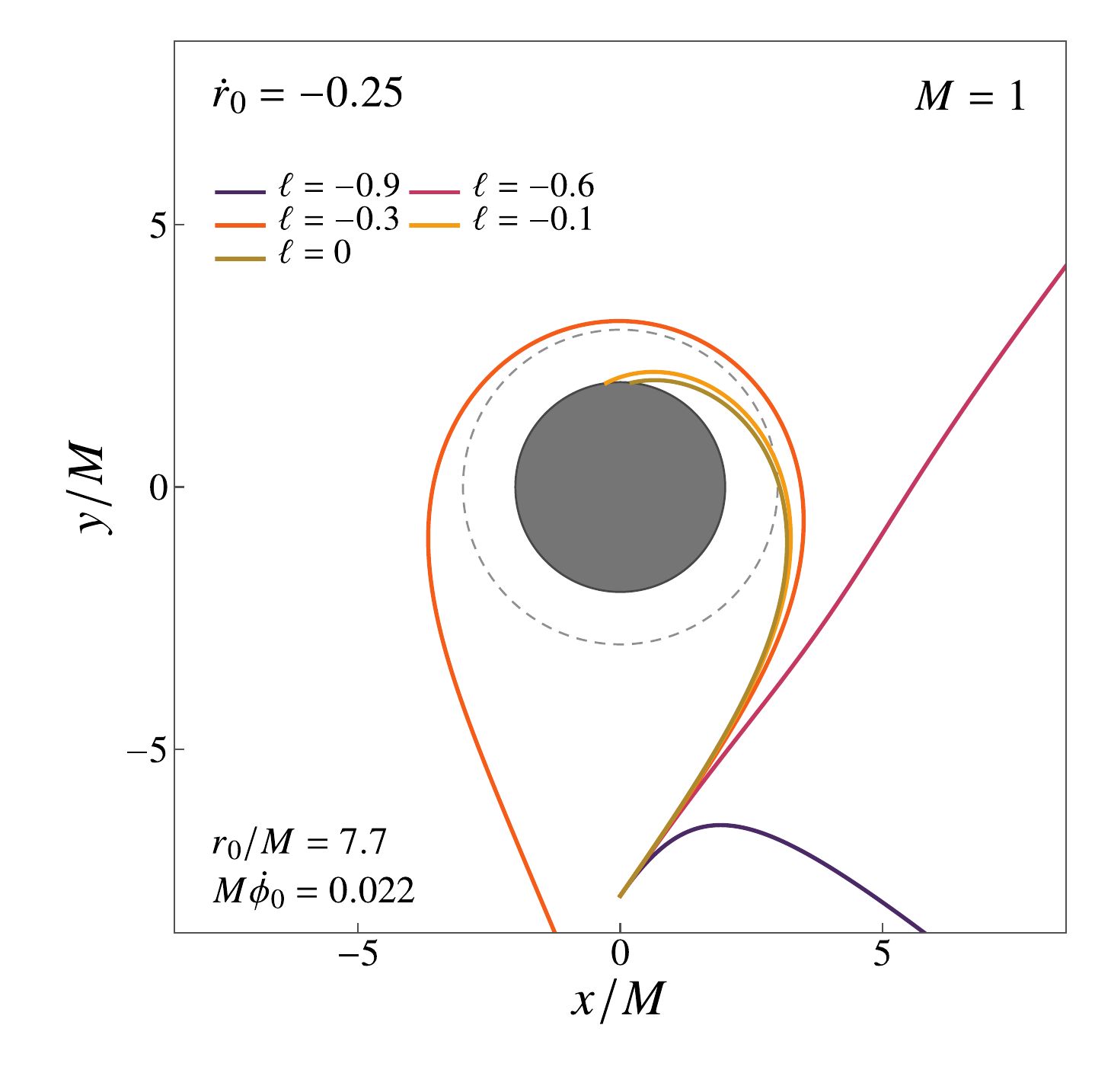}
\caption{Null trajectories with $v_{r0}=-0.25$. The rays with $\ell=-0.9,-0.6,-0.3$ scatter, while those with $\ell=-0.1,0$ are captured. The transition is fixed by the impact parameters in Table~\ref{tab:sv-null-impact}.}
\label{fig:sv-null025}
\end{figure}

\begin{figure}[tbp]
\centering
\includegraphics[width=0.58\linewidth]{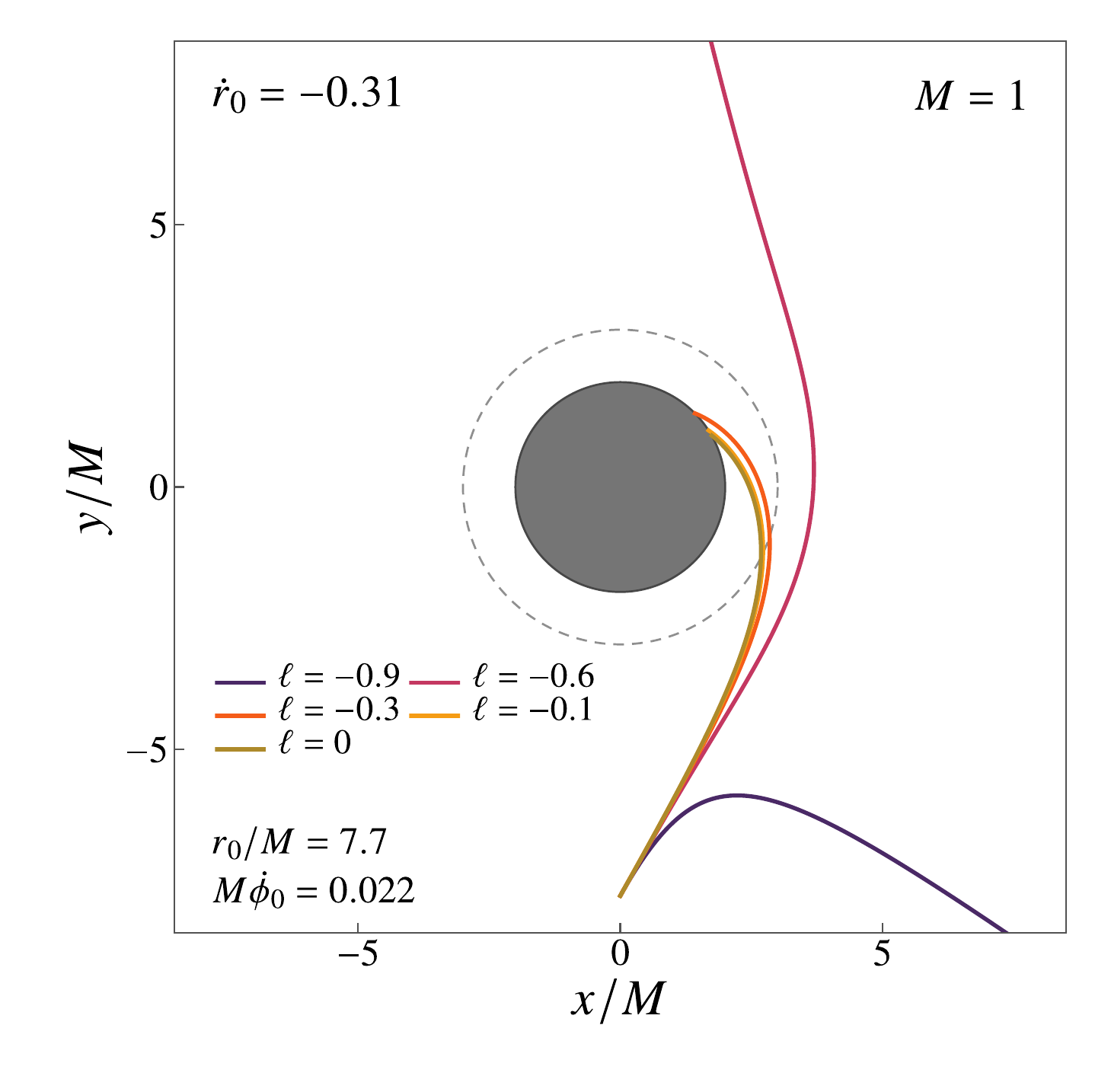}
\caption{Null trajectories with $v_{r0}=-0.31$. The rays with $\ell=-0.9,-0.6$ scatter, whereas $\ell=-0.3,-0.1,0$ give capture. The larger inward coordinate velocity decreases $b_{\gamma}$ in every column of Table~\ref{tab:sv-null-impact}.}
\label{fig:sv-null031}
\end{figure}

Figs.~\ref{fig:sv-null018}--\ref{fig:sv-null031} show the four trajectory families individually. We use $x=R\cos\phi$ and $y=R\sin\phi$ to represent the areal radius and the physical azimuth in a plane. The gray disk marks $R_{\mathrm{h}}=2M$, and the dashed circle marks $R_{\mathrm{ph}}=3M$. This is a coordinate representation: proper radial lengths in the exterior follow from $\sqrt{\gamma}\,\mathrm{d}r/\sqrt{A}$, and the displayed plane does not constitute an isometric embedding. Captured rays terminate at the horizon, while scattered rays are followed until they return to the chosen outer sphere. Also, the conservation of Eq.~\eqref{eq:sv-conserved-potential} is monitored throughout the integration. The plotting windows emphasize the passage through the central region; the integrations retain the complete exterior paths up to their stated endpoints.

We now fix the incident ray by taking $E=1$ and $b_{\gamma}=6M$, and launch it at the common areal radius $R_{0}=20M$ with $\phi_{0}=3\pi/2$. The corresponding coordinate radius is $r_{0}=\sqrt{R_{0}^{2}-a^{2}}$, and the inward radial velocity becomes
\begin{equation}
v_{r0}=-\frac{1}{\sqrt{\gamma}}\sqrt{1-b_{\gamma}^{2}\frac{1-2M/R_{0}}{R_{0}^{2}}}.
\label{eq:sv-fixed-impact-launch}
\end{equation}

\begin{figure}[tbp]
\centering
\includegraphics[width=\linewidth]{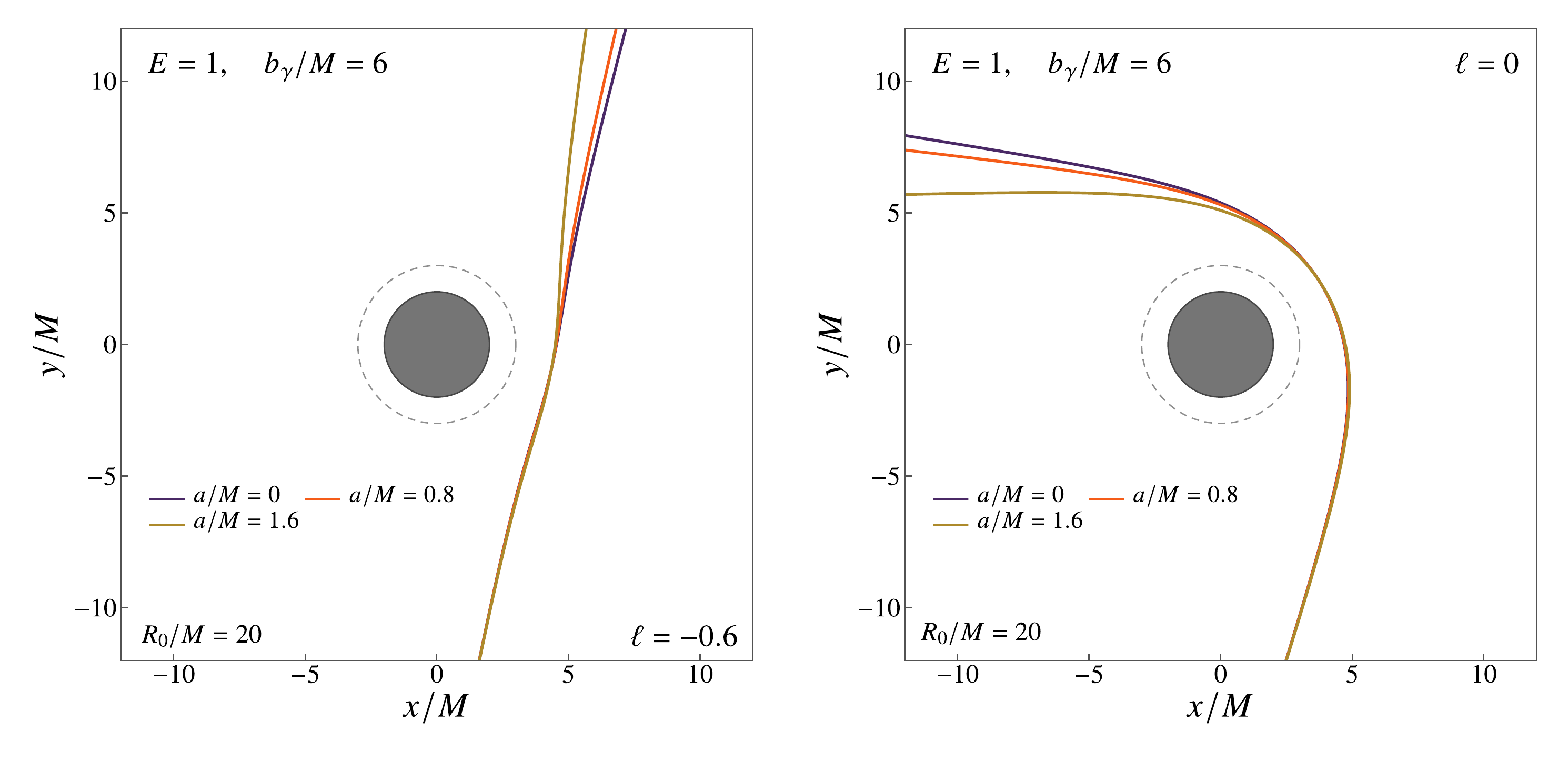}
\caption{Scattered null trajectories at fixed $E=1$, $b_{\gamma}=6M$, $R_{0}=20M$, and $\phi_{0}=3\pi/2$. The panels compare $\ell=-0.6$ and $\ell=0$ for $a/M=0,0.8,1.6$. All rays have the same local launch angle and the same areal closest approach, $R_{\min}\simeq4.4534M$. The factor $\sqrt{\gamma}$ reduces the angular sweep for negative $\ell$, while increasing $a$ enhances it. The central portion of each path is displayed within $|x|,|y|\leq12M$; the integrations extend to the common outer sphere at $R=20M$.}
\label{fig:sv-fixed-null}
\end{figure}

These data give the same local launch angle and the same areal turning point for every displayed geometry. Equation~\eqref{eq:sv-null-turning} yields $R_{\min}\simeq4.4534M$, while $r_{\min}$ depends on $a$. Figure~\ref{fig:sv-fixed-null} separates the influence of the two deformations. At fixed $a$, the angular sweep is proportional to $\sqrt{\gamma}$. At fixed $\gamma$, increasing $a$ increases the sweep through the radial conversion between $r$ and $R$.

\subsection{Weak and strong angular deflection}\label{subsec:sv-geodesic-deflection}

The transition from weak bending to long excursions near the photon sphere follows from the exact azimuthal integral~\cite{Perlick2004,Tsukamoto2021}. Since $\mathrm{d}r=\mathrm{d}R/\sqrt{1-a^{2}/R^{2}}$ in the right exterior, the complete angular sweep of a reflected ray arriving from and returning to infinity is
\begin{equation}
\Delta\phi(b_{\gamma})=2\sqrt{\gamma}\int_{R_{\min}}^{\infty}\frac{b_{\gamma}\,\mathrm{d}R}{R^{2}\sqrt{1-a^{2}/R^{2}}\sqrt{1-b_{\gamma}^{2}(1-2M/R)/R^{2}}}.
\label{eq:sv-exact-null-sweep}
\end{equation}

The reference angle must account for the asymptotic spatial geometry. At large radius, the equatorial spatial metric approaches $\gamma\,\mathrm{d}R^{2}+R^{2}\mathrm{d}\phi^{2}$, whose locally unfolded azimuth is $\psi=\phi/\sqrt{\gamma}$. A complete straight ray in this asymptotic cone sweeps $\pi\sqrt{\gamma}$ in the physical azimuth. We define in this sense the coordinate angular excess by
\begin{equation}
\alpha_{\mathrm{cone}}(b_{\gamma})=\Delta\phi(b_{\gamma})-\pi\sqrt{\gamma},\qquad \Delta\phi-\pi=\pi\left(\sqrt{\gamma}-1\right)+\alpha_{\mathrm{cone}},\qquad \Delta\psi-\pi=\frac{\alpha_{\mathrm{cone}}}{\sqrt{\gamma}}.
\label{eq:sv-cone-reference}
\end{equation}
This quantity measures accumulated azimuth relative to the asymptotic cone. A lensing angle measured by a specified source and observer also involves their local ray directions and positions~\cite{Perlick2004,Ishihara2016}. A static observer at any exterior radius measures $\sin\beta=b_{\gamma}\sqrt{A}/R$. Keeping this distinction explicit is particularly useful when $\gamma\ne1$, since an angular offset relative to $\pi$ is already present in the asymptotic reference geometry.

For $b_{\gamma}\gg M,a$, expansion of Eq.~\eqref{eq:sv-exact-null-sweep} gives
\begin{equation}
\alpha_{\mathrm{cone}}=\sqrt{\gamma}\left[\frac{4M}{b_{\gamma}}+\frac{\pi(15M^{2}+a^{2})}{4b_{\gamma}^{2}}+\frac{128M^{3}+8Ma^{2}}{3b_{\gamma}^{3}}+\mathcal{O}\!\left(\frac{M^{4}+M^{2}a^{2}+a^{4}}{b_{\gamma}^{4}}\right)\right].
\label{eq:sv-weak-bending}
\end{equation}

The $a=0$ coefficients reproduce the Schwarzschild expansion multiplied by $\sqrt{\gamma}$~\cite{IyerPetters2007,Casana2018}. The leading bounce correction is quadratic in $a$ and adds to the angular excess. There is an additional distinction between the asymptotic cone and the complete zero mass geometry: setting $M=0$ at finite $a$ leaves a curved ultrastatic wormhole~\cite{SimpsonVisser2019}. For a reflected ray with $b_{\gamma}>a$, its exact sweep is
\begin{equation}
\Delta\phi_{0}(b_{\gamma})=2\sqrt{\gamma}\,K\!\left(\frac{a^{2}}{b_{\gamma}^{2}}\right),\qquad K(m)=\int_{0}^{\pi/2}\frac{\mathrm{d}\zeta}{\sqrt{1-m\sin^{2}\zeta}},\qquad \alpha_{0}=\Delta\phi-\Delta\phi_{0}.
\label{eq:sv-zero-mass-reference}
\end{equation}
$\alpha_{0}$ subtracts the full bending of the same zero mass bounce geometry. In its weak expansion, the pure $\pi a^{2}/(4b_{\gamma}^{2})$ contribution in Eq.~\eqref{eq:sv-weak-bending} cancels, while the terms containing $M$ remain. The figures use $\alpha_{\mathrm{cone}}$, so that the mass and bounce contributions are maintained relative to one common asymptotic angular reference.

\begin{figure}[tbp]
\centering
\includegraphics[width=1\linewidth]{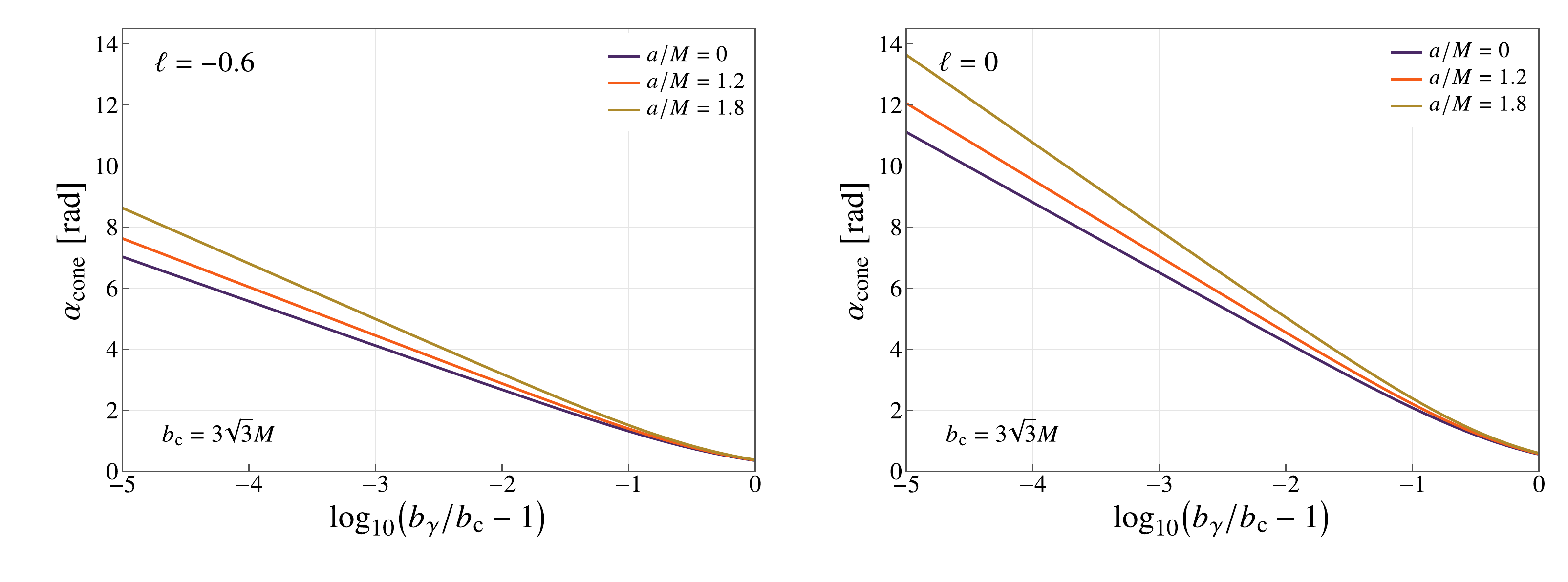}
\caption{Coordinate angular excess $\alpha_{\mathrm{cone}}$ from Eq.~\eqref{eq:sv-exact-null-sweep}, plotted against $\log_{10}\delta$, where $\delta=b_{\gamma}/b_{\mathrm{c}}-1$. The panels compare $\ell=-0.6$ and $\ell=0$ for $a/M=0,1.2,1.8$. The reference is the asymptotic cone. Near capture, the logarithmic coefficient is $\overline{a}=\sqrt{\gamma}/\sqrt{1-a^{2}/(9M^{2})}$.}
\label{fig:sv-bending}
\end{figure}

Close to the capture threshold, let $\delta=b_{\gamma}/b_{\mathrm{c}}-1>0$ and introduce $q_{a}=a^{2}/(9M^{2})$. The nondegenerate exterior maximum produces the usual logarithmic strong deflection behavior~\cite{Bozza2002,Tsukamoto2017,Tsukamoto2021},
\begin{equation}
\begin{aligned}
\alpha_{\mathrm{cone}}&=-\overline{a}\ln\delta+\overline{b}_{\mathrm{cone}}+\mathcal{O}(\delta\ln\delta),\\ \overline{a}&=\frac{\sqrt{\gamma}}{\sqrt{1-q_{a}}},\qquad \overline{b}_{\mathrm{cone}}=\sqrt{\gamma}\left[\mathcal{J}(q_{a})+\frac{\ln6}{\sqrt{1-q_{a}}}-\pi\right],
\end{aligned}
\label{eq:sv-strong-bending}
\end{equation}
where the finite regular integral is
\begin{equation}
\mathcal{J}(q_{a})=\int_{0}^{1}\left[\frac{2\sqrt{3}}{\sqrt{1+2z}\sqrt{1-q_{a}z^{2}}}-\frac{2}{\sqrt{1-q_{a}}}\right]\frac{\mathrm{d}z}{1-z}.
\label{eq:sv-strong-regular-integral}
\end{equation}
The two terms cancel at $z=1$, making the integral convergent throughout the black--bounce segment. At $a=0$, $\mathcal{J}(0)=\ln[36(7-4\sqrt{3})]$, and Eq.~\eqref{eq:sv-strong-bending} reduces to $\alpha_{\mathrm{cone}}=\sqrt{\gamma}\{-\ln\delta+\ln[216(7-4\sqrt{3})]-\pi\}+\mathcal{O}(\delta\ln\delta)$. At $\ell=0$, the coefficients recover the Simpson--Visser result~\cite{Tsukamoto2021}. Subtracting the complete zero mass geometry instead changes only the finite constant, giving $\overline{b}_{0}=\overline{b}_{\mathrm{cone}}-\sqrt{\gamma}[2K(q_{a}/3)-\pi]$. Figure~\ref{fig:sv-bending} shows the exact integral across the strong and moderate bending regimes. Increasing $a$ enhances the logarithmic coefficient, whereas negative $\ell$ lowers it at a prescribed $a/M$.

\subsection{Photon orbit instability}\label{subsec:sv-geodesic-photon-instability}

The location of the photon sphere and the rate at which a nearby ray leaves it encode different aspects of the geometry. In this view, by linearizing the null radial equation about $r_{\mathrm{ph}}$ at the critical conserved quantities gives
\begin{equation}
\frac{\mathrm{d}^{2}\delta r}{\mathrm{d}\lambda^{2}}=\frac{E^{2}}{3\gamma M^{2}}\left(1-\frac{a^{2}}{9M^{2}}\right)\delta r.
\label{eq:sv-affine-instability}
\end{equation}

On the circular orbit, $\mathrm{d}t/\mathrm{d}\lambda=3E$. Moreover, the instability exponent in normalized Killing time and the orbital angular frequency therefore take the form~\cite{Cardoso2009}
\begin{equation}
\delta r\propto\exp(\lambda_{t}t),\qquad \lambda_{t}=\frac{\sqrt{1-a^{2}/(9M^{2})}}{3\sqrt{3}\,M\sqrt{\gamma}},\qquad \Omega_{\mathrm{ph}}=\left.\frac{\mathrm{d}\phi}{\mathrm{d}t}\right|_{\mathrm{ph}}=\frac{1}{3\sqrt{3}\,M},\qquad \frac{\Omega_{\mathrm{ph}}}{\lambda_{t}}=\overline{a}.
\label{eq:sv-photon-frequencies}
\end{equation}

\begin{figure}[tbp]
\centering
\includegraphics[width=0.6\linewidth]{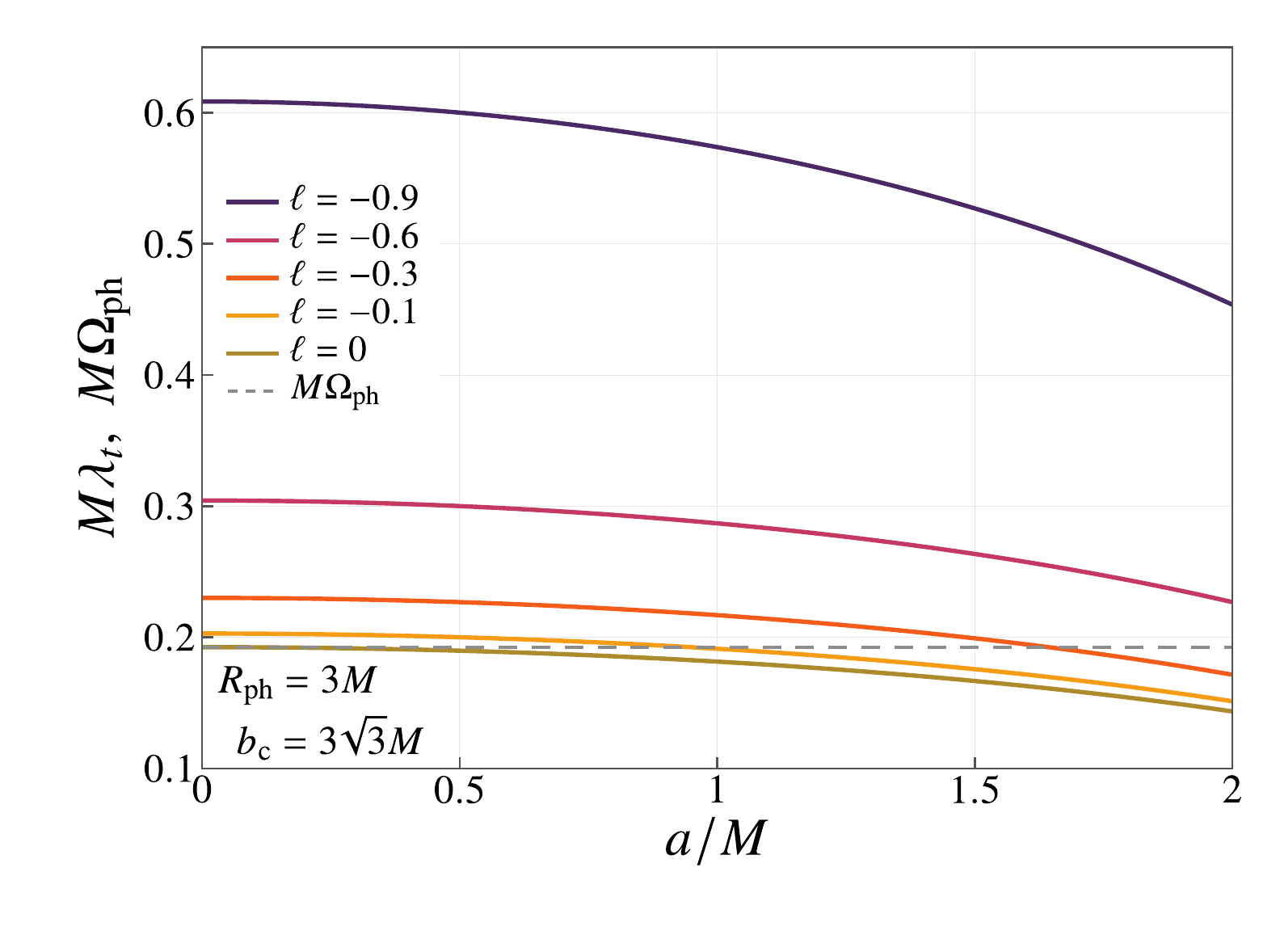}
\caption{Dimensionless photon orbit instability exponent $M\lambda_{t}$ as a function of $a/M$ at fixed lapse scale, for the indicated values of $\ell$. The dashed horizontal line gives $M\Omega_{\mathrm{ph}}=1/(3\sqrt{3})$. The instability decreases as $a$ grows and increases as $\ell$ becomes more negative, although the areal photon sphere radius and $b_{\mathrm{c}}/M$ remain unchanged. The endpoints $a/M=0$ and $a/M=2$ are shown as limiting geometries.}
\label{fig:sv-instability}
\end{figure}

The growth factor during one complete azimuthal revolution is $\exp(2\pi\lambda_{t}/\Omega_{\mathrm{ph}})=\exp(2\pi/\overline{a})$. In this regard, a larger $a$ increases the residence time near the circular null orbit, while decreasing $\ell$ shortens it. The unchanged orbital frequency and the modified instability exponent explain how the same capture threshold can coexist with different angular excursions. Figure~\ref{fig:sv-instability} displays this dependence. The equality $\Omega_{\mathrm{ph}}/\lambda_{t}=\overline{a}$ also connects the local instability to the strong deflection logarithm.

\subsection{Circular timelike orbits}\label{subsec:sv-geodesic-circular-orbits}

For massive particles, the effective potential is $V_{1}=A(1+L^{2}/R^{2})$. A circular orbit at $R=R_{\mathrm{c}}$ satisfies $V_{1}=E_{\mathrm{c}}^{2}$ and $\mathrm{d}V_{1}/\mathrm{d}r=0$. Since $\mathrm{d}R/\mathrm{d}r\ne0$ in the exterior, these conditions give
\begin{equation}
E_{\mathrm{c}}^{2}=\frac{(R_{\mathrm{c}}-2M)^{2}}{R_{\mathrm{c}}(R_{\mathrm{c}}-3M)},\qquad L_{\mathrm{c}}^{2}=\frac{MR_{\mathrm{c}}^{2}}{R_{\mathrm{c}}-3M},\qquad R_{\mathrm{c}}>3M.
\label{eq:sv-timelike-circular-integrals}
\end{equation}

The circular energy and angular momentum retain their Schwarzschild expressions when written in the areal radius~\cite{Darwin1959,SimpsonVisser2019}. Their radial stability follows from
\begin{equation}
\left.\frac{\mathrm{d}^{2}V_{1}}{\mathrm{d}r^{2}}\right|_{\mathrm{c}}=\left(1-\frac{a^{2}}{R_{\mathrm{c}}^{2}}\right)\frac{2M(R_{\mathrm{c}}-6M)}{R_{\mathrm{c}}^{3}(R_{\mathrm{c}}-3M)},\qquad R_{\mathrm{ISCO}}=6M,\qquad R_{\mathrm{mb}}=4M.
\label{eq:sv-circular-stability}
\end{equation}

Notice that $R_{\mathrm{c}}>6M$ accounts for stable circular motion, whereas $3M<R_{\mathrm{c}}<6M$ gives radial instability. The marginally bound orbit has $E_{\mathrm{c}}=1$ at $R_{\mathrm{mb}}=4M$. In the original radial coordinate, these radii are $r_{\mathrm{ISCO}}=\sqrt{36M^{2}-a^{2}}$ and $r_{\mathrm{mb}}=\sqrt{16M^{2}-a^{2}}$. Their change with $a$ reflects the relation between the coordinate and areal radii, while the physical circumferences remain fixed at a given $M$.

For a small radial oscillation around a stable circular orbit, linearization of Eq.~\eqref{eq:sv-numerical-system} and conversion to Killing time yield the epicyclic frequency~\cite{Biesiada2003},
\begin{equation}
\Omega_{\mathrm{c}}^{2}=\frac{M}{R_{\mathrm{c}}^{3}},\qquad \kappa_{t}^{2}=\frac{M(R_{\mathrm{c}}-6M)}{\gamma R_{\mathrm{c}}^{4}}\left(1-\frac{a^{2}}{R_{\mathrm{c}}^{2}}\right),\qquad \frac{\Omega_{\mathrm{c}}}{\kappa_{t}}=\frac{\sqrt{\gamma}}{\sqrt{1-a^{2}/R_{\mathrm{c}}^{2}}\sqrt{1-6M/R_{\mathrm{c}}}}.
\label{eq:sv-epicyclic-frequencies}
\end{equation}
The azimuthal frequency is unchanged at fixed areal radius. Both the bounce parameter and the bumblebee deformation enter the radial frequency. In the small eccentricity limit, the azimuthal angle accumulated between successive periapses is $2\pi\Omega_{\mathrm{c}}/\kappa_{t}$, and the coordinate periapsis advance relative to one full revolution is $2\pi(\Omega_{\mathrm{c}}/\kappa_{t}-1)$.


\subsection{Bound orbits and periapsis advance}\label{subsec:sv-geodesic-bound-orbits}

For finite eccentricity, an exact description of the bound exterior motion follows from a radial anomaly $\chi$, introduced in the areal radius as in the Schwarzschild construction~\cite{Darwin1959,Cutler1994,Hopper2016}:
\begin{equation}
\begin{aligned}
\frac{R(\chi)}{M}&=\frac{p}{1+e\cos\chi},\qquad r(\chi)=\sqrt{R^{2}(\chi)-a^{2}},\\ E^{2}&=\frac{(p-2)^{2}-4e^{2}}{p(p-3-e^{2})},\qquad \frac{L^{2}}{M^{2}}=\frac{p^{2}}{p-3-e^{2}}.
\end{aligned}
\label{eq:sv-bound-anomaly}
\end{equation}
Here, $0<e<1$ and $p>6+2e$ describe a noncircular stable bound orbit; the limit $e\to0$ recovers small radial oscillations. Notice that the dimensionless semilatus rectum is $p$, while the periapsis and apoapsis are $R_{\mathrm{p}}=pM/(1+e)$ and $R_{\mathrm{a}}=pM/(1-e)$. Their coordinate values follow from $r_{\mathrm{p,a}}=\sqrt{R_{\mathrm{p,a}}^{2}-a^{2}}$. The constants in Eq.~\eqref{eq:sv-bound-anomaly} depend only on these areal turning points. Substitution into the radial and angular first integrals gives
\begin{equation}
\frac{\mathrm{d}\phi}{\mathrm{d}\chi}=\frac{\sqrt{\gamma}\sqrt{p}}{\sqrt{p-6-2e\cos\chi}\sqrt{1-\dfrac{a^{2}(1+e\cos\chi)^{2}}{p^{2}M^{2}}}},\qquad \Phi_{r}=\int_{0}^{2\pi}\frac{\mathrm{d}\phi}{\mathrm{d}\chi}\,\mathrm{d}\chi,\qquad \Delta\phi_{\mathrm{p}}=\Phi_{r}-2\pi.
\label{eq:sv-bound-angular-period}
\end{equation}

At fixed $(p,e,a/M)$, $\Phi_{r}$ is proportional to $\sqrt{\gamma}$. The factor containing $a$ varies around the orbit and increases the accumulated azimuth at a prescribed $\gamma$. In particular, the complete elliptic integral expression of Schwarzschild motion is recovered only at $a=0$,
\begin{equation}
\left.\Phi_{r}\right|_{a=0}=4\sqrt{\gamma}\sqrt{\frac{p}{p-6+2e}}\,K\!\left(\frac{4e}{p-6+2e}\right).
\label{eq:sv-bound-schwarzschild-limit}
\end{equation}

The separatrix remains $p=6+2e$ throughout the black--bounce part because $R$ is monotonic and $1-a^{2}/R^{2}>0$ in the exterior. As it is approached from above, the orbit spends an increasing part of each radial period near an unstable circular trajectory, producing the familiar zoom--whirl behavior~\cite{Cutler1994,LevinPerezGiz2008}. In other words, the bounce and bumblebee factors modify the angular accumulation without shifting this boundary in $(p,e)$.

Moreover, in the weak field limit $p\gg1$ at fixed $a/M$ and $e$, Eq.~\eqref{eq:sv-bound-angular-period} yields
\begin{equation}
\Delta\phi_{\mathrm{p}}=2\pi\left(\sqrt{\gamma}-1\right)+\frac{6\pi\sqrt{\gamma}}{p}+\frac{\pi\sqrt{\gamma}}{2p^{2}}\left[3(18+e^{2})+\frac{a^{2}}{M^{2}}(2+e^{2})\right]+\mathcal{O}(p^{-3}).
\label{eq:sv-weak-periapsis}
\end{equation}
The first term follows from the angular periodicity of the asymptotic cone. The second is the leading relativistic contribution, and the leading bounce correction enters at order $p^{-2}$. At negative $\ell$, the conical contribution can make the total coordinate advance negative for sufficiently distant orbits, even though the mass and bounce corrections are positive. An additional important point is that the excess relative to the asymptotic conical period is instead $\Phi_{r}-2\pi\sqrt{\gamma}$.

Following this, the coordinate radial period can also be obtained without integrating the time equation through radial turning points. Defining $H_{\chi}=1-a^{2}(1+e\cos\chi)^{2}/(p^{2}M^{2})$, we find
\begin{equation}
\frac{\mathrm{d}t}{\mathrm{d}\chi}=\frac{M\sqrt{\gamma}\,p^{2}\sqrt{(p-2)^{2}-4e^{2}}}{(1+e\cos\chi)^{2}(p-2-2e\cos\chi)\sqrt{p-6-2e\cos\chi}\sqrt{H_{\chi}}},\qquad T_{r}=\int_{0}^{2\pi}\frac{\mathrm{d}t}{\mathrm{d}\chi}\,\mathrm{d}\chi.
\label{eq:sv-bound-time-period}
\end{equation}

\begin{figure}[tbp]
\centering
\includegraphics[width=\linewidth]{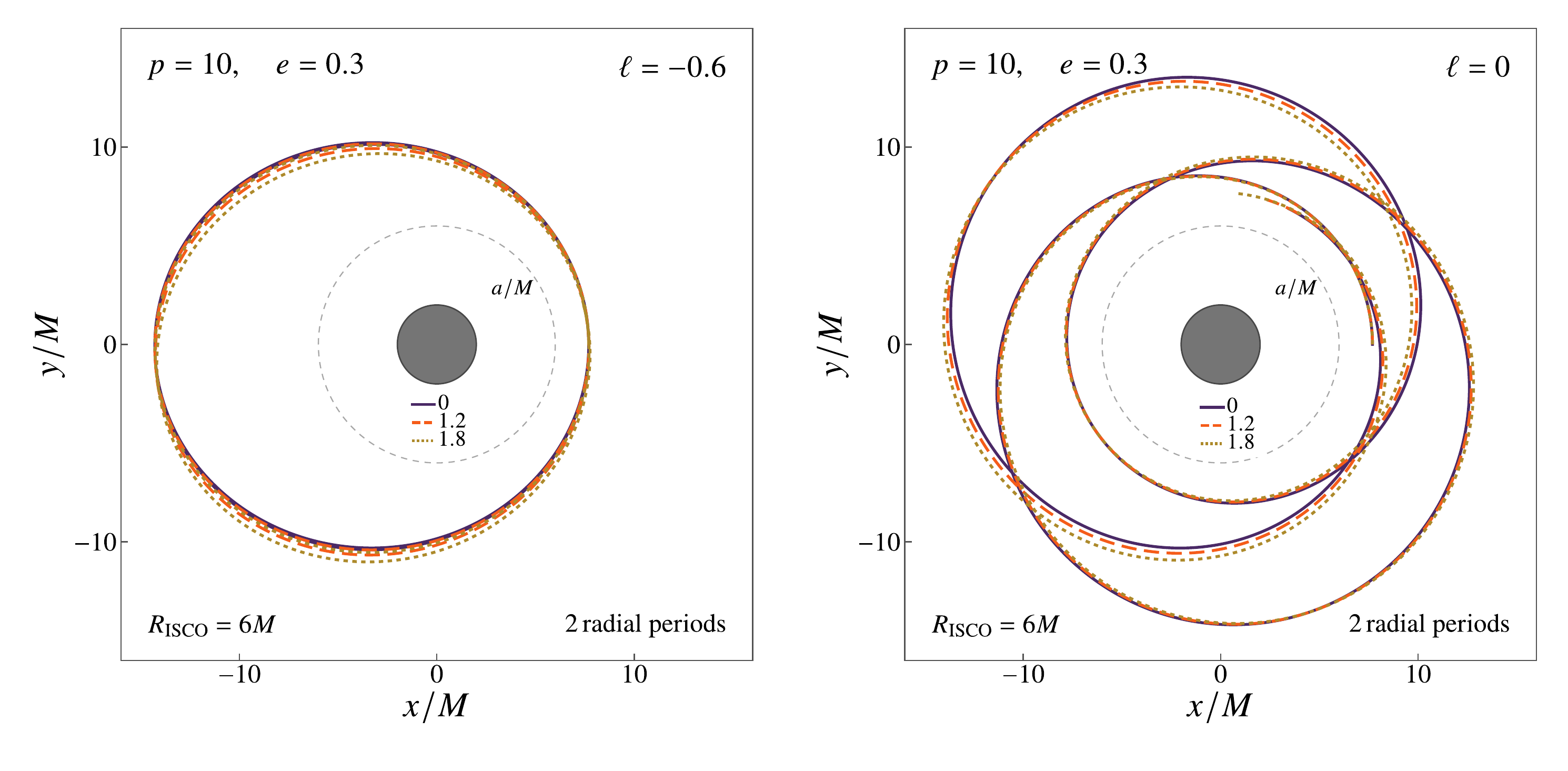}
\caption{Bound timelike trajectories over two radial periods, with $p=10$, $e=0.3$, and $\phi(0)=0$. The panels compare $\ell=-0.6$ and $\ell=0$ for $a/M=0,1.2,1.8$. All curves share $R_{\mathrm{p}}=10M/1.3$, $R_{\mathrm{a}}=10M/0.7$, $E$, and $L$. The gray disk marks $R_{\mathrm{h}}=2M$, and the dashed circle marks $R_{\mathrm{ISCO}}=6M$. The plotted coordinates are $x=R\cos\phi$ and $y=R\sin\phi$.}
\label{fig:sv-bound-orbits}
\end{figure}

The fundamental coordinate time frequencies are $\Omega_{r}=2\pi/T_{r}$ and $\Omega_{\phi}=\Phi_{r}/T_{r}$. At fixed $(M,p,e,a/M)$, both $T_{r}$ and $\Phi_{r}$ scale with $\sqrt{\gamma}$; then, $\Omega_{r}\propto\gamma^{-1/2}$ while the mean azimuthal frequency $\Omega_{\phi}$ is independent of $\gamma$. Their circular limits reproduce Eq.~\eqref{eq:sv-epicyclic-frequencies}. Figure~\ref{fig:sv-bound-orbits} displays two radial periods for $p=10$ and $e=0.3$. All trajectories have the same areal turning points, energy, and angular momentum, while their accumulated azimuth depends on both deformations. For these turning radii, the bounce correction is modest, so the trajectories remain close while their periapsis directions gradually separate.

The null and timelike trajectories distinguish the areal locations of critical orbits from their angular and temporal properties. At fixed lapse scale, the capture threshold, the photon sphere radius, and the areal marginally bound and marginally stable radii retain their Schwarzschild values. The bending coefficients, photon orbit instability, radial epicyclic frequency, and bound orbit periods contain the additional information carried by $a$ and $\ell$. In summary, the stability statements established here concern test particle orbits; the stability of the reconstructed supporting fields is determined by their coupled perturbation equations.


\section{Apparent size and observational features}\label{sec:sv-shadow}

The null geodesic analysis in Sec.~\ref{sec:sv-geodesic-motion} gives an areal photon sphere radius $R_{\mathrm{ph}}=3M$ and a capture threshold $b_{\mathrm{c}}=3\sqrt{3}M$, both independent of $a$ and $\ell$ at fixed normalized lapse mass $M$. We use these results to determine the apparent boundary seen by a static observer. In same section, the shadow refers to an independent, minimally coupled Maxwell probe in geometric optics.


\subsection{Shadow measured by a static observer}\label{subsec:sv-shadow-static-observer}

\begin{figure}[tbp]
\centering
\includegraphics[width=1.05\linewidth]{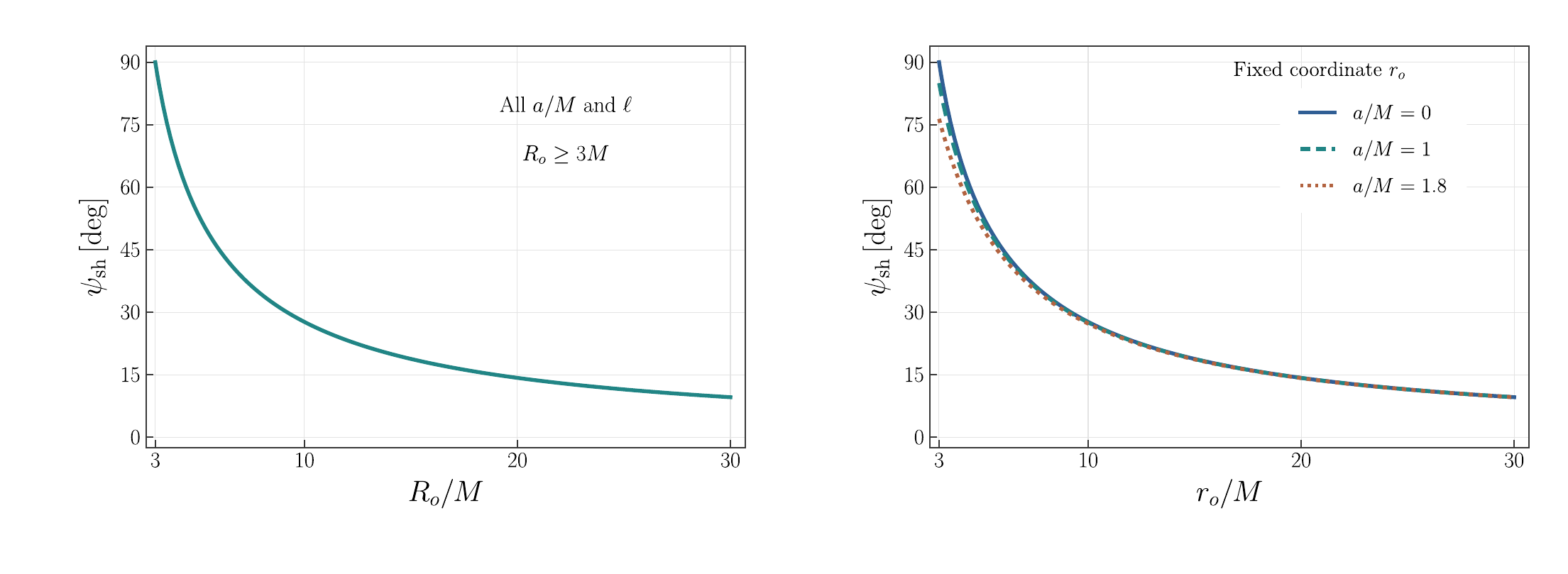}
\caption{Angular half radius of the metric shadow for static observers outside the photon sphere. The left panel uses $R_o/M$ and gives a common profile for all admissible $a$ and $\ell$. The right panel holds $r_o/M$ fixed and compares $a/M=0,1,1.8$; the separation arises from $R_o=\sqrt{r_o^{2}+a^{2}}$. Every curve is independent of $\ell$.}
\label{fig:sv-shadow-finite-distance}
\end{figure}

Let a static observer be located at $R_o=\sqrt{r_o^{2}+a^{2}}>3M$, with $A_o=1-2M/R_o$. The local angle relation derived in Sec.~\ref{sec:sv-geodesic-motion} gives the shadow half angle upon setting the photon impact parameter to $b_{\mathrm{c}}$~\cite{PerlickTsupko2022}:
\begin{equation}
\sin\psi_{\mathrm{sh}}=\frac{b_{\mathrm{c}}\sqrt{A_o}}{R_o},\qquad \psi_{\mathrm{sh}}=\arcsin\!\left[\frac{3\sqrt{3}M}{R_o}\sqrt{1-\frac{2M}{R_o}}\right],\qquad R_o>3M.
\label{eq:sv-shadow-half-angle}
\end{equation}
At fixed areal position $R_o$, neither deformation changes the angle. It approaches $\pi/2$ at the photon sphere and decreases outward. At fixed original coordinate position $r_o$, changing $a$ changes $R_o$ and hence the measured angle, as shown in Fig.~\ref{fig:sv-shadow-finite-distance} and Table~\ref{tab:sv-shadow-finite-values}. 

\begin{table}[tbp]
\centering
\caption{Finite observer comparison at fixed coordinate position $r_o=10M$. Here, $R_{\mathrm{sh}}=R_o\sin\psi_{\mathrm{sh}}$ is the radius on the areal screen. The entries are independent of $\ell$; $a=0$ is the limiting exterior reference.}
\label{tab:sv-shadow-finite-values}
\begin{tabular}{cccc}
\toprule
$a/M$ & $R_o/M$ & $\psi_{\mathrm{sh}}$ [degrees] & $R_{\mathrm{sh}}/M$ \\
\midrule
$0$ & $10.000000$ & $27.694561$ & $4.647580$ \\
$1$ & $10.049876$ & $27.563946$ & $4.650462$ \\
$1.8$ & $10.160709$ & $27.278142$ & $4.656760$ \\
\bottomrule
\end{tabular}
\end{table}

\begin{figure}[tbp]
\centering
\includegraphics[width=0.6\linewidth]{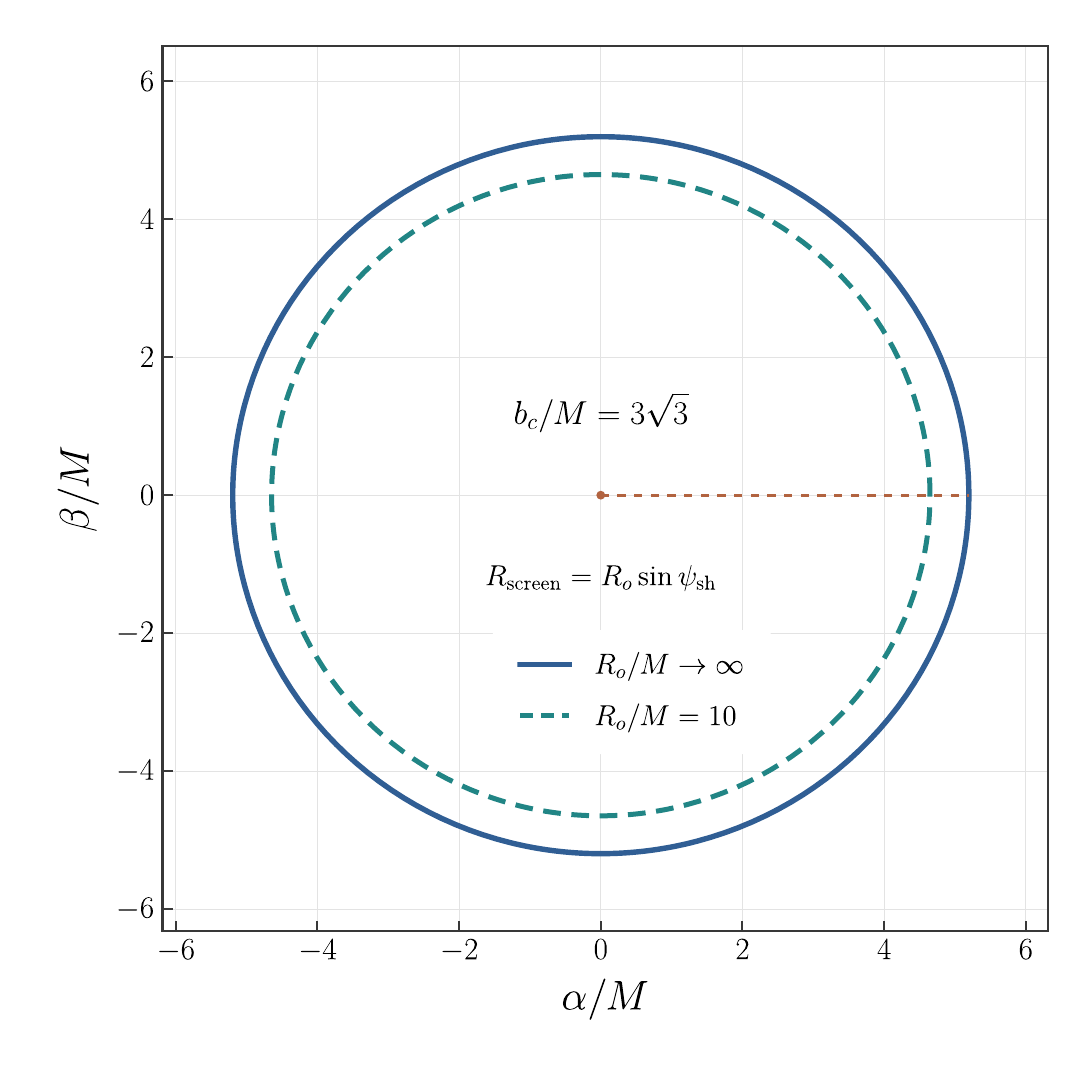}
\caption{Circular capture boundary on the areal screen, in units of $M$. The distant observer radius is $3\sqrt{3}$, and the finite screen at $R_o=10M$ has radius $3\sqrt{3}\sqrt{0.8}$. Each contour is common to the displayed metric black hole domain at the stated areal observer position.}
\label{fig:sv-shadow-circle}
\end{figure}

Furthermore, for an orthographic areal screen with azimuth $\chi$, define $\alpha=R_o\sin\psi_{\mathrm{sh}}\cos\chi$ and $\beta=R_o\sin\psi_{\mathrm{sh}}\sin\chi$. The boundary is circular, with
\begin{equation}
\alpha^{2}+\beta^{2}=R_{\mathrm{sh}}^{2},\qquad R_{\mathrm{sh}}=b_{\mathrm{c}}\sqrt{1-\frac{2M}{R_o}},\qquad \lim_{R_o\to\infty}R_{\mathrm{sh}}=3\sqrt{3}M.
\label{eq:sv-shadow-screen}
\end{equation}
Figure~\ref{fig:sv-shadow-circle} compares the finite distance and asymptotic radii at fixed areal observer position. With $D_A=R_o$ denoting the distance used for this areal screen and $\vartheta=M/D_A$, the angular diameter has the expansion
\begin{equation}
\theta_{\mathrm{sh}}=2\psi_{\mathrm{sh}}=6\sqrt{3}\,\vartheta\left[1-\vartheta+4\vartheta^{2}+\mathcal{O}(\vartheta^{3})\right].
\label{eq:sv-shadow-angular-expansion}
\end{equation}
This distance convention matters because the spatial asymptotics are conical. If the same angle is instead normalized by $r_o$, then $\theta_{\mathrm{sh}}/(6\sqrt{3}M/r_o)-1=-M/r_o+(4M^{2}-a^{2}/2)/r_o^{2}+\mathcal{O}((M/r_o)^{3})$ at fixed $a/M$. The apparent $a$ dependence comes from using $r_o$ in place of the observer's areal position $R_o$.

\subsection{Orbital information beyond shadow size}\label{subsec:sv-shadow-photon sphere}

\begin{figure}[tbp]
\centering
\includegraphics[width=0.65\linewidth]{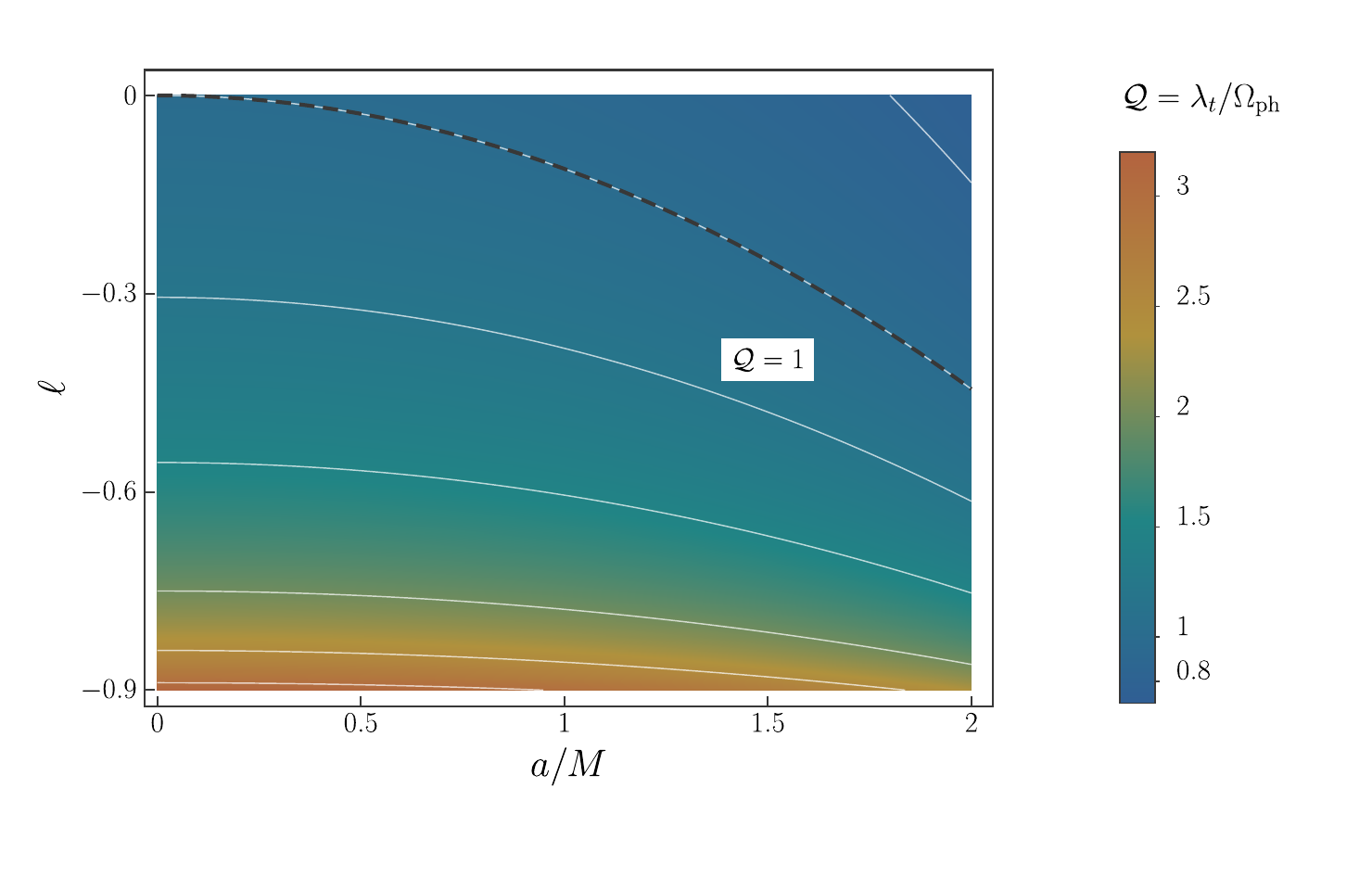}
\caption{Photon orbit instability relative to its orbital frequency, $\mathcal{Q}=\lambda_t/\Omega_{\mathrm{ph}}$. The contour $\mathcal{Q}=1$ follows $\ell=-a^{2}/(9M^{2})$ and gives the Schwarzschild ratio at fixed $M$. Below this curve the instability is larger, and above it the instability is smaller. The edges $a/M=0,2$ and $\ell=0$ are limiting geometries.}
\label{fig:sv-shadow-instability}
\end{figure}

The static shadow size does not exhaust the information carried by null trajectories. In particular, the photon orbit instability retains information about the deformations. Combining the instability exponent and orbital frequency, obtained in the previous section, gives
\begin{equation}
\mathcal{Q}=\frac{\lambda_t}{\Omega_{\mathrm{ph}}}=\sqrt{\frac{1-a^{2}/(9M^{2})}{1+\ell}},\qquad \mathcal{Q}=1\ \Longleftrightarrow\ \ell=-\frac{a^{2}}{9M^{2}}.
\label{eq:sv-shadow-instability-ratio}
\end{equation}
In this regard, different values of $a$ and $\ell$ can have the Schwarzschild value of $\mathcal{Q}$ at the same capture size. Figure~\ref{fig:sv-shadow-instability} displays this compensation curve. It describes the metric black hole domain; the source regularity and kinetic conditions established earlier must also hold for a sourced configuration.

\subsection{Observational comparison with Sgr A* and M87*}\label{subsec:sv-shadow-observations}

The observational normalization is supplied by the angular gravitational scale inferred from stellar dynamics. For a circular timelike orbit in the present metric,
\begin{equation}
\Omega_{\mathrm{orb}}^{2}=\frac{\mathrm{d}A/\mathrm{d}r}{\mathrm{d}(R^{2})/\mathrm{d}r}=\frac{M}{R^{3}}.
\label{eq:sv-shadow-dynamical-mass}
\end{equation}
Thereby, the leading orbital mass is the same lapse parameter that enters Eq.~\eqref{eq:sv-shadow-angular-expansion}. We use the published angular scales as external inputs to a geometric test. Their stellar fits and the EHT image to shadow calibrations were obtained within GR based models.

\begin{figure}[tbp]
\centering
\includegraphics[width=0.6\linewidth]{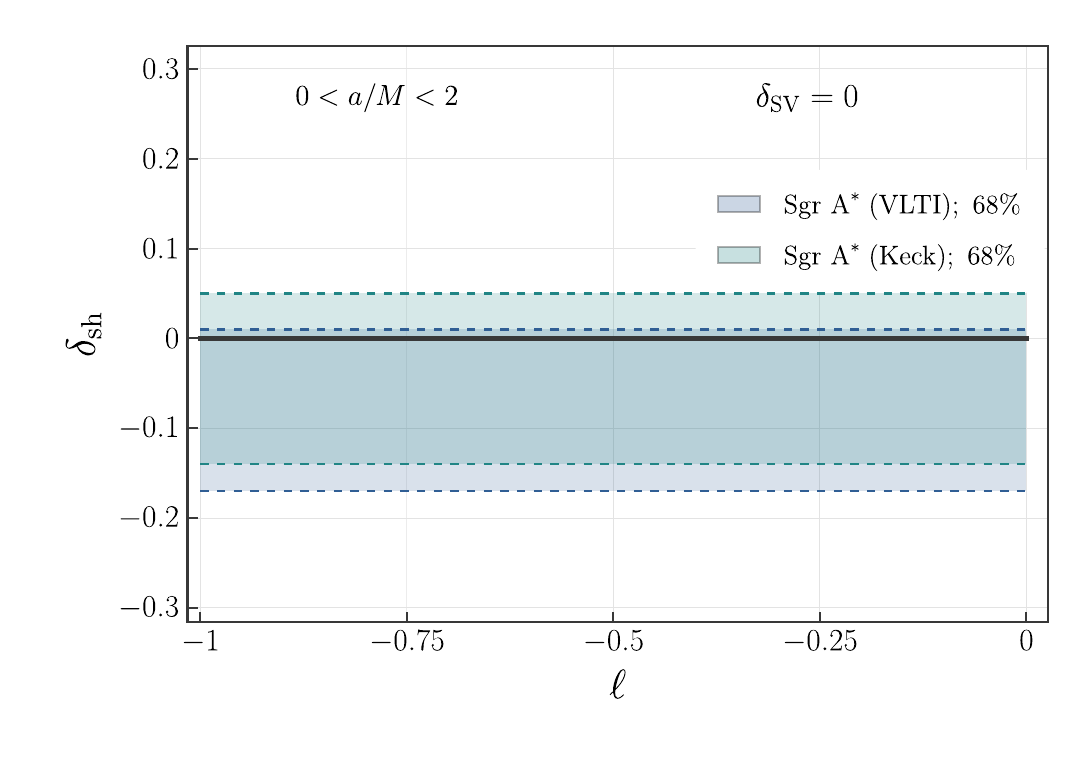}
\caption{Sgr A* shadow diameter intervals from the alternative VLTI and Keck priors, together with the distant metric prediction $\delta_{\mathrm{sh}}^{\mathrm{SV}}=0$. The displayed coupling dependence is exactly flat for every $0<a/M<2$ at fixed $\vartheta_{\mathrm{dyn}}$. The shaded bands retain their separate 68\% interpretation.}
\label{fig:sv-shadow-sgra}
\end{figure}

\begin{figure}[tbp]
\centering
\includegraphics[width=0.6\linewidth]{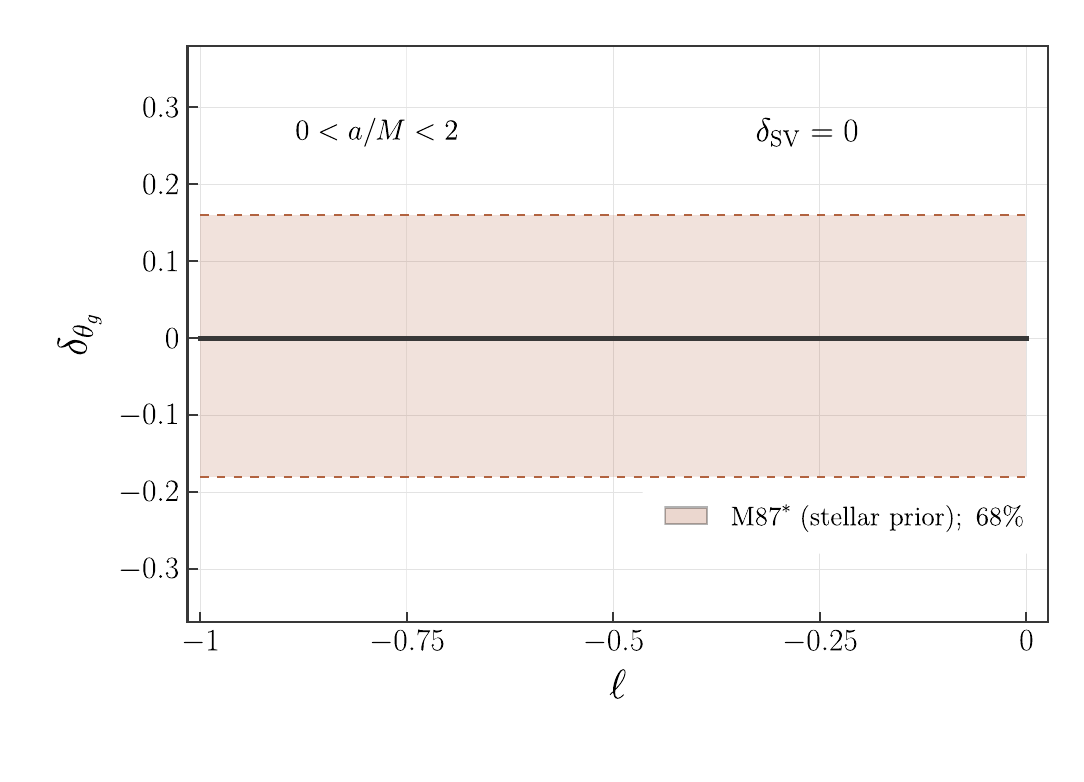}
\caption{M87* angular mass consistency interval for the stellar prior, $-0.18\leq\delta_{\theta_g}\leq0.16$, compared with the common geometric prediction. }
\label{fig:sv-shadow-m87}
\end{figure}

\begin{figure}[tbp]
\centering
\includegraphics[width=0.8\linewidth]{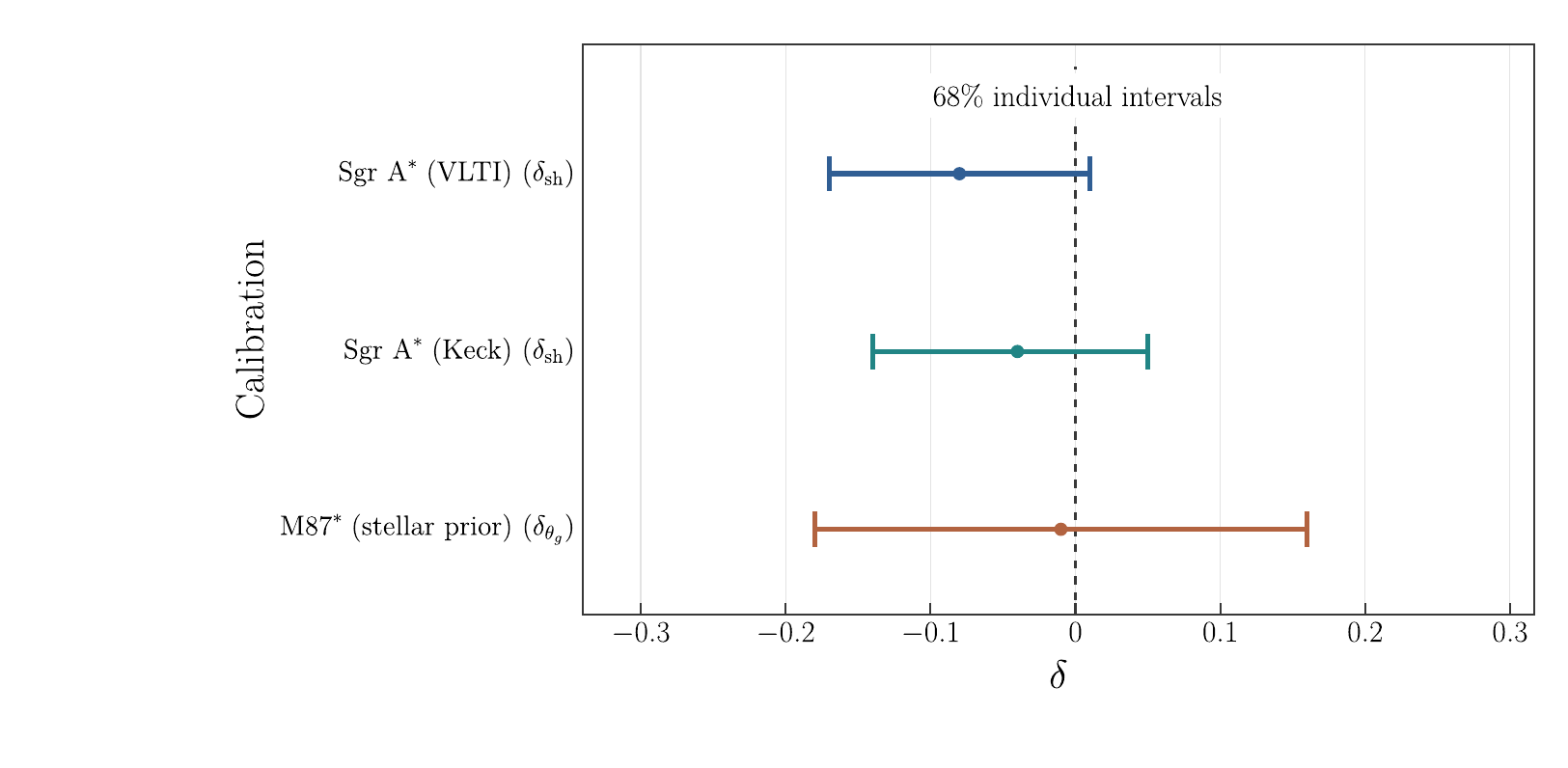}
\caption{Reported central values and individual 68\% intervals for the three settings in Table~\ref{tab:sv-shadow-observations}. The vertical reference line marks the common zero prediction. The Sgr A* rows describe a shadow diameter statistic and the M87* row an angular mass statistic.}
\label{fig:sv-shadow-observational-comparison}
\end{figure}

For Sgr A*, the EHT comparison is formulated in terms of the inferred shadow diameter~\cite{EHTSgrA2022VI}, whereas the M87* statistic compares a calibrated angular mass with its stellar dynamical counterpart~\cite{EHTM872019VI}:
\begin{equation}
\delta_{\mathrm{sh}} = \frac{\theta_{\mathrm{sh}}}{6\sqrt{3}\,\vartheta_{\mathrm{dyn}}} - 1,\qquad \delta_{\theta_g}=\frac{\vartheta_{\mathrm{EHT}}}{\vartheta_{\mathrm{dyn}}}-1.
\label{eq:sv-shadow-observational-statistics}
\end{equation}
Table~\ref{tab:sv-shadow-observations} retains the separate published definitions and their 68\% intervals. The VLTI and Keck rows represent alternative priors for the same Sgr A* image, and are not independent observations to be multiplied into a joint likelihood. At the distant observer accuracy appropriate to these sources, identifying $\vartheta_{\mathrm{dyn}}=M/D_A$ gives
\begin{equation}
\delta_{\mathrm{sh}}^{\mathrm{SV}}=0,\qquad \delta_{\theta_g}^{\mathrm{SV}}=0,\qquad \left.\frac{\partial\delta^{\mathrm{SV}}}{\partial(a/M)}\right|_{\vartheta_{\mathrm{dyn}}}=0,\qquad \left.\frac{\partial\delta^{\mathrm{SV}}}{\partial\ell}\right|_{\vartheta_{\mathrm{dyn}}}=0.
\label{eq:sv-shadow-observational-degeneracy}
\end{equation}
Here, the M87* equality refers to the geometric angular mass comparison under the adopted image calibration. The finite distance correction to the Sgr A* diameter statistic is $-\vartheta+4\vartheta^{2}+\mathcal{O}(\vartheta^{3})$, which is negligible on these astronomical scales and remains independent of both parameters. All three intervals contain the common zero prediction. Figures~\ref{fig:sv-shadow-sgra} and~\ref{fig:sv-shadow-m87} display this agreement along the coupling axis, and Fig.~\ref{fig:sv-shadow-observational-comparison} places the reported statistics side by side. Within this comparison, the diameter calibrates $M/D_A$ and supplies no separate determination of $a$ or $\ell$.

\begin{table}[tbp]
\centering
\caption{Published angular scales and individual 68\% intervals used in the geometric comparison~\cite{EHTSgrA2022VI,EHTM872019VI}. The two uncertainties on the Sgr A* angular scales are statistical and systematic. Sgr A* uses $\delta_{\mathrm{sh}}$; M87* uses $\delta_{\theta_g}$.}
\label{tab:sv-shadow-observations}
\small
\setlength{\tabcolsep}{5pt}
\begin{tabular}{lccc}
\toprule
Source and prior & $\vartheta_{\mathrm{dyn}}$ [$\mu\mathrm{as}$] & Published deviation & 68\% interval \\
\midrule
Sgr A*, VLTI & $5.125\pm0.009\pm0.020$ & $\delta_{\mathrm{sh}}=-0.08\pm0.09$ & $[-0.17,0.01]$ \\
Sgr A*, Keck & $4.92\pm0.03\pm0.01$ & $\delta_{\mathrm{sh}}=-0.04^{+0.09}_{-0.10}$ & $[-0.14,0.05]$ \\
M87*, stellar & $3.62^{+0.60}_{-0.34}$ & $\delta_{\theta_g}=-0.01\pm0.17$ & $[-0.18,0.16]$ \\
\bottomrule
\end{tabular}
\end{table}

Moreover, the brightness ring must be distinguished from this geometric boundary. The Sgr A* analysis reports a ring diameter of $51.8\pm2.3\,\mu\mathrm{as}$~\cite{EHTSgrA2022IV}, while its shadow diameter is $48.9^{+5.2}_{-5.1}\,\mu\mathrm{as}$~\cite{EHTSgrA2022VI}. The 2017--2021 M87* study gives a mean emission ring diameter of $43.9\pm0.6\,\mu\mathrm{as}$~\cite{EHTM872025Variability}. This ring precision cannot replace the uncertainty of the angular mass statistic. In the present geometry, the null trajectory equation is
\begin{equation}
\frac{\mathrm{d}\phi}{\mathrm{d}R}=\pm\frac{\sqrt{\gamma}\,b_\gamma}{R^{2}\sqrt{1-a^{2}/R^{2}}\sqrt{1-b_\gamma^{2}A(R)/R^{2}}}.
\label{eq:sv-shadow-ray-transfer}
\end{equation}
The azimuth accumulated by a ray keeps both deformations even when its critical impact parameter agrees with Schwarzschild. Together with the instability in Eq.~\eqref{eq:sv-shadow-instability-ratio}, this relation supplies the geometric input for studying higher order images and the transfer of emission near the photon sphere. Notice that such a calculation is required before the ring data can be assigned a likelihood for $(a,\ell)$.

\subsection{Conditional sensitivities from proper radial distance}\label{subsec:sv-shadow-conditional}

\begin{figure}[tbp]
\centering
\includegraphics[width=0.6\linewidth]{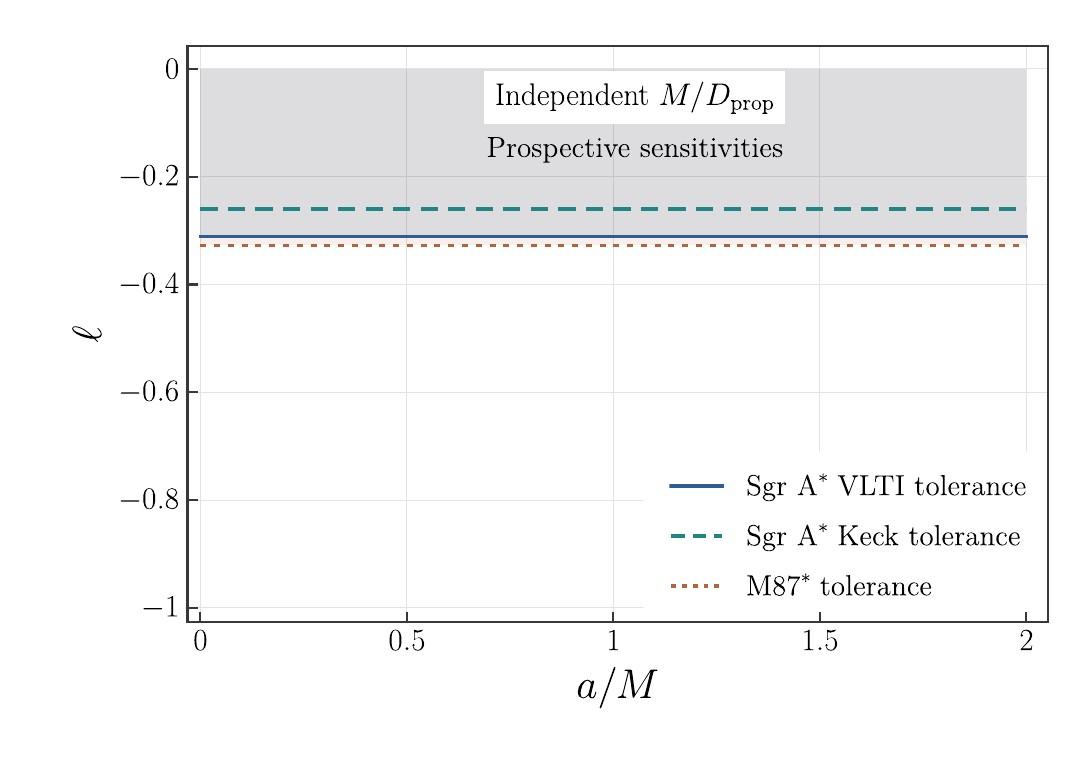}
\caption{Conditional proper distance sensitivity in the metric black hole domain. The separate tolerances yield lower boundaries at $\ell=-0.3111$ (VLTI), $-0.2604$ (Keck), and $-0.3276$ (M87*, stellar), with the segment restricting the upper edge to $\ell<0$. }
\label{fig:sv-shadow-conditional}
\end{figure}

A conditional coupling sensitivity becomes possible if a second distance normalization is measured independently. For a fixed reference sphere $R_*>2M$, define the spatial proper radial length
\begin{equation}
D_{\mathrm{prop}}(R_o;R_*)=\sqrt{\gamma}\int_{R_*}^{R_o}\frac{\mathrm{d}R}{\sqrt{(1-2M/R)(1-a^{2}/R^{2})}},\qquad \lim_{R_o\to\infty}\frac{D_{\mathrm{prop}}}{R_o}=\sqrt{\gamma}.
\label{eq:sv-shadow-proper-distance}
\end{equation}
The choice of the finite reference contributes only an additive constant to this asymptotic limit. With an independently scale $\vartheta_{\mathrm{prop}}=M/D_{\mathrm{prop}}$, the leading angular diameter would instead give
\begin{equation}
\Delta_{\mathrm{prop}}=\frac{\theta_{\mathrm{sh}}}{6\sqrt{3}\,\vartheta_{\mathrm{prop}}}-1\longrightarrow\sqrt{1+\ell}-1,\qquad \ell=(1+\Delta_{\mathrm{prop}})^{2}-1.
\label{eq:sv-shadow-conditional-distance}
\end{equation}
For a hypothetical fractional tolerance $\Delta_-\leq\Delta_{\mathrm{prop}}\leq\Delta_+$, with $\Delta_->-1$, the corresponding asymptotic sensitivity is
\begin{equation}
(1+\Delta_-)^{2}-1\leq\ell\leq(1+\Delta_+)^{2}-1,\qquad -1<\ell<0,\qquad 0<a/M<2.
\label{eq:sv-shadow-conditional-interval}
\end{equation}
Table~\ref{tab:sv-shadow-conditional} uses the three published fractional ranges only as illustrative tolerances for this additional measurement. Figure~\ref{fig:sv-shadow-conditional} shows the resulting horizontal strips in the $(a/M,\ell)$ plane. Their independence of $a$ follows from the asymptotic ratio in Eq.~\eqref{eq:sv-shadow-proper-distance}.

\begin{table}[tbp]
\centering
\caption{Illustrative sensitivity to an independent calibration of $M/D_{\mathrm{prop}}$. Each fractional tolerance is transformed separately through Eq.~\eqref{eq:sv-shadow-conditional-interval}; the final column intersects that transformation with the reconstructed scenario. Every row retains $0<a/M<2$. }
\label{tab:sv-shadow-conditional}
\small
\begin{tabular}{lccc}
\toprule
Tolerance & $[\Delta_-,\Delta_+]$ & Algebraic $\ell$ interval &  Intersection \\
\midrule
Sgr A*, VLTI & $[-0.17,0.01]$ & $[-0.3111,0.0201]$ & $[-0.3111,0)$ \\
Sgr A*, Keck & $[-0.14,0.05]$ & $[-0.2604,0.1025]$ & $[-0.2604,0)$ \\
M87*, stellar & $[-0.18,0.16]$ & $[-0.3276,0.3456]$ & $[-0.3276,0)$ \\
\bottomrule
\end{tabular}
\end{table}

\section{Quasinormal modes and the dynamics of the supporting fields}
\label{sec:quasinormal}

The dynamical content of the solution follows from the action in Eq.~\eqref{eq:action}. Its perturbations carry information about the minimum sphere, the orientation of the bumblebee field and the constitutive response of the magnetic source. These contributions enter at different levels: the metric determines propagation for an additional minimally coupled field, the nonminimal curvature coupling changes the characteristics of the tensor and bumblebee modes, and the derivatives of $\mathscr L$ determine the electromagnetic response. We develop each of these equations explicitly in what follows. This organization also gives a precise stability statement, since decay of a probe (test) field and stability of the fields supporting the geometry refer to different dynamical questions.


\subsection{Harmonic decomposition and radiative boundary conditions}
\label{subsec:qnm_harmonics}

Initially, we retain the time normalization of Eq.~\eqref{eq:final-lv-metric}, for which $g_{tt}\to-1$ at either asymptotic end, and write
\begin{equation}
\mathrm{d}s^2 = q_{ab}\mathrm{d}x^a\mathrm{d}x^b + \Sigma^2\Omega_{AB}\mathrm{d}x^A\mathrm{d}x^B, \qquad q_{ab}\mathrm{d}x^a\mathrm{d}x^b = -A\mathrm{d}t^2 + \frac{\gamma}{A}\mathrm{d}r^2.
\label{eq:qnm_base}
\end{equation}
Lower case Latin indices refer to the stationary radial plane, and upper case Latin indices refer to the unit sphere. Their covariant derivatives are $D_a$ and $\mathcal D_A$. Scalar harmonics satisfy $\mathcal D^A\mathcal D_A Y_{jm}=-JY_{jm}$, with $J=j(j+1)$. The axial harmonics and their symmetric derivatives are
\begin{equation}
S_A=\epsilon_A{}^B\mathcal D_B Y_{jm},\qquad
S_{AB}=\mathcal D_{(A}S_{B)},\qquad
\mathcal D^AS_A=0.
\label{eq:qnm_harmonics}
\end{equation}
A scalar fluctuation has parity $(-1)^j$, while $S_A$ has parity $(-1)^{j+1}$. In this manner, the scalar sector has its own spherical decomposition and the axial sector contains metric and vector amplitudes. In a magnetic background, contraction with $F_{AB}=q\epsilon_{AB}$ exchanges the electromagnetic harmonic types. An axial metric amplitude couples to a polar electromagnetic amplitude, and an axial electromagnetic amplitude couples to a polar metric amplitude. This pairing follows directly from the magnetic stress tensor and is familiar in the perturbation theory of nonlinear electromagnetic black holes \cite{MorenoSarbach2003,Chaverra2016,Nomura2020,AraujoFilho:2024xhm,AraujoFilho:2024lsi,Filho:2023qxu}.

For a field whose characteristics coincide with those of $g_{\mu\nu}$, introduce the tortoise coordinate $x$ through
\begin{equation}
\frac{\mathrm{d}x}{\mathrm{d}r}=\frac{\sqrt\gamma}{A}.
\label{eq:qnm_tortoise}
\end{equation}
In the black--bounce exterior $0<a<2M$, the positive metric horizon is $r_h=\sqrt{4M^2-a^2}$. The limits $r\to r_h^+$ and $r\to+\infty$ give $x\to-\infty$ and $x\to+\infty$. With time dependence $\exp(-\mathrm{i}\omega t)$, the radiative conditions are
\begin{equation}
\psi\sim\mathrm{e}^{-\mathrm{i}\omega x}\quad(x\to-\infty),
\qquad \psi\sim\mathrm{e}^{+\mathrm{i}\omega x}\quad(x\to+\infty).
\label{eq:qnm_bc}
\end{equation}
They describe absorption at the future horizon and radiation toward infinity. A decaying resonance has $\operatorname{Im}\omega<0$. For a static wormhole, $a>2M$, $x$ spans the two asymptotic ends and the same respective phases represent radiation toward the left and right infinities. At $a=2M$, the metric horizon is degenerate; the numerical compactification used below applies to the nondegenerate exterior. The asymptotic radial deformation is maintained in both the phase and the potential, so all frequencies are measured relative to the normalized Killing time $t$.


\subsection{Scalar propagation and the source scalar}
\label{subsec:qnm_scalar}

An additional massless scalar $\chi$ gives a direct measure of the geometry through $\Box\chi=0$. Its decomposition,
\begin{equation}
\chi(t,r,\theta,\varphi)=\frac{1}{\Sigma(r)}
\sum_{jm}\psi_{jm}(t,r)Y_{jm}(\theta,\varphi),
\label{eq:qnm_scalar_ansatz}
\end{equation}
removes the first radial derivative after Eq.~\eqref{eq:qnm_tortoise} is used. Substitution into the wave equation gives
\begin{equation}
\left[-\partial_t^2 + \partial_x^2-V_{\mathrm{s}}(r)\right]\psi_{jm} = 0, \qquad V_{\mathrm{s}} = \frac{AJ}{\Sigma^2}+\frac{1}{\Sigma}\frac{\mathrm{d}^2\Sigma}{\mathrm{d}x^2} =A\left[\frac{J}{\Sigma^2} + \frac{(A\Sigma')'}{\gamma\Sigma}\right].
\label{eq:qnm_scalar_master}
\end{equation}
Here and below, a prime denotes $\mathrm{d}/\mathrm{d}r$. The explicit derivatives $\Sigma'=r/\Sigma$ and $A'=2Mr/\Sigma^3$ yield
\begin{equation}
V_{\mathrm{s}}= A\left[\frac{J}{\Sigma^2}+\frac{2Mr^2}{\gamma\Sigma^5} +\frac{Aa^2}{\gamma\Sigma^4}\right].
\label{eq:qnm_scalar_potential}
\end{equation}
Every term is nonnegative in a static exterior for $M>0$, $a>0$ and $\gamma>0$. This property supplies an analytic mode stability result for the probe. If $\operatorname{Im}\omega>0$, the radiative phases in Eq.~\eqref{eq:qnm_bc} decay at both radial endpoints. Multiplication of the radial equation by $\psi^*$ and integration give
\begin{equation}
\int\left(\left|\frac{\mathrm{d}\psi}{\mathrm{d}x}\right|^2 +V_{\mathrm{s}}|\psi|^2\right)\mathrm{d}x =\omega^2\int|\psi|^2\mathrm{d}x.
\label{eq:qnm_probe_energy}
\end{equation}
The left hand side is real and nonnegative. Its imaginary part forces $\operatorname{Re}\omega=0$ for a growing mode, and its real part then excludes $\omega^2<0$. In other words, the massless scalar probe has no exponentially growing normal mode. The argument also applies to the two outgoing ends of a static wormhole.

The scalar $\phi$ appearing in the action has a different fluctuation equation. Varying Eq.~\eqref{eq:scalar-eom} about the reconstructed background gives
\begin{equation}
\left(\Box+U_{,\phi\phi}\right)\delta\phi
+\delta(\Box)\phi_0=0,
\qquad \delta(\Box)\phi_0=-h^{\mu\nu}\nabla_\mu\nabla_\nu\phi_0
-\left(\nabla_\mu h^{\mu\nu}-\frac12\nabla^\nu h\right)\nabla_\nu\phi_0,
\label{eq:qnm_source_scalar}
\end{equation}
where $h=g^{\mu\nu}h_{\mu\nu}$. For the potential in Eq.~\eqref{eq:fixed-action-potential}, its second derivative is explicit:
\begin{equation}
U_{,\phi\phi} = \frac{2\alpha}{\gamma}
\left(4\cos^3z\sin^2z - \cos^5z\right),
\qquad z=\arctan(r/a).
\label{eq:qnm_scalar_mass}
\end{equation}
This fluctuation belongs to the polar system and interacts with its metric constraints. The phantom kinetic term also gives its quadratic action a negative kinetic coefficient. This energetic property and a classical exponentially growing mode are distinct statements. Equation~\eqref{eq:qnm_probe_energy} proves stability of $\chi$ (the radial anomaly introduced in Eq. (\ref{eq:sv-bound-anomaly}) ), while Eq.~\eqref{eq:qnm_source_scalar} identifies the scalar equation that participates in the dynamics of the supporting theory.


\subsection{Electromagnetic vector fields and the magnetic response}
\label{subsec:qnm_em}

For an additional ordinary Maxwell field, an axial potential $\delta\mathcal A_A = a(t,r)S_A$ obeys a single wave equation. With the usual harmonic normalization, its canonical amplitude can be chosen as $a$, and the potential is
\begin{equation}
\left[-\partial_t^2 + \partial_x^2-V_{\mathrm{M}}\right]a = 0, \qquad V_{\mathrm{M}} = \frac{AJ}{\Sigma^2},\qquad j\geq1.
\label{eq:qnm_maxwell_probe}
\end{equation}
Its nonnegative exterior potential gives the same integral proof as Eq.~\eqref{eq:qnm_probe_energy}. This field is an auxiliary probe. The electromagnetic field supporting the solution is governed by the fixed function $\mathscr L(\mathcal F)$ in Eq.~\eqref{eq:fixed-action-lagrangian}, and its response involves two independent derivatives,
\begin{equation}
p=\mathscr L_{\mathcal F},\qquad
P=\mathscr L_{\mathcal F}+2\mathcal F\mathscr L_{\mathcal F\mathcal F}.
\label{eq:qnm_ned_response}
\end{equation}
We use $p$ exclusively for the constitutive coefficient in this section. Both functions are explicit derivatives of the reconstructed source. They can also be evaluated along the geometry through
\begin{equation}
p(r)=\frac{\mathscr L'(r)}{\mathcal F'(r)},\qquad P(r)=p(r)+2\mathcal F(r)\frac{p'(r)}{\mathcal F'(r)},
\label{eq:qnm_ned_chain}
\end{equation}
with regular endpoint limits at $r=0$. Their fixed action representation uses $\mathcal F=\beta^2\cos^4z/2$, so these response functions have the same dependence on $z$ for every allowed $a$.

In addition, the electromagnetic quadratic action at fixed metric follows by expanding the invariant to second order. If $f_{\mu\nu}=2\nabla_{[\mu}\delta\mathcal A_{\nu]}$, then
\begin{equation}
S_{\mathrm{em}}^{(2)} =  - \int\sqrt{-g}\,\mathrm{d}^4x \left[\frac{p}{4}f_{\mu\nu}f^{\mu\nu} +\frac{\mathscr L_{\mathcal F\mathcal F}}{8} \left(F^{\mu\nu}f_{\mu\nu}\right)^2\right].
\label{eq:qnm_ned_action}
\end{equation}
For an axial electromagnetic harmonic, the angular integration gives
\begin{equation}
S_{\mathrm{em,ax}}^{(2)} = \frac{J}{2}\int\mathrm{d}t\,\mathrm{d}x \left[p\left((\partial_t a)^2-(\partial_x a)^2\right) -\frac{AJ}{\Sigma^2}P a^2\right].
\label{eq:qnm_ned_axial_action}
\end{equation}
At $p>0$, the canonical amplitude $\Psi_{\mathrm{ax}}=\sqrt p\,a$ yields
\begin{equation}
\left[-\partial_t^2 + \partial_x^2-V_{\mathrm{em,ax}}\right]\Psi_{\mathrm{ax}}=0, \qquad V_{\mathrm{em,ax}} = \frac{AJ}{\Sigma^2}\frac{P}{p} +\frac{1}{\sqrt p}\frac{\mathrm{d}^2\sqrt p}{\mathrm{d}x^2}.
\label{eq:qnm_ned_axial_potential}
\end{equation}
The ordinary electromagnetic polarization is described by polar electromagnetic harmonics on the magnetic sphere. If $f_{tx}$ is their radial electric amplitude, its canonical variable and potential are
\begin{equation}
\Psi_{\mathrm{pol}}=\frac{\Sigma^2\sqrt p}{A}f_{tx},\qquad
V_{\mathrm{em,pol}}=\frac{AJ}{\Sigma^2}
+\frac{1}{p^{-1/2}}\frac{\mathrm{d}^2p^{-1/2}}{\mathrm{d}x^2}.
\label{eq:qnm_ned_polar_potential}
\end{equation}
These two radial equations describe electromagnetic fluctuations on a fixed geometry. The ordinary polarization is the electromagnetic member of the coupled axial metric system developed below.

The ordinary fixed metric operator has a useful exact factorization. With $s = p^{-1/2}$ and $\mathcal Q = \partial_x-s_{,x}/s$,
\begin{equation}
-\partial_x^2+V_{\mathrm{em,pol}}
=\mathcal Q^\dagger\mathcal Q+\frac{AJ}{\Sigma^2}.
\label{eq:qnm_ned_factorization}
\end{equation}
For $p>0$ and $A\geq0$, it excludes growing electromagnetic modes with the radial decay conditions. The extraordinary polarization depends on $P$ as well. In a local orthonormal frame aligned with the magnetic radial direction, its principal dispersion relation is
\begin{equation}
\Omega^2=k_{\hat r}^2+\frac{P}{p}\left(k_{\hat\theta}^2+k_{\hat\varphi}^2\right).
\label{eq:qnm_extraordinary_dispersion}
\end{equation}
The corresponding inverse optical tensor is proportional to
\begin{equation}
\mathcal G_{\mathrm{em}}^{\mu\nu}
=p g^{\mu\nu}+\mathscr L_{\mathcal F\mathcal F}F^{\mu\alpha}F^\nu{}_{\alpha},
\label{eq:qnm_optical_metric}
\end{equation}
whose orthonormal entries are $(-p,p,P,P)$. This expresses directly the role of the angular constitutive response \cite{Novello2000,Abalos2015}.

For the mixed case considered here, Eq.~\eqref{eq:fixed-action-weak-field} gives $c_{1/2}>0$ and $c_1>0$. To control the derivatives, set $s_0=(2\mathcal F/\beta^2)^{1/4}$. The explicit source is analytic in $s_0$ near zero, since $z=\pi/2-\arcsin s_0$ there. Its convergent Taylor series gives the differentiable fractional power expansion in Eq.~\eqref{eq:fixed-action-weak-field}. Differentiating this series yields
\begin{equation}
p = \frac{c_{1/2}}{2\sqrt{\mathcal F}}-c_1 + \mathcal O(\mathcal F^{1/4}),
\qquad P=-c_1+\mathcal O(\mathcal F^{1/4}).
\label{eq:qnm_ned_endpoint}
\end{equation}
For $p>0$ and $P<0$ throughout a sufficiently distant finite radial interval. A transverse wave packet localized there has $\Omega^2<0$ at sufficiently large angular wave number. Its local growth rate and the corresponding stationary time rate for a predominantly angular packet are
\begin{equation}
\Gamma_{\mathrm{loc}}\simeq\frac{j}{\Sigma}\sqrt{-\frac{P}{p}},
\qquad \Gamma_t\simeq\sqrt A\,\Gamma_{\mathrm{loc}}.
\label{eq:qnm_local_growth}
\end{equation}
At any fixed exterior point with $P<0$, the growth rate increases with the angular wave number. Remarkably, this gives the short wavelength instability of evolution in the physical time. The optical tensor retains an indefinite signature with the radial direction as its distinguished direction when $P<0$; its cone is incompatible with evolution on the spacelike slices used by the metric fields. This is an angular gradient instability of the full supporting theory. Minimal electromagnetic coupling gives metric variations at lower derivative order in the electromagnetic equation, and electromagnetic stress gives at most first derivatives of its potential in the metric equation. The extraordinary electromagnetic factor therefore remains in the principal determinant of the coupled system. The conclusion concerns finite points of the exterior, where all source coefficients are finite, and follows independently of the zero invariant endpoint itself.

A radial variational construction makes the growth explicit for the fixed metric electromagnetic operator. Take $\Psi=\sqrt p\,\chi$ with real compactly supported $\chi$. Integration by parts cancels the derivative term in Eq.~\eqref{eq:qnm_ned_axial_potential} and gives
\begin{equation}
\frac{\langle\Psi,(-\partial_x^2+V_{\mathrm{em,ax}})\Psi\rangle} {\langle\Psi,\Psi\rangle} = \frac{\displaystyle\int\left[p(\partial_x\chi)^2+\frac{AJ}{\Sigma^2}P\chi^2\right]\mathrm{d}x} {\displaystyle\int p\chi^2\mathrm{d}x}.
\label{eq:qnm_ned_rayleigh}
\end{equation}
On any exterior interval with $p>0$ and $P<0$, the angular contribution dominates the positive derivative contribution as $j$ increases. A negative quotient gives a bound state with $\omega^2=-\Gamma^2$ and a growing solution $\exp(\Gamma t)$. Its decaying radial amplitudes satisfy the analytic continuation of the radiative conditions. In general lines, the construction proves a mode instability of the fixed metric electromagnetic equation, while Eq.~\eqref{eq:qnm_extraordinary_dispersion} establishes the local growing characteristic in the coupled theory.


\subsection{Axial tensor perturbations and the dynamical features}
\label{subsec:qnm_tensor}

The physical tensor perturbation is obtained by varying the bumblebee field together with the metric. For $j\geq2$, choose the angular gauge $h_{AB}=0$ and write the complete amplitudes in the axial metric parity as
\begin{equation}
h_{aA}=h_a(t,r)S_A,\qquad
\delta B_A=\beta_B(t,r)S_A,\qquad
\delta\mathcal A_a=a_a(t,r)Y_{jm},\qquad
\delta\mathcal A_A=a_E(t,r)\mathcal D_A Y_{jm}.
\label{eq:qnm_coupled_ansatz}
\end{equation}
The amplitudes $\beta_B$ and $a_E$ are distinct from the fixed electromagnetic scale $\beta$. In this parity block, $\delta\phi=\delta\lambda=\delta\mathcal F=0$. The linear norm variation $2b^\mu\delta B_\mu-h^{\mu\nu}b_\mu b_\nu$ vanishes identically. This verifies the multiplier constraint while allowing the transverse fildt to fluctuate. The dipole $j=1$ has $S_{AB}=0$ and describes vector and rotational degrees of freedom; propagating tensor harmonics begin at $j=2$.

The electromagnetic gauge invariant and the background vector strength are
\begin{equation}
e_a=D_a a_E-a_a,\qquad
\mathcal E_a=e_a-\frac{q}{\Sigma^2}h_a,\qquad
f_{ab}=-2D_{[a}e_{b]},\qquad
\mathcal H_{ab}=2D_{[a}b_{b]}.
\label{eq:qnm_coupled_invariants}
\end{equation}
The axial metric curvature is conveniently represented by
\begin{equation}
\mathfrak F_{ab}=2D_{[a}\left(\frac{h_{b]}}{\Sigma^2}\right),\qquad
\mathfrak E_a=-\frac{1}{2\Sigma^2}D^b\left(\Sigma^4\mathfrak F_{ba}\right)
+\frac{J-2}{2\Sigma^2}h_a.
\label{eq:qnm_curvature_operator}
\end{equation}
Direct variation of the Ricci tensor gives
\begin{equation}
\delta R_{aA}=(R_\Omega h_a+\mathfrak E_a)S_A,\qquad
\delta R_{AB}=(D_a h^a)S_{AB},\qquad \delta R=0,
\label{eq:qnm_ricci_variation}
\end{equation}
The angular background coefficient is
\begin{equation}
R_\Omega=R^\theta{}_\theta=
\frac{\gamma-A(\Sigma'^2+\Sigma\Sigma'')-A'\Sigma\Sigma'}{\gamma\Sigma^2}
=\frac{\ell}{\gamma\Sigma^2}+\frac{2Ma^2}{\gamma\Sigma^5}.
\label{eq:qnm_angular_ricci}
\end{equation}
The source functions $p(r)$, $\lambda(r)$, $b_a(r)$ and $\mathcal H_{ab}(r)$ are those of the exact background.

The ordinary electromagnetic equations in this coupled system follow from the variation of Eq.~\eqref{eq:em-eom}:
\begin{equation}
D_b\left(\Sigma^2p f^{ba}\right)+Jp\mathcal E^a=0,
\qquad D_a(p\mathcal E^a)=0.
\label{eq:qnm_coupled_em}
\end{equation}
The second equation is also the differential consequence of the first, and provides a constraint check. Varying the bumblebee equation gives
\begin{equation}
\begin{split}
0={}&D_a\left(D^a\beta_B-\mathcal H^{ab}h_b\right)
-\left(\frac{J}{\Sigma^2}+\lambda-\eta R_\Omega\right)\beta_B
+\mathcal J^a h_a+\eta b^a\mathfrak E_a,\\
\mathcal J^a={}&\lambda b^a-\eta R^a{}_b b^b,\qquad \eta=\frac{\xi}{\kappa}.
\end{split}
\label{eq:qnm_coupled_bumblebee}
\end{equation}
In particular, the curvature determined multiplier remains in the fluctuation equation. Its contribution is fixed by the same field realization that determined the background.

The ordered variation of the curvature stress in Eq.~\eqref{eq:curvature-stress} admits a compact reduction. Put $m_j=J-2$. The full metric equations and their angular constraint are
\begin{equation}
\begin{split}
\gamma\mathfrak E_a={}&\kappa\left[-\mathcal H_a{}^bD_b\beta_B
+\lambda b_a\beta_B+\frac{qp}{\Sigma^2}\mathcal E_a\right]
+\frac{\xi m_j}{2\Sigma^2}\left(b^2h_a-b_a b^c h_c\right)
+\xi\mathcal C_a^{\beta_B},\\
0={}&D_a\left[(q^{ab}+\xi b^a b^b)h_b-\xi b^a\beta_B\right].
\end{split}
\label{eq:qnm_coupled_metric}
\end{equation}
Both residual operators are explicit. With a dot denoting $\partial_t$, they read
\begin{align}
\mathcal C_t^{\beta_B}={}&\frac{1}{2\gamma}\bigg\{
\left[\frac{2Au(\Sigma'^2+\Sigma\Sigma'')}{\Sigma^2}-Au''
+u\left(\frac{\gamma m_j}{\Sigma^2}-\frac{2A'\Sigma'}{\Sigma}-A''\right)\right]\beta_B
-A(2u'\beta_B'+u\beta_B'')\bigg\}\nonumber\\
&+\frac{1}{2\sqrt\gamma}\left[w\dot\beta_B'
+\left(w'+\frac{2w\Sigma'}{\Sigma}-\frac{wA'}{A}\right)\dot\beta_B\right],
\label{eq:qnm_Ct}\\
\mathcal C_r^{\beta_B}={}&\frac{w}{2\sqrt\gamma}\left[
\frac{4\Sigma''}{\Sigma}+\frac{\gamma m_j}{A\Sigma^2}
+\frac{2A'\Sigma'}{A\Sigma}+\frac{A''}{A}\right]\beta_B\nonumber\\
&+\frac{1}{2A}\left[\left(\frac{2u\Sigma'}{\Sigma}-u'\right)\dot\beta_B-u\dot\beta_B'\right]
+\frac{\sqrt\gamma\,w}{2A^2}\ddot\beta_B.
\label{eq:qnm_Cr}
\end{align}
Here, $u=u_0\cosh(kz)$ and $w=\sqrt{u^2+b^2A}$, with the positive root fixed in the background. Equations~\eqref{eq:qnm_coupled_em}, \eqref{eq:qnm_coupled_bumblebee} and \eqref{eq:qnm_coupled_metric}, together with Eqs.~\eqref{eq:qnm_Ct} and \eqref{eq:qnm_Cr}, are a closed linear differential system. Fourier substitution $\partial_t\to-\mathrm{i}\omega$ gives the coupled radial ordinary differential equations and constraints. All coefficients are explicit functions of $r$. The static coordinate poles at $A=0$ are handled by the regular ingoing coordinates of Eq.~\eqref{eq:final-lv-ingoing} or by the common time introduced below.

The reduction keeps the metric variation of both vector indices and every connection term in the two ordered covariant derivatives. The accompanying Mathematica files derive (and check) the Ricci identities, the electromagnetic and vector equations, and the curvature operators directly. The latter were derived at $j=2,3$ and checked independently at $j=4$. Their dependence on $J$ is linear, as expected from two angular derivatives. This gives naturally a component verification of the complete axial equations in addition to the background xAct verification. The coupling between tensor and bumblebee amplitudes is also central to the perturbation analysis of other bumblebee geometries \cite{Mai2024,Liu:2026cxs}; the profiles and source functions in the present system are maintained explicitly.

The physical propagation cones can be derived independently from the principal quadratic action. In a local inertial frame, let $a,b$ temporarily label the three coordinates transverse to the $y$ polarization, choose the metric amplitude $h_a=h_{ay}$, and define $f^h_{ab}=2\partial_{[a}h_{b]}$ and $W_a=b^c f^h_{ac}$. Up to a divergence, the curvature coupling uses the identity $b^\mu b^\nu R_{\mu\nu}=(\nabla_\mu b^\mu)^2-\nabla_\mu b^\nu\nabla_\nu b^\mu$. The principal action then becomes
\begin{equation}
\mathcal L_{\mathrm{prin}}^{(2)}
=-\frac{f^h_{ab}f_h^{ab}}{8\kappa}
-\frac{\xi W_aW^a}{4\kappa}
+\frac{\xi\partial_a\beta_B W^a}{2\kappa}
-\frac12\partial_a\beta_B\partial^a\beta_B.
\label{eq:qnm_principal_action}
\end{equation}
Eliminating the gauge direction and factoring the determinant gives two characteristic factors,
\begin{equation}
\begin{split}
\mathcal P_T&=g^{\mu\nu}k_\mu k_\nu+\xi(b^\mu k_\mu)^2,\\
\mathcal P_B&=d\,g^{\mu\nu}k_\mu k_\nu
+\frac{\xi^2}{2\kappa}(b^\mu k_\mu)^2,
\qquad d=\gamma-\frac{\xi^2b^2}{2\kappa}.
\end{split}
\label{eq:qnm_characteristic_factors}
\end{equation}
The ordinary electromagnetic member supplies the factor $g^{\mu\nu}k_\mu k_\nu$. At $d>0$, the tensor and transverse bumblebee inverse characteristic metrics are
\begin{equation}
G_T^{\mu\nu}=g^{\mu\nu}+\xi b^\mu b^\nu,
\qquad
G_B^{\mu\nu}=g^{\mu\nu}+\frac{\xi^2}{2\kappa d}b^\mu b^\nu.
\label{eq:qnm_characteristic_metrics}
\end{equation}
A local frame with $b^{\hat0}=0$ makes the kinetic conditions transparent. The tensor speed squared along thhis case is $\gamma$, the transverse bumblebee speed squared is $\gamma/d$, and its reduced kinetic coefficient is $d/\gamma$. Thus $\gamma>0$ and $d>0$ give positive principal kinetic energy and real characteristic speeds in this parity block. For $\eta=-k^2/2<0$, they strengthen the background condition to
\begin{equation}
-\frac{1}{1+k^2/4}<\ell<0.
\label{eq:qnm_kinetic_interval}
\end{equation}
These conditions refer to the axial tensor and vector modes; the scalar kinetic sign remains that of Eq.~\eqref{eq:action}.

A common time coordinate adapted to the spacelike case is
\begin{equation}
\mathrm{d}T=\mathrm{d}t+\frac{\sqrt\gamma\,u}{Aw}\mathrm{d}r
=\mathrm{d}v-\frac{\sqrt\gamma\,b^2}{w(w+u)}\mathrm{d}r.
\label{eq:qnm_common_time}
\end{equation}
It is regular across the metric horizon and satisfies $b^T=0$ and $g^{TT}=-b^2/w^2<0$. Equation~\eqref{eq:qnm_characteristic_metrics} then gives $G_T^{TT}=g^{TT}$ and $G_B^{TT}=g^{TT}$. With $\gamma,d,p>0$, these slices support a common local evolution for the axial metric block and its ordinary electromagnetic member.

The temporal component separates their characteristic horizons. In the radial direction,
\begin{equation}
G_T^{rr}=f_T,\qquad G_B^{rr}=\frac{f_B}{d},\qquad
 g^{rr}=\frac{A}{\gamma},\qquad
f_T=A+\frac{\xi u^2}{\gamma},\qquad
f_B=A+\frac{\xi^2u^2}{2\kappa\gamma}.
\label{eq:qnm_horizon_functions}
\end{equation}
The ordinary horizon is $A=0$, the tensor horizon is $f_T=0$, and the transverse bumblebee horizon is $f_B=0$. An asymptotically timelike tensor Killing vector imposes
\begin{equation}
f_T(+\infty)>0
\quad\Longleftrightarrow\quad
\gamma+\xi u_\infty^2>0,
\qquad u_\infty=\nu a\cosh(k\pi/2).
\label{eq:qnm_tensor_asymptotic}
\end{equation}
For $\xi<0$, this places at least one tensor horizon outside a nondegenerate metric horizon. Since $f_B$ increases with $r>0$, a transverse bumblebee horizon exists exactly when
\begin{equation}
f_B(0)=1-2\alpha a^2+\frac{\xi^2\nu^2a^2}{2\kappa\gamma}<0.
\label{eq:qnm_vector_horizon_condition}
\end{equation}
In that case it lies inside $r_h$. Equality describes a degenerate vector horizon at the minimum sphere, and a positive value describes a vector cone that crosses the minimum sphere without a radial horizon. Radiative conditions for the physical coupled system are formulated in its characteristic eigenvectors, with absorption at each future characteristic horizon and radiation at the asymptotic end. Notice that the exact amplitudes and constraints in Eq.~\eqref{eq:qnm_coupled_metric} determine their interaction.

For geometric optics, inversion of Eq.~\eqref{eq:qnm_characteristic_metrics} is especially useful:
\begin{equation}
(G_T^{-1})_{\mu\nu}=g_{\mu\nu}-\frac{\xi}{\gamma}b_\mu b_\nu,
\qquad
(G_B^{-1})_{\mu\nu}=g_{\mu\nu}-\frac{\xi^2}{2\kappa\gamma}b_\mu b_\nu.
\label{eq:qnm_covariant_characteristics}
\end{equation}
After a radial change of time coordinate that preserves $\partial_t$, each stationary characteristic line element is diagonal,
\begin{equation}
\mathrm{d}s_I^2=-f_I\mathrm{d}t_I^2+\frac{D_I}{f_I}\mathrm{d}r^2+\Sigma^2\mathrm{d}\Omega^2,
\qquad
(f_I,D_I)=(A,\gamma),(f_T,1),(f_B,d).
\label{eq:qnm_diagonal_cones}
\end{equation}
The index $I$ labels the ordinary, tensor and bumblebee families. Their diagonal times are explicit:
\begin{equation}
\mathrm{d}t_I=\mathrm{d}t+
\frac{c_I\sqrt\gamma\,uw}{Af_I}\mathrm{d}r,
\qquad c_{\mathrm O}=0,\quad c_T=\frac{\xi}{\gamma},\quad
c_B=\frac{\xi^2}{2\kappa\gamma},\qquad f_I=A+c_Iu^2.
\label{eq:qnm_characteristic_times}
\end{equation}
A separated characteristic wave has an ingoing phase $\exp[-\mathrm{i}\omega(t_I+x_I)]$ at its future horizon and an outgoing phase $\exp[-\mathrm{i}\omega(t_I-x_I)]$ at infinity, where $\mathrm{d}x_I/\mathrm{d}r=\sqrt{D_I}/f_I$. These phases supply the leading radiative eigenvectors. The radial time shifts preserve the Killing frequency while retaining the condensate tilt in the phase. Define $\mathcal U_I=f_I/\Sigma^2$ and select an exterior unstable circular ray at $r_{c,I}$ by $\mathcal U_I'=0$, $\mathcal U_I''<0$. Its angular frequency and radial instability exponent, measured in the original Killing time, are
\begin{equation}
\Omega_I=\sqrt{\mathcal U_I(r_{c,I})},\qquad
\Lambda_I^2=-\left.\frac{f_I\Sigma^2}{2D_I}\mathcal U_I''\right|_{r_{c,I}}.
\label{eq:qnm_ray_frequencies}
\end{equation}
The ordinary cone gives the explicit values
\begin{equation}
\Sigma_{c,\mathrm{O}}=3M,\qquad
\Omega_{\mathrm{O}}=\frac{1}{3\sqrt3\,M},\qquad
\Lambda_{\mathrm{O}}=\frac{\sqrt{1-a^2/(9M^2)}}{3\sqrt{3\gamma}\,M}.
\label{eq:qnm_ordinary_ray}
\end{equation}
Under adiabatic separation of an isolated nondegenerate characteristic family, the leading barrier resonance estimate is
\begin{equation}
\omega_{Ijn}^{(\mathrm{eik})}=
\left(j+\frac12\right)\Omega_I
-\mathrm{i}\left(n+\frac12\right)\Lambda_I.
\label{eq:qnm_eikonal_estimate}
\end{equation}
It follows by expanding the leading angular potential $J\mathcal U_I$ about its maximum in the characteristic tortoise coordinate $\mathrm{d}x_I/\mathrm{d}r=\sqrt{D_I}/f_I$. The curvature of that maximum supplies Eq.~\eqref{eq:qnm_ray_frequencies}, in agreement with the circular ray construction \cite{Cardoso2009}. The numerical values below are explicitly leading geometric optics estimates. In general lines, as we have been sumarizing at the end of the sections, their computation uses the actual characteristic factors; finite angular momentum corrections and mode conversion are encoded in the complete coupled equations above.


\subsection{Numerical spectra and characteristic scales within one action}
\label{subsec:qnm_numerics}

A meaningful parameter comparison holds the matter functions fixed. We use $\kappa=1$ units, choose $b=1$, and take
\begin{equation}
\xi=-\frac15,\qquad \alpha=\frac12,\qquad \beta=1,\qquad
\nu=\frac45,\qquad \gamma=\frac45,\qquad d=\frac{39}{50}.
\label{eq:qnm_numerical_action}
\end{equation}
The family has $M=a^3/2$, $|q|=a^2$ and $u_0=4a/5$. Its vector reality condition is $w(0)^2=1-9a^2/25>0$. We select $a=1.15,1.30,1.50$, all in the black--bounce range and all satisfying Eq.~\eqref{eq:qnm_tensor_asymptotic}. The metric cone is fixed by $\gamma$, while the asymptotic orientation changes according to the boundary data in Eq.~\eqref{eq:fixed-action-boundary-data}. Every entry in the following tables turns out to belong to the same action, with its asymptotic features specified by the corresponding $a$.

\begin{figure}[tbp]
\centering
\includegraphics[width=\linewidth]{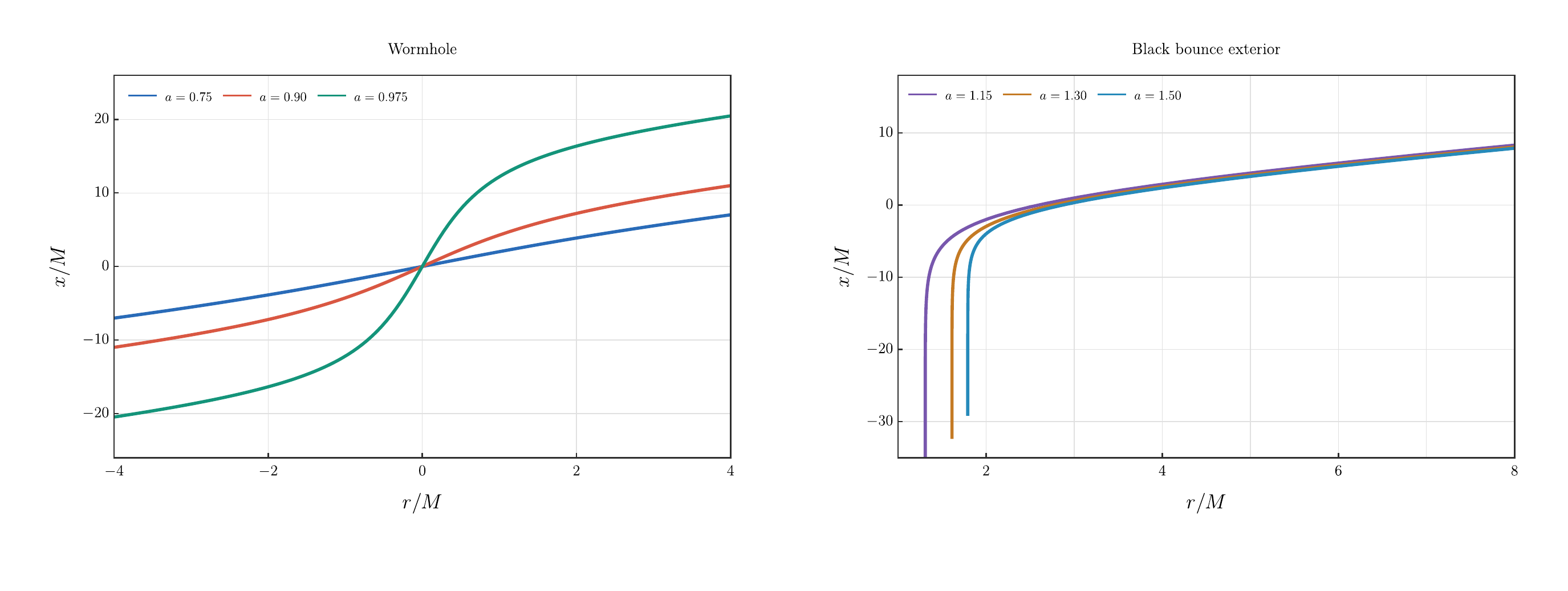}
\caption{Tortoise coordinate $x/M$ versus $r/M$ for the fixed action family of Eq.~(170). The wormholes (left) have $a=0.75,0.90,0.975$ and $x(0)=0$. The black--bounce exteriors (right) have $a=1.15,1.30,1.50$ and $x(r_{\mathrm{ph}})=0$, where $r_{\mathrm{ph}}=\sqrt{9M^2-a^2}$.}
\label{fig:sv-tortoise-x}
\end{figure}

To illustrate the scattering geometry across the transition within the same fixed action, we add the wormhole values $a=0.75,0.90,0.975$ to the black--bounce values selected above. Figure~\ref{fig:sv-tortoise-x} shows that the tortoise map steepens near the wormhole throat as $a$ approaches the null minimum value $a_*=1$ from below. In the black--bounce exterior, the metric horizon maps to $x\to-\infty$.

\begin{figure}[tbp]
\centering
\includegraphics[width=\linewidth]{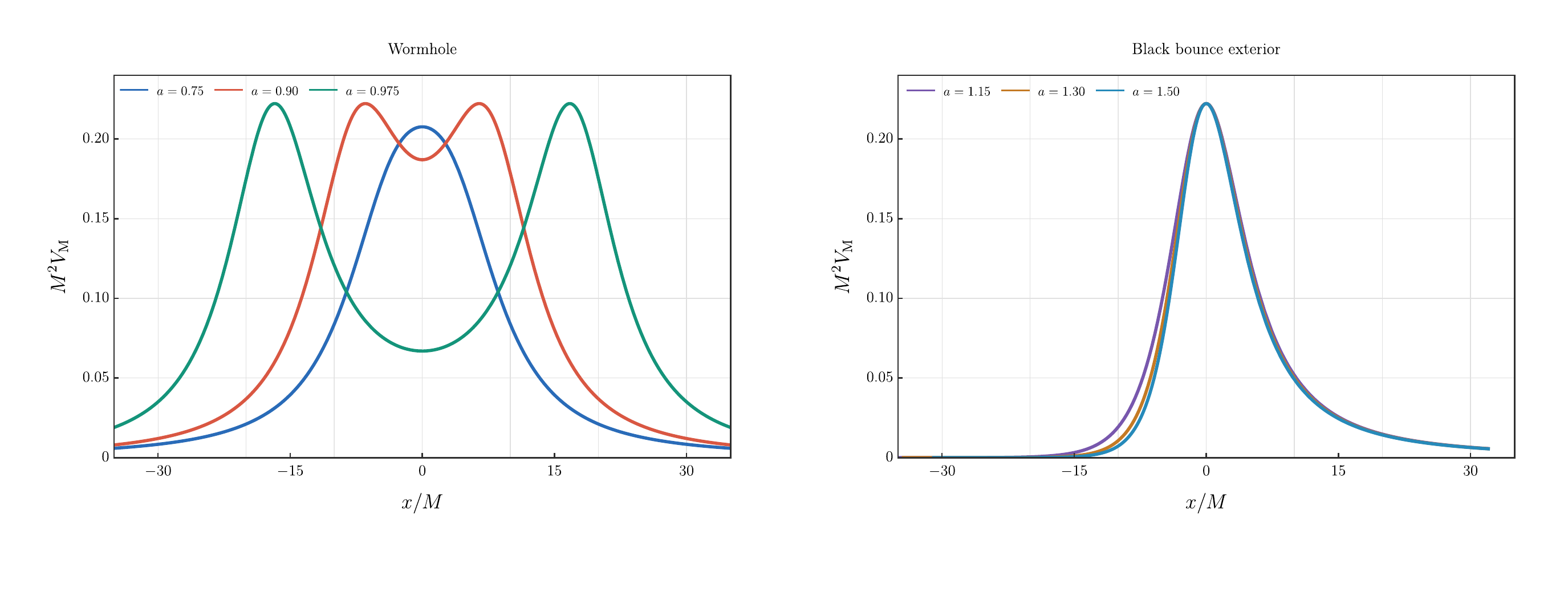}
\caption{Ordinary Maxwell probe potential $M^2V_{\mathrm M}$ of Eq.~(133) versus $x/M$, for $j=2$ and $\gamma=4/5$. The left panel shows wormholes and the right panel shows black--bounce exteriors; the values of $a$ are indicated in the legends.}
\label{fig:sv-maxwell-x}
\end{figure}

\begin{figure}[tbp]
\centering
\includegraphics[width=\linewidth]{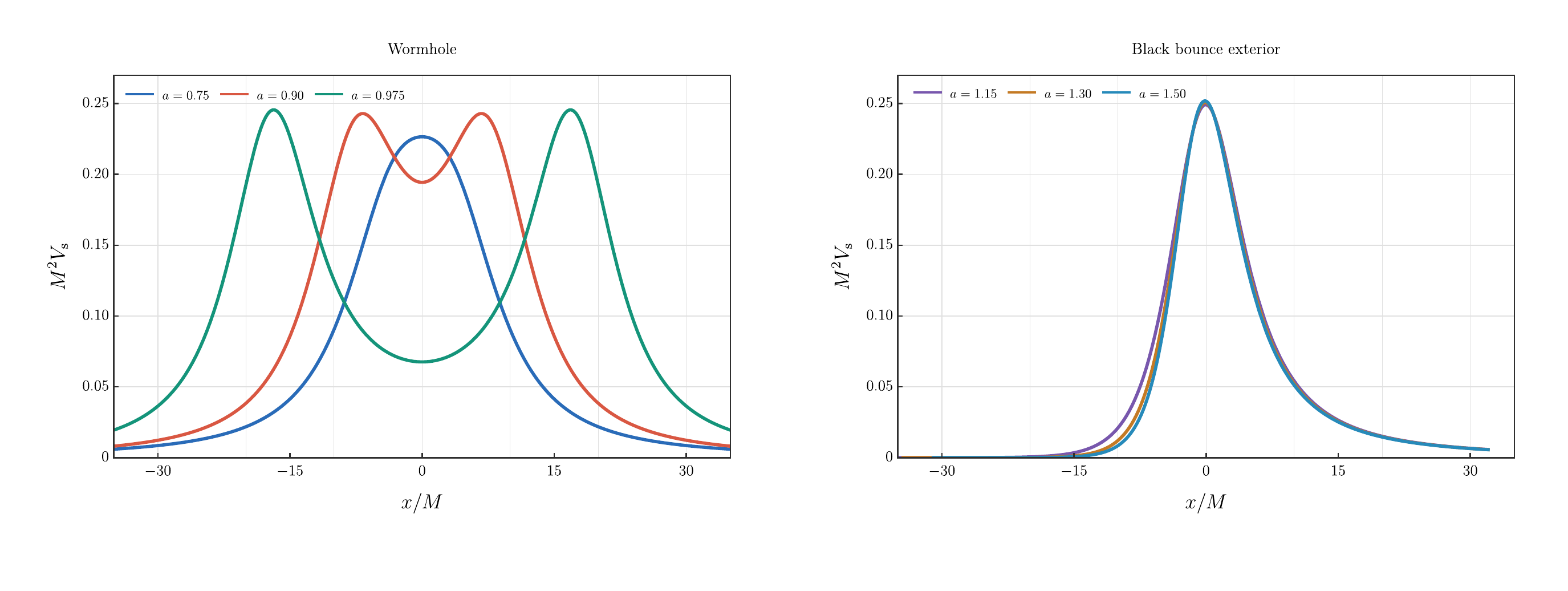}
\caption{Additional massless scalar probe potential $M^2V_{\mathrm s}$ of Eq.~(129) versus $x/M$, for $j=2$ and $\gamma=4/5$. The parameter values and panel assignments match Fig.~\ref{fig:sv-maxwell-x}.}
\label{fig:sv-scalar-x}
\end{figure}

Figures~\ref{fig:sv-maxwell-x} and \ref{fig:sv-scalar-x} show the ordinary Maxwell and massless scalar probe potentials as functions of $x/M$. At $a=0.75$, each wormhole profile has a single maximum at the throat. At $a=0.90$ and $0.975$, two barriers enclose a central well, and their separation grows toward $a_*=1$. Repeated reflections within this cavity can produce delayed pulses, with longer travel times for the wider cavity, i.e., the presence of echos, similar whaht it was recently reported in the literature \cite{Santos:2025xbk}. Each displayed black--bounce exterior has a single barrier between the metric horizon and spatial infinity.

The scalar and ordinary Maxwell probe spectra are obtained from a polynomial eigenvalue problem. Use $y = \Sigma/M$, $\delta = ‹a/M$ and $\widehat\omega = M\omega$. The radial derivative is
\begin{equation}
\frac{\mathrm{d}}{\mathrm{d}(x/M)} = 
\frac{1-2/y}{\sqrt\gamma}\sqrt{1-\frac{\delta^2}{y^2}}
\frac{\mathrm{d}}{\mathrm{d}y}.
\label{eq:qnm_spectral_derivative}
\end{equation}
The compact variable $X=1-2/y$ maps the metric horizon to $X=0$ and infinity to $X=1$. Introduce
\begin{equation}
 g(X)=\left[\frac{\mathrm{d}X}{\mathrm{d}(x/M)}\right]^2 = \frac{X^2(1-X)^4}{4\gamma}
\left[1-\frac{\delta^2(1-X)^2}{4}\right],\qquad h_0=\frac{\sqrt{1-\delta^2/4}}{2\sqrt\gamma}.
\label{eq:qnm_compact_functions}
\end{equation}
The radiative phases are included in the unknown through
\begin{equation}
\psi(X) = \mathrm{e}^{\mathrm{i}\widehat\omega s(X)}\chi(X),\qquad
s(X)=-\frac{\ln X}{h_0} - 2\sqrt\gamma\ln(1-X)+\frac{2\sqrt\gamma}{1-X}.
\label{eq:qnm_spectral_phase}
\end{equation}
Its logarithm at the horizon supplies the ingoing phase; its pole and logarithm at infinity supply the outgoing phase. The remaining regular equation is
\begin{equation}
\begin{split}
g\chi_{,XX} + \left(\frac{g_{,X}}2+2\mathrm{i}\widehat\omega g s_{,X}\right)\chi_{,X}
+\bigg[\mathrm{i}\widehat\omega\left(gs_{,XX}+\frac{g_{,X}s_{,X}}2\right)
+\widehat\omega^2(1-gs_{,X}^2)-\widehat V\bigg]\chi=0,
\end{split}
\label{eq:qnm_spectral_regular}
\end{equation}
where $\widehat V=M^2V$. Division by $X(1-X)^2$ gives finite endpoint limits. Evaluation at the Chebyshev points $X_k=[1+\cos(\pi k/N)]/2$, $k=0,\ldots,N$, with the derivative matrices gives
\begin{equation}
(\mathsf A_0+\widehat\omega\mathsf A_1+\widehat\omega^2\mathsf A_2)\boldsymbol\chi=0.
\label{eq:qnm_quadratic_eigenproblem}
\end{equation}
The regular endpoint limits impose the radiative conditions as part of this equation. The quadratic problem is solved by the equivalent generalized eigenvalue system
\begin{equation}
\begin{pmatrix}0&\mathsf I\\-\mathsf A_0&-\mathsf A_1\end{pmatrix}
\begin{pmatrix}\boldsymbol\chi\\\widehat\omega\boldsymbol\chi\end{pmatrix}
=\widehat\omega
\begin{pmatrix}\mathsf I&0\\0&\mathsf A_2\end{pmatrix}
\begin{pmatrix}\boldsymbol\chi\\\widehat\omega\boldsymbol\chi\end{pmatrix}.
\label{eq:qnm_linear_eigenproblem}
\end{equation}
This approach follows the spectral construction of Jansen \cite{Jansen2017}. We use $45$ digit arithmetic at $N=30,45,60,75$, and follow the fundamental oscillatory root across all four resolutions. An independent third order WKB calculation locates the initial root and provides an approximation comparison \cite{IyerWill1987,KonoplyaZhidenkoZinhailo2019}. For the twelve reported frequencies, the largest change from $N=30$ to $45$ is $5.92\times10^{-12}$, and the largest change from $N=60$ to $75$ is $1.30\times10^{-18}$. These changes measure spectral convergence. The eight digits displayed in Table~\ref{tab:probe_qnm} are relative to the resolution tests. The separate Schwarzschild limit $\gamma=1$, $a=0$, $j=2$ gives $M\omega=0.4836438722-0.0967587760\mathrm{i}$ for the scalar and $M\omega=0.4575955116-0.0950044258\mathrm{i}$ for Maxwell.

\begin{table}[tb]
\centering
\caption{Fundamental oscillatory resonances of the auxiliary massless scalar and ordinary Maxwell fields. The frequencies are obtained with $45$ digit arithmetic at spectral resolution $N=75$ and tracked through $N=30,45,60,75$. The parameters satisfy $\kappa=b=1$, $\xi=-1/5$, $\alpha=1/2$, $\beta=1$ and $\nu=4/5$. Every displayed entry belongs to the same matter action.}
\label{tab:probe_qnm}
\begin{tabular}{ccccc}
\toprule
$a$ & $a/M$ & $j$ & $M\omega_{\mathrm{scalar}}$ & $M\omega_{\mathrm{Maxwell}}$ \\
\midrule
1.15 & 1.512287 & 2 & $0.48573966-0.09351776\mathrm{i}$ & $0.45792375-0.09100373\mathrm{i}$ \\
1.15 & 1.512287 & 3 & $0.67690305-0.09323534\mathrm{i}$ & $0.65698566-0.09191618\mathrm{i}$ \\
1.30 & 1.183432 & 2 & $0.48628829-0.09948711\mathrm{i}$ & $0.45667596-0.09710975\mathrm{i}$ \\
1.30 & 1.183432 & 3 & $0.67728187-0.09919099\mathrm{i}$ & $0.65615490-0.09795363\mathrm{i}$ \\
1.50 & 0.888889 & 2 & $0.48651225-0.10342047\mathrm{i}$ & $0.45564023-0.10105367\mathrm{i}$ \\
1.50 & 0.888889 & 3 & $0.67743199-0.10310274\mathrm{i}$ & $0.65546237-0.10188487\mathrm{i}$ \\
\bottomrule
\end{tabular}
\end{table}

\begin{figure}[tb]
\centering\includegraphics[width=\linewidth]{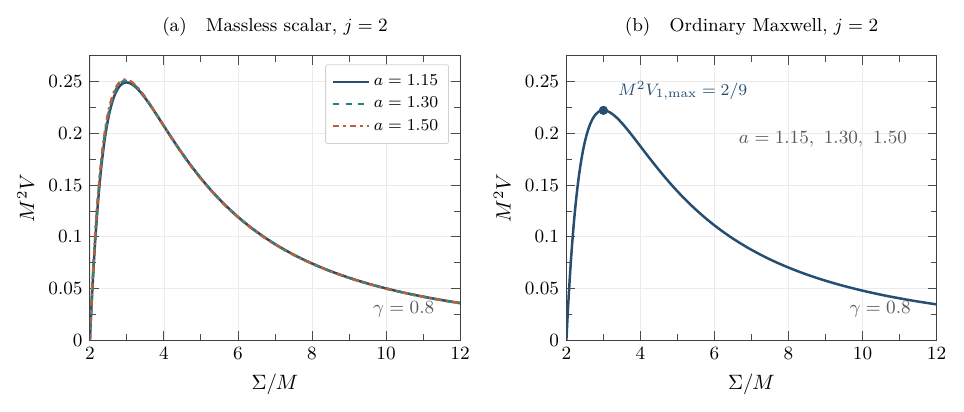}
\caption{Dimensionless scalar and ordinary Maxwell probe potentials at $j=2$, for the three solutions of Eq.~\eqref{eq:qnm_numerical_action}. The radial variable is the areal radius $\Sigma/M$. The metric horizon is at $\Sigma/M=2$. }
\label{fig:qnm_probe_potentials}
\end{figure}

\begin{figure}[tb]
\centering\includegraphics[width=\linewidth]{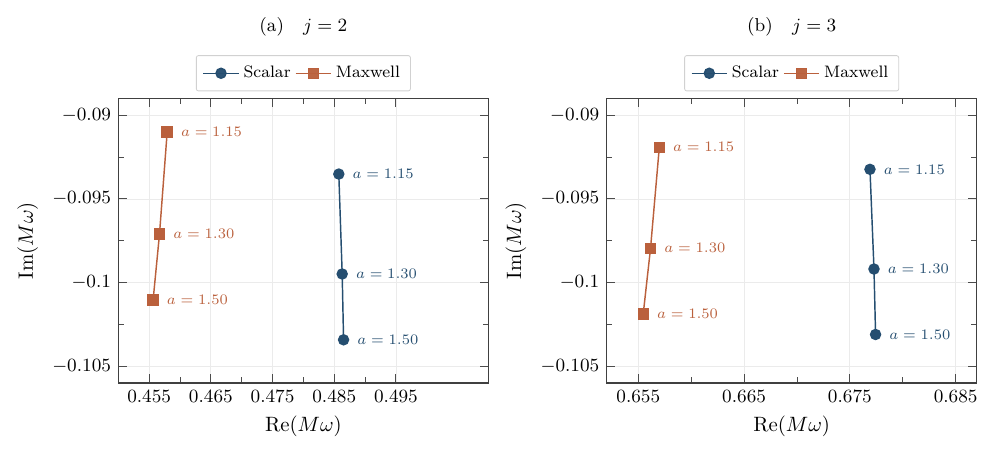}
\caption{Fundamental probe resonances at $j=2$ and $j=3$. Circles denote the auxiliary massless scalar and squares the ordinary Maxwell field. Blue identifies the scalar and orange the Maxwell field; labels identify $a=1.15,1.30,1.50$ within one action. The vertical coordinate is $\operatorname{Im}(M\omega)$, so the distance below the real axis gives the damping rate. }
\label{fig:qnm_probe_spectrum}
\end{figure}

Figure~\ref{fig:qnm_probe_potentials} displays the two probe barriers, and Fig.~\ref{fig:qnm_probe_spectrum} displays their resonances. Along the chosen family, the scalar oscillation frequency changes mildly in units of $M$, while its damping rate increases as $a/M$ decreases. The ordinary Maxwell oscillation frequency decreases mildly and its damping rate also increases. These variations follow the two explicit potentials. The Maxwell potential as a function of $y=\Sigma/M$ is $M^2V_{\mathrm M}=J(1-2/y)/y^2$, and is identical for the three solutions. Its spectral variation comes from the $a/M$ dependence of the tortoise map in Eq.~\eqref{eq:qnm_spectral_derivative}. Their common large $j$ scale is Eq.~\eqref{eq:qnm_ordinary_ray}; the scalar derivative term supplies a finite $j$ difference from Maxwell. Changing $a$ along this family changes $M$, the magnetic flux and the asymptotic configuration together, as required by the fixed source functions.


The physical tensor and transverse bumblebee families have the characteristic scales in Eq.~\eqref{eq:qnm_ray_frequencies}. Table~\ref{tab:qnm_horizons} gives their horizons, and Table~\ref{tab:qnm_eikonal} gives the leading estimates of Eq.~\eqref{eq:qnm_eikonal_estimate}. The root locations and derivatives were computed with $50$ digit arithmetic, with an independent calculation of the same stationary rays. The precision of a root calculation and the accuracy of an eikonal approximation are separate quantities. In Table~\ref{tab:qnm_eikonal}, the displayed precision describes evaluation of the leading formula; its finite $j$ corrections are determined by Eqs.~\eqref{eq:qnm_coupled_em}--\eqref{eq:qnm_Cr}. In this sense, the table shows characteristic resonance estimates, and Table~\ref{tab:probe_qnm} gives numerically resolved probe eigenfrequencies.

\begin{table}[tb]
\centering
\caption{Characteristic horizons for the fixed action in Eq.~\eqref{eq:qnm_numerical_action}. The tensor horizon lies outside the ordinary metric horizon, and the transverse bumblebee horizon lies inside it. Radii are normalized by the corresponding $M$.}
\label{tab:qnm_horizons}
\begin{tabular}{crrr}\toprule
$a$ & $r_h/M$ & $r_T/M$ & $r_B/M$ \\
\midrule
1.15 & 1.308811 & 2.491334 & 1.231875 \\
1.30 & 1.612293 & 3.668818 & 1.522793 \\
1.50 & 1.791613 & 8.687944 & 1.673043 \\
\bottomrule\end{tabular}

\end{table}
\begin{table}[tb]
\centering
\caption{Leading characteristic geometric optics estimates for the actual tensor and transverse bumblebee eigenfamilies, using Eq.~\eqref{eq:qnm_eikonal_estimate} with $n=0$ and adiabatic separation. These are evaluations of the leading formula at $j=10,20$. }
\label{tab:qnm_eikonal}
\begin{tabular}{ccrr}\toprule
$a$ & Mode & $M\omega^{\mathrm{eik}}_{j=10,0}$ & $M\omega^{\mathrm{eik}}_{j=20,0}$ \\
\midrule
1.15 & Tensor & $1.015112-0.033519\,\mathrm{i}$ & $1.981886-0.033519\,\mathrm{i}$ \\
1.15 & Bumblebee & $2.116627-0.099735\,\mathrm{i}$ & $4.132462-0.099735\,\mathrm{i}$ \\
1.30 & Tensor & $0.647248-0.018774\,\mathrm{i}$ & $1.263674-0.018774\,\mathrm{i}$ \\
1.30 & Bumblebee & $2.155496-0.109005\,\mathrm{i}$ & $4.208349-0.109005\,\mathrm{i}$ \\
1.50 & Tensor & $0.181960-0.003411\,\mathrm{i}$ & $0.355256-0.003411\,\mathrm{i}$ \\
1.50 & Bumblebee & $2.216778-0.117892\,\mathrm{i}$ & $4.327995-0.117892\,\mathrm{i}$ \\
\bottomrule\end{tabular}
\end{table}

\begin{figure}[tb]
\centering\includegraphics[width=\linewidth]{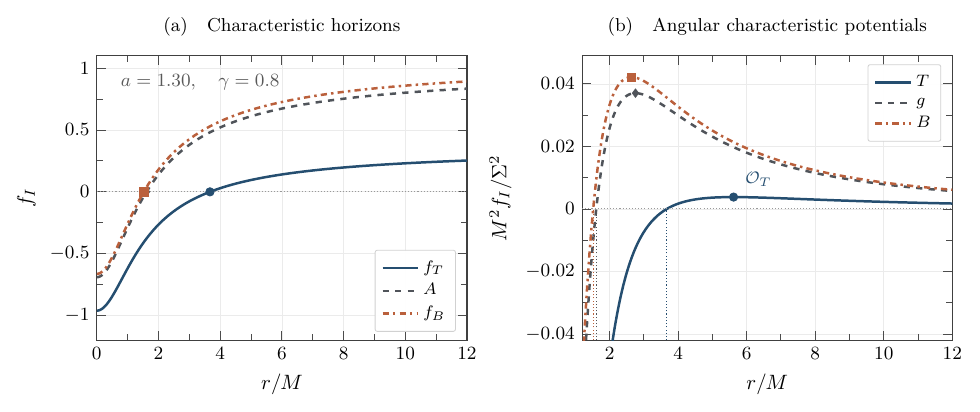}
\caption{Characteristic horizon functions and trapping potentials for $a=1.30$. The three radial functions are $f_T$, $A$ and $f_B$, and the trapping functions are $M^2f_I/\Sigma^2$. Their zeroes identify the tensor, ordinary and transverse bumblebee horizons; their exterior maxima determine the circular rays used in the leading eikonal estimates.}
\label{fig:qnm_characteristics}
\end{figure}

The separation of the tensor and metric horizons is substantial for these cased under consideration. At $a=1.50$, the tensor horizon is $r_T\simeq14.6609$, whereas $r_h\simeq3.02335$, and its unstable circular ray is at $r_{c,T}\simeq21.9799$. Its smaller $\Omega_T$ and $\Lambda_T$ give a slower oscillation and weaker leading damping in the normalized Killing time. The transverse bumblebee family has its horizon inside the metric horizon and its trapping radius close to the ordinary circular ray. Figure~\ref{fig:qnm_characteristics} displays both the horizon functions and the characteristic trapping potentials for $a=1.30$. These shifts are generated by the temporal configuration in Eq.~\eqref{eq:qnm_horizon_functions}; a constant radial deformation alone would omit this propagation information.

\subsection{Stability of the reconstructed field configuration}
\label{subsec:qnm_stability}

\begin{figure}[tb]
\centering\includegraphics[width=\linewidth]{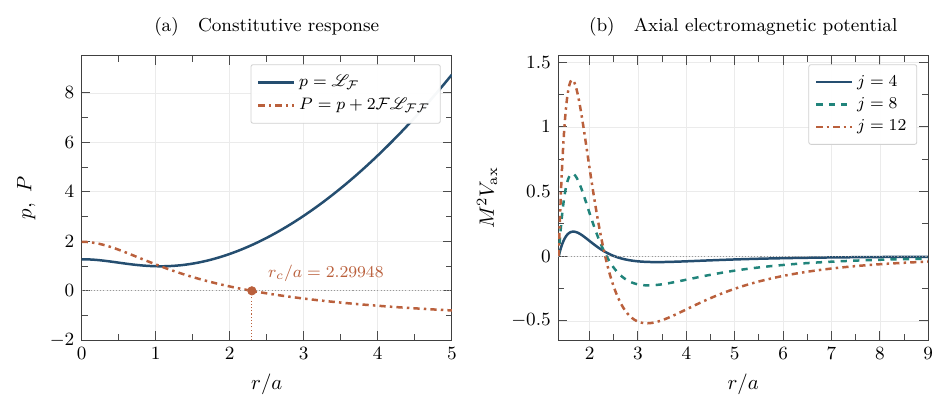}
\caption{The fixed action electromagnetic response and its extraordinary axial potential. The left panel shows $p$ and $P$; the zero of $P$ is marked at $r/a\simeq2.29948$. The right panel shows $M^2V_{\mathrm{em,ax}}$ for $a=1.30$ and $j=4,8,12$. The growing angular contribution produces a negative exterior well. }
\label{fig:qnm_constitutive}
\end{figure}

The same fixed action gives a quantitative electromagnetic instability diagnostic. For Eq.~\eqref{eq:qnm_numerical_action}, let $k=\sqrt{2/5}$, $c=\cosh(kz)$ and $s=\sinh(kz)$. Differentiation of the explicit source gives
\begin{equation}
p(z)=\frac32\cos z - \frac4{25}\mathcal B(z),\qquad \mathcal B(z) = \frac75c^2+\frac{12}{5}s^2-\tan^2z\,c^2-2k\tan z\,sc.
\label{eq:qnm_explicit_p}
\end{equation}
Its throat values are $p(0)=319/250$ and $P(0)=4953/2500$. Positivity of $p$ can be established throughout the invariant interval. On $0\leq z\leq11/10$, $\mathcal B\leq(7/5)(1+3s^2)$ gives $p\geq0.07516$. On $11/10\leq z<\pi/2$,
\begin{equation}
\frac{\mathcal B}{c^2}\leq \frac{19}{5} -\tan^2(11/10) - 2k\tan(11/10)\tanh(11k/10)<-1.5554,
\label{eq:qnm_p_positive_bound}
\end{equation}
so $p>3\cos z/2>0$. Evenness extends the result to the other end. Numerical minimization gives $p_{\min}\simeq0.9921555$ at $z\simeq0.8177153$. The first zero of $P$ is at $z_c\simeq1.1605858$, or $r_c/a\simeq2.2994771$, and its limiting value is
\begin{equation}
P(+\infty)=-\frac{56}{125}\cosh\!\left(\sqrt{\frac25}\,\pi\right)<0.
\label{eq:qnm_P_limit}
\end{equation}
Figure~\ref{fig:qnm_constitutive} shows the response functions and the extraordinary electromagnetic potential. The negative angular contribution grows in magnitude with $J$, in accordance with Eq.~\eqref{eq:qnm_ned_rayleigh}.

For a constructive bound choose $\chi(r)=\sin[\pi(r-12a)/(8a)]$ on $12a\leq r\leq20a$, with zero continuation outside. This compact interval lies outside every metric, tensor and bumblebee horizon in the sample and inside the region with $P<0$. It tests a finite annulus of the common exterior. Let us define
\begin{equation}
\begin{split}
\mathcal N& = \int_{12a}^{20a}\frac{pA}{\sqrt\gamma} \left(\frac{\mathrm{d}\chi}{\mathrm{d}r}\right)^2\mathrm{d}r,\\ \mathcal D&=\int_{12a}^{20a}\frac{\sqrt\gamma(-P)}{\Sigma^2}\chi^2\mathrm{d}r,\qquad \mathcal W=\int_{12a}^{20a}\frac{\sqrt\gamma\,p}{A}\chi^2\mathrm{d}r.
\end{split}
\label{eq:qnm_trial_integrals}
\end{equation}
All three integrals are positive. The trial quotient is $(\mathcal N-J\mathcal D)/\mathcal W$, and the variational principle yields
\begin{equation}
J > \frac{\mathcal N}{\mathcal D} \quad\Longrightarrow\quad \Gamma\geq\sqrt{\frac{J\mathcal D-\mathcal N}{\mathcal W}}.
\label{eq:qnm_growth_bound}
\end{equation}
Table~\ref{tab:qnm_instability} gives sufficient angular indices and lower bounds evaluated with $40$ digit arithmetic. The chosen indices establish an instability with this particular trial function; they give no information about the first unstable harmonic. These bounds belong to the fixed metric electromagnetic master equation. 

\begin{table}[tb]
\centering
\caption{Constructive electromagnetic instability bounds for one action. Each displayed $j$ satisfies Eq.~\eqref{eq:qnm_growth_bound}. The growth rate refers to $\exp(\Gamma t)$ in the length units of Eq.~\eqref{eq:qnm_numerical_action}. The entries are variational lower bounds for the fixed metric extraordinary electromagnetic operator, with the last column rounded down.}
\label{tab:qnm_instability}
\begin{tabular}{cccc}
\toprule
$a$ & $\mathcal N/\mathcal D$ & sufficient $j$ & lower bound on $\Gamma$\\
\midrule
1.15 & 3213.3995 & 57 & 0.06032\\
1.30 & 3136.0038 & 56 & 0.04098\\
1.50 & 3018.0676 & 55 & 0.03661\\
\bottomrule
\end{tabular}
\end{table}

The nonnegative scalar and ordinary Maxwell probe potentials establish their mode stability. The physical axial metric system has positive principal kinetic coefficients when $\gamma,d,p>0$, and the characteristic calculation identifies its propagation horizons and leading resonance scales. The finite $j$ dynamics are governed by the complete constrained equations in Eqs.~\eqref{eq:qnm_coupled_em}--\eqref{eq:qnm_Cr}; the numerical probe spectra and the leading characteristic estimates keep the distinct meanings specified in their tables.

In summary, we have seen that the full reconstructed solution has an extraordinary electromagnetic angular instability for $\eta<0$ and $\nu>0$. Its growing characteristic occurs in a finite annulus outside all mode horizons and survives the coupling to the other fields. The phantom scalar adds an independent negative kinetic coefficient in the polar system.


\section{Conclusion}
\label{sec:conclusion}

We constructed an exact field realization of the Simpson--Visser geometry in metric bumblebee gravity. A phantom scalar, minimally coupled magnetic nonlinear electrodynamics, and a spacelike configuration of fixed norm supplied the required sources. The independent vector equation determined a local multiplier through the curvature and related the temporal setting profile to the electromagnetic action. We reconstructed the scalar potential and electromagnetic Lagrangian explicitly and verified the complete background equations and constraint identities.

The temporal configuration admitted a smooth continuation across the metric horizons in the stated ingoing domain, while its contraction with the horizon generator excluded a smooth stationary extension to the bifurcation sphere. The fixed combinations $\alpha=M/a^{3}$, $\beta=|q|/a^{2}$, and $\nu=u_{0}/a$ identified one matter action with a continuous family of traversable wormholes, null minimum configurations, and black--bounces. Along this family, the lapse mass parameter, magnetic flux, and asymptotic orientation varied together. The scalar and effective bumblebee contributions accounted for the radial null energy condition violation, and the magnetic contribution to that contraction vanished.

In the black hole exterior, the metric geodesics retained the areal photon sphere radius $3M$, the critical impact parameter $3\sqrt{3}M$, the marginally bound radius $4M$, and the innermost stable circular orbit radius $6M$ at fixed lapse mass. The bound orbit separatrix also retained its Schwarzschild form. The bounce and bumblebee parameters modified the weak and strong angular deflection, the photon orbit instability, the radial epicyclic frequency, and the angular and temporal periods of bound motion. The distinction between fixed coordinate launch velocities and fixed conserved quantities explained the different capture regimes in the numerical trajectories.

The static observer calculation yielded the same metric shadow as Schwarzschild at fixed $M$ and areal observer position. The adopted Sgr A* and M87* intervals contained the common geometric prediction, but supplied no separate determination of $a$ or $\ell$. This comparison used the published angular mass priors and image calibrations.

We derived a closed constrained axial system for the metric, transverse bumblebee, and ordinary electromagnetic amplitudes. Its principal action yielded distinct propagation cones under the stated kinetic positivity conditions. In the numerical configurations examined, the tensor horizon lay outside the metric horizon and the transverse bumblebee horizon lay inside it. Chebyshev spectral collocation produced converged fundamental auxiliary massless scalar and ordinary Maxwell probe frequencies, with independent WKB comparisons and Schwarzschild limit checks. Their dimensionless damping rates increased as $a/M$ decreased along the selected fixed action family. The tensor and bumblebee resonance scales were obtained only at leading geometric optics order; the complete coupled equations specified their dynamics at finite angular momentum.

The nonnegative scalar and ordinary Maxwell probe potentials excluded exponentially growing probe modes. The phantom source scalar retained a negative kinetic coefficient, but an exponentially growing scalar mode was not demonstrated. For $\eta<0$ and $\nu>0$, the supporting electromagnetic action yielded $p>0$ and $P<0$ in a finite exterior region, which generated an angular gradient instability in the extraordinary polarization. A compactly supported variational construction established growing modes of the electromagnetic operator on the fixed geometry and supplied sufficient angular indices and lower bounds on their growth rates. The same angular characteristic persisted in the coupled principal determinant, and the source instability survived interaction with the remaining fields.

Finally, the corresponding field reconstruction in metric-affine bumblebee gravity and the influence of nonmetricity on the propagation and stability of the supporting fields remained unresolved.


\section*{Acknowledgments}
\hspace{0.5cm} A. A. Araújo Filho is supported by Conselho Nacional de Desenvolvimento Cient\'{\i}fico e Tecnol\'{o}gico (CNPq) -- [150223/2025-0]. Also, the author thanks C. F. S. Pereira for encouraging the work that led to this new solution in bumblebee gravity.

\section*{Data Availability Statement}

Data associated with this study consist of the analytical expressions and numerical figures presented in the manuscript. No additional dataset is required to reproduce the analytical results. The code used to generate the analytical/numerical results and figures is available from the corresponding author upon reasonable request.

\bibliography{main}

\begin{thebibliography}{74}%
\makeatletter
\providecommand \@ifxundefined [1]{%
 \@ifx{#1\undefined}
}%
\providecommand \@ifnum [1]{%
 \ifnum #1\expandafter \@firstoftwo
 \else \expandafter \@secondoftwo
 \fi
}%
\providecommand \@ifx [1]{%
 \ifx #1\expandafter \@firstoftwo
 \else \expandafter \@secondoftwo
 \fi
}%
\providecommand \natexlab [1]{#1}%
\providecommand \enquote  [1]{``#1''}%
\providecommand \bibnamefont  [1]{#1}%
\providecommand \bibfnamefont [1]{#1}%
\providecommand \citenamefont [1]{#1}%
\providecommand \href@noop [0]{\@secondoftwo}%
\providecommand \href [0]{\begingroup \@sanitize@url \@href}%
\providecommand \@href[1]{\@@startlink{#1}\@@href}%
\providecommand \@@href[1]{\endgroup#1\@@endlink}%
\providecommand \@sanitize@url [0]{\catcode `\\12\catcode `\$12\catcode
  `\&12\catcode `\#12\catcode `\^12\catcode `\_12\catcode `\%12\relax}%
\providecommand \@@startlink[1]{}%
\providecommand \@@endlink[0]{}%
\providecommand \url  [0]{\begingroup\@sanitize@url \@url }%
\providecommand \@url [1]{\endgroup\@href {#1}{\urlprefix }}%
\providecommand \urlprefix  [0]{URL }%
\providecommand \Eprint [0]{\href }%
\providecommand \doibase [0]{http://dx.doi.org/}%
\providecommand \selectlanguage [0]{\@gobble}%
\providecommand \bibinfo  [0]{\@secondoftwo}%
\providecommand \bibfield  [0]{\@secondoftwo}%
\providecommand \translation [1]{[#1]}%
\providecommand \BibitemOpen [0]{}%
\providecommand \bibitemStop [0]{}%
\providecommand \bibitemNoStop [0]{.\EOS\space}%
\providecommand \EOS [0]{\spacefactor3000\relax}%
\providecommand \BibitemShut  [1]{\csname bibitem#1\endcsname}%
\let\auto@bib@innerbib\@empty
\bibitem [{\citenamefont {Kosteleck{\'y}}\ and\ \citenamefont
  {Samuel}(1989)}]{KosteleckySamuel1989}%
  \BibitemOpen
  \bibfield  {author} {\bibinfo {author} {\bibfnamefont {V.~Alan}\ \bibnamefont
  {Kosteleck{\'y}}}\ and\ \bibinfo {author} {\bibfnamefont {Stuart}\
  \bibnamefont {Samuel}},\ }\bibfield  {title} {\enquote {\bibinfo {title}
  {Spontaneous breaking of {Lorentz} symmetry in string theory},}\ }\href
  {\doibase 10.1103/PhysRevD.39.683} {\bibfield  {journal} {\bibinfo  {journal}
  {Physical Review D}\ }\textbf {\bibinfo {volume} {39}},\ \bibinfo {pages}
  {683--685} (\bibinfo {year} {1989})}\BibitemShut {NoStop}%
\bibitem [{\citenamefont {Colladay}\ and\ \citenamefont
  {Kosteleck{\'y}}(1997)}]{ColladayKostelecky1997}%
  \BibitemOpen
  \bibfield  {author} {\bibinfo {author} {\bibfnamefont {Don}\ \bibnamefont
  {Colladay}}\ and\ \bibinfo {author} {\bibfnamefont {V.~Alan}\ \bibnamefont
  {Kosteleck{\'y}}},\ }\bibfield  {title} {\enquote {\bibinfo {title} {{CPT}
  violation and the standard model},}\ }\href {\doibase
  10.1103/PhysRevD.55.6760} {\bibfield  {journal} {\bibinfo  {journal}
  {Physical Review D}\ }\textbf {\bibinfo {volume} {55}},\ \bibinfo {pages}
  {6760--6774} (\bibinfo {year} {1997})},\ \Eprint
  {http://arxiv.org/abs/hep-ph/9703464} {arXiv:hep-ph/9703464 [hep-ph]}
  \BibitemShut {NoStop}%
\bibitem [{\citenamefont {Colladay}\ and\ \citenamefont
  {Kosteleck{\'y}}(1998)}]{ColladayKostelecky1998}%
  \BibitemOpen
  \bibfield  {author} {\bibinfo {author} {\bibfnamefont {Don}\ \bibnamefont
  {Colladay}}\ and\ \bibinfo {author} {\bibfnamefont {V.~Alan}\ \bibnamefont
  {Kosteleck{\'y}}},\ }\bibfield  {title} {\enquote {\bibinfo {title}
  {{Lorentz}-violating extension of the standard model},}\ }\href {\doibase
  10.1103/PhysRevD.58.116002} {\bibfield  {journal} {\bibinfo  {journal}
  {Physical Review D}\ }\textbf {\bibinfo {volume} {58}},\ \bibinfo {pages}
  {116002} (\bibinfo {year} {1998})},\ \Eprint
  {http://arxiv.org/abs/hep-ph/9809521} {arXiv:hep-ph/9809521 [hep-ph]}
  \BibitemShut {NoStop}%
\bibitem [{\citenamefont {Kosteleck{\'y}}(2004)}]{Kostelecky2004}%
  \BibitemOpen
  \bibfield  {author} {\bibinfo {author} {\bibfnamefont {V.~Alan}\ \bibnamefont
  {Kosteleck{\'y}}},\ }\bibfield  {title} {\enquote {\bibinfo {title} {Gravity,
  {Lorentz} violation, and the standard model},}\ }\href {\doibase
  10.1103/PhysRevD.69.105009} {\bibfield  {journal} {\bibinfo  {journal}
  {Physical Review D}\ }\textbf {\bibinfo {volume} {69}},\ \bibinfo {pages}
  {105009} (\bibinfo {year} {2004})},\ \Eprint
  {http://arxiv.org/abs/hep-th/0312310} {arXiv:hep-th/0312310 [hep-th]}
  \BibitemShut {NoStop}%
\bibitem [{\citenamefont {Bailey}\ and\ \citenamefont
  {Kosteleck{\'y}}(2006)}]{BaileyKostelecky2006}%
  \BibitemOpen
  \bibfield  {author} {\bibinfo {author} {\bibfnamefont {Quentin~G.}\
  \bibnamefont {Bailey}}\ and\ \bibinfo {author} {\bibfnamefont {V.~Alan}\
  \bibnamefont {Kosteleck{\'y}}},\ }\bibfield  {title} {\enquote {\bibinfo
  {title} {Signals for {Lorentz} violation in post-{Newtonian} gravity},}\
  }\href {\doibase 10.1103/PhysRevD.74.045001} {\bibfield  {journal} {\bibinfo
  {journal} {Physical Review D}\ }\textbf {\bibinfo {volume} {74}},\ \bibinfo
  {pages} {045001} (\bibinfo {year} {2006})},\ \Eprint
  {http://arxiv.org/abs/gr-qc/0603030} {arXiv:gr-qc/0603030 [gr-qc]}
  \BibitemShut {NoStop}%
\bibitem [{\citenamefont {Bluhm}\ and\ \citenamefont
  {Kosteleck{\'y}}(2005)}]{BluhmKostelecky2005}%
  \BibitemOpen
  \bibfield  {author} {\bibinfo {author} {\bibfnamefont {Robert}\ \bibnamefont
  {Bluhm}}\ and\ \bibinfo {author} {\bibfnamefont {V.~Alan}\ \bibnamefont
  {Kosteleck{\'y}}},\ }\bibfield  {title} {\enquote {\bibinfo {title}
  {Spontaneous {Lorentz} violation, {Nambu--Goldstone} modes, and gravity},}\
  }\href {\doibase 10.1103/PhysRevD.71.065008} {\bibfield  {journal} {\bibinfo
  {journal} {Physical Review D}\ }\textbf {\bibinfo {volume} {71}},\ \bibinfo
  {pages} {065008} (\bibinfo {year} {2005})},\ \Eprint
  {http://arxiv.org/abs/hep-th/0412320} {arXiv:hep-th/0412320 [hep-th]}
  \BibitemShut {NoStop}%
\bibitem [{\citenamefont {Bluhm}\ \emph
  {et~al.}(2008{\natexlab{a}})\citenamefont {Bluhm}, \citenamefont {Fung},\
  and\ \citenamefont {Kosteleck{\'y}}}]{BluhmFungKostelecky2008}%
  \BibitemOpen
  \bibfield  {author} {\bibinfo {author} {\bibfnamefont {Robert}\ \bibnamefont
  {Bluhm}}, \bibinfo {author} {\bibfnamefont {Shu-Hong}\ \bibnamefont {Fung}},
  \ and\ \bibinfo {author} {\bibfnamefont {V.~Alan}\ \bibnamefont
  {Kosteleck{\'y}}},\ }\bibfield  {title} {\enquote {\bibinfo {title}
  {Spontaneous {Lorentz} and diffeomorphism violation, massive modes, and
  gravity},}\ }\href {\doibase 10.1103/PhysRevD.77.065020} {\bibfield
  {journal} {\bibinfo  {journal} {Physical Review D}\ }\textbf {\bibinfo
  {volume} {77}},\ \bibinfo {pages} {065020} (\bibinfo {year}
  {2008}{\natexlab{a}})},\ \Eprint {http://arxiv.org/abs/0712.4119}
  {arXiv:0712.4119 [hep-th]} \BibitemShut {NoStop}%
\bibitem [{\citenamefont {Bluhm}\ \emph
  {et~al.}(2008{\natexlab{b}})\citenamefont {Bluhm}, \citenamefont {Gagne},
  \citenamefont {Potting},\ and\ \citenamefont
  {Vrublevskis}}]{BluhmGagnePottingVrublevskis2008}%
  \BibitemOpen
  \bibfield  {author} {\bibinfo {author} {\bibfnamefont {Robert}\ \bibnamefont
  {Bluhm}}, \bibinfo {author} {\bibfnamefont {Nolan~L.}\ \bibnamefont {Gagne}},
  \bibinfo {author} {\bibfnamefont {Robertus}\ \bibnamefont {Potting}}, \ and\
  \bibinfo {author} {\bibfnamefont {Arturs}\ \bibnamefont {Vrublevskis}},\
  }\bibfield  {title} {\enquote {\bibinfo {title} {Constraints and stability in
  vector theories with spontaneous {Lorentz} violation},}\ }\href {\doibase
  10.1103/PhysRevD.77.125007} {\bibfield  {journal} {\bibinfo  {journal}
  {Physical Review D}\ }\textbf {\bibinfo {volume} {77}},\ \bibinfo {pages}
  {125007} (\bibinfo {year} {2008}{\natexlab{b}})},\ \bibinfo {note} {erratum:
  Physical Review D 79, 029902 (2009)},\ \Eprint
  {http://arxiv.org/abs/0802.4071} {arXiv:0802.4071 [hep-th]} \BibitemShut
  {NoStop}%
\bibitem [{\citenamefont {Casana}\ \emph {et~al.}(2018)\citenamefont {Casana},
  \citenamefont {Cavalcante}, \citenamefont {Poulis},\ and\ \citenamefont
  {Santos}}]{Casana2018}%
  \BibitemOpen
  \bibfield  {author} {\bibinfo {author} {\bibfnamefont {R.}~\bibnamefont
  {Casana}}, \bibinfo {author} {\bibfnamefont {A.}~\bibnamefont {Cavalcante}},
  \bibinfo {author} {\bibfnamefont {F.~P.}\ \bibnamefont {Poulis}}, \ and\
  \bibinfo {author} {\bibfnamefont {E.~B.}\ \bibnamefont {Santos}},\ }\bibfield
   {title} {\enquote {\bibinfo {title} {Exact {Schwarzschild}-like solution in
  a bumblebee gravity model},}\ }\href {\doibase 10.1103/PhysRevD.97.104001}
  {\bibfield  {journal} {\bibinfo  {journal} {Physical Review D}\ }\textbf
  {\bibinfo {volume} {97}},\ \bibinfo {pages} {104001} (\bibinfo {year}
  {2018})},\ \Eprint {http://arxiv.org/abs/1711.02273} {arXiv:1711.02273
  [gr-qc]} \BibitemShut {NoStop}%
\bibitem [{\citenamefont {Ara{\'u}jo~Filho}\ \emph
  {et~al.}(2026{\natexlab{a}})\citenamefont {Ara{\'u}jo~Filho}, \citenamefont
  {Heidari}, \citenamefont {Lobo}, \citenamefont {Shi},\ and\ \citenamefont
  {Lobo}}]{AraujoFilho:2025rvn}%
  \BibitemOpen
  \bibfield  {author} {\bibinfo {author} {\bibfnamefont {A.~A.}\ \bibnamefont
  {Ara{\'u}jo~Filho}}, \bibinfo {author} {\bibfnamefont {N.}~\bibnamefont
  {Heidari}}, \bibinfo {author} {\bibfnamefont {Iarley~P.}\ \bibnamefont
  {Lobo}}, \bibinfo {author} {\bibfnamefont {Yuxuan}\ \bibnamefont {Shi}}, \
  and\ \bibinfo {author} {\bibfnamefont {Francisco S.~N.}\ \bibnamefont
  {Lobo}},\ }\bibfield  {title} {\enquote {\bibinfo {title} {{The flight of the
  bumblebee in a non-commutative geometry: A new black hole solution}},}\
  }\href {\doibase 10.1016/j.aop.2026.170487} {\bibfield  {journal} {\bibinfo
  {journal} {Annals Phys.}\ }\textbf {\bibinfo {volume} {490}},\ \bibinfo
  {pages} {170487} (\bibinfo {year} {2026}{\natexlab{a}})},\ \Eprint
  {http://arxiv.org/abs/2509.17867} {arXiv:2509.17867 [gr-qc]} \BibitemShut
  {NoStop}%
\bibitem [{\citenamefont {Maluf}\ and\ \citenamefont
  {Neves}(2021)}]{MalufNeves2021}%
  \BibitemOpen
  \bibfield  {author} {\bibinfo {author} {\bibfnamefont {R.~V.}\ \bibnamefont
  {Maluf}}\ and\ \bibinfo {author} {\bibfnamefont {Juliano C.~S.}\ \bibnamefont
  {Neves}},\ }\bibfield  {title} {\enquote {\bibinfo {title} {Black holes with
  a cosmological constant in bumblebee gravity},}\ }\href {\doibase
  10.1103/PhysRevD.103.044002} {\bibfield  {journal} {\bibinfo  {journal}
  {Physical Review D}\ }\textbf {\bibinfo {volume} {103}},\ \bibinfo {pages}
  {044002} (\bibinfo {year} {2021})},\ \Eprint
  {http://arxiv.org/abs/2011.12841} {arXiv:2011.12841 [gr-qc]} \BibitemShut
  {NoStop}%
\bibitem [{\citenamefont {Xu}\ \emph {et~al.}(2023)\citenamefont {Xu},
  \citenamefont {Liang},\ and\ \citenamefont {Shao}}]{XuLiangShao2023}%
  \BibitemOpen
  \bibfield  {author} {\bibinfo {author} {\bibfnamefont {Rui}\ \bibnamefont
  {Xu}}, \bibinfo {author} {\bibfnamefont {Dicong}\ \bibnamefont {Liang}}, \
  and\ \bibinfo {author} {\bibfnamefont {Lijing}\ \bibnamefont {Shao}},\
  }\bibfield  {title} {\enquote {\bibinfo {title} {Static spherical vacuum
  solutions in the bumblebee gravity model},}\ }\href {\doibase
  10.1103/PhysRevD.107.024011} {\bibfield  {journal} {\bibinfo  {journal}
  {Physical Review D}\ }\textbf {\bibinfo {volume} {107}},\ \bibinfo {pages}
  {024011} (\bibinfo {year} {2023})},\ \Eprint
  {http://arxiv.org/abs/2209.02209} {arXiv:2209.02209 [gr-qc]} \BibitemShut
  {NoStop}%
\bibitem [{\citenamefont {Liu}\ \emph {et~al.}(2025)\citenamefont {Liu},
  \citenamefont {Guo}, \citenamefont {Wei},\ and\ \citenamefont
  {Liu}}]{Liu2025Charged}%
  \BibitemOpen
  \bibfield  {author} {\bibinfo {author} {\bibfnamefont {Jia-Zhou}\
  \bibnamefont {Liu}}, \bibinfo {author} {\bibfnamefont {Wen-Di}\ \bibnamefont
  {Guo}}, \bibinfo {author} {\bibfnamefont {Shao-Wen}\ \bibnamefont {Wei}}, \
  and\ \bibinfo {author} {\bibfnamefont {Yu-Xiao}\ \bibnamefont {Liu}},\
  }\bibfield  {title} {\enquote {\bibinfo {title} {Charged spherically
  symmetric and slowly rotating charged black hole solutions in bumblebee
  gravity},}\ }\href {\doibase 10.1140/epjc/s10052-025-13859-x} {\bibfield
  {journal} {\bibinfo  {journal} {European Physical Journal C}\ }\textbf
  {\bibinfo {volume} {85}},\ \bibinfo {pages} {145} (\bibinfo {year} {2025})},\
  \Eprint {http://arxiv.org/abs/2407.08396} {arXiv:2407.08396 [gr-qc]}
  \BibitemShut {NoStop}%
\bibitem [{\citenamefont {Liu}\ \emph {et~al.}(2026{\natexlab{a}})\citenamefont
  {Liu}, \citenamefont {Wu}, \citenamefont {Wei},\ and\ \citenamefont
  {Liu}}]{Liu2026MixedVEVs}%
  \BibitemOpen
  \bibfield  {author} {\bibinfo {author} {\bibfnamefont {Jia-Zhou}\
  \bibnamefont {Liu}}, \bibinfo {author} {\bibfnamefont {Shan-Ping}\
  \bibnamefont {Wu}}, \bibinfo {author} {\bibfnamefont {Shao-Wen}\ \bibnamefont
  {Wei}}, \ and\ \bibinfo {author} {\bibfnamefont {Yu-Xiao}\ \bibnamefont
  {Liu}},\ }\bibfield  {title} {\enquote {\bibinfo {title} {Exact black hole
  solutions in bumblebee gravity with lightlike or spacelike {VEVs}},}\ }\href
  {\doibase 10.1007/s11433-026-2961-8} {\bibfield  {journal} {\bibinfo
  {journal} {Science China Physics, Mechanics \& Astronomy}\ }\textbf {\bibinfo
  {volume} {69}},\ \bibinfo {pages} {270411} (\bibinfo {year}
  {2026}{\natexlab{a}})},\ \Eprint {http://arxiv.org/abs/2510.16731}
  {arXiv:2510.16731 [gr-qc]} \BibitemShut {NoStop}%
\bibitem [{\citenamefont {Zhu}\ and\ \citenamefont
  {Li}(2026)}]{ZhuLi2026Classification}%
  \BibitemOpen
  \bibfield  {author} {\bibinfo {author} {\bibfnamefont {Jie}\ \bibnamefont
  {Zhu}}\ and\ \bibinfo {author} {\bibfnamefont {Hao}\ \bibnamefont {Li}},\
  }\bibfield  {title} {\enquote {\bibinfo {title} {Full classification of
  static spherical vacuum solutions to bumblebee gravity with general
  {VEVs}},}\ }\href {\doibase 10.1016/j.physletb.2026.140396} {\bibfield
  {journal} {\bibinfo  {journal} {Physics Letters B}\ }\textbf {\bibinfo
  {volume} {876}},\ \bibinfo {pages} {140396} (\bibinfo {year} {2026})},\
  \Eprint {http://arxiv.org/abs/2511.03231} {arXiv:2511.03231 [gr-qc]}
  \BibitemShut {NoStop}%
\bibitem [{\citenamefont {Delhom}\ \emph {et~al.}(2021)\citenamefont {Delhom},
  \citenamefont {Nascimento}, \citenamefont {Olmo}, \citenamefont {Petrov},\
  and\ \citenamefont {Porf{\'i}rio}}]{Delhom2021}%
  \BibitemOpen
  \bibfield  {author} {\bibinfo {author} {\bibfnamefont {Adri{\`a}}\
  \bibnamefont {Delhom}}, \bibinfo {author} {\bibfnamefont {J.~R.}\
  \bibnamefont {Nascimento}}, \bibinfo {author} {\bibfnamefont {Gonzalo~J.}\
  \bibnamefont {Olmo}}, \bibinfo {author} {\bibfnamefont {A.~Yu.}\ \bibnamefont
  {Petrov}}, \ and\ \bibinfo {author} {\bibfnamefont {Paulo~J.}\ \bibnamefont
  {Porf{\'i}rio}},\ }\bibfield  {title} {\enquote {\bibinfo {title}
  {Metric-affine bumblebee gravity: classical aspects},}\ }\href {\doibase
  10.1140/epjc/s10052-021-09083-y} {\bibfield  {journal} {\bibinfo  {journal}
  {European Physical Journal C}\ }\textbf {\bibinfo {volume} {81}},\ \bibinfo
  {pages} {287} (\bibinfo {year} {2021})},\ \Eprint
  {http://arxiv.org/abs/1911.11605} {arXiv:1911.11605 [hep-th]} \BibitemShut
  {NoStop}%
\bibitem [{\citenamefont {Ara{\'u}jo~Filho}(2026)}]{AraujoFilho:2026oiy}%
  \BibitemOpen
  \bibfield  {author} {\bibinfo {author} {\bibfnamefont {A.~A.}\ \bibnamefont
  {Ara{\'u}jo~Filho}},\ }\bibfield  {title} {\enquote {\bibinfo {title}
  {{Gravitational waves in metric{\textendash}affine bumblebee gravity}},}\
  }\href {\doibase 10.1140/epjc/s10052-026-15814-w} {\bibfield  {journal}
  {\bibinfo  {journal} {Eur. Phys. J. C}\ }\textbf {\bibinfo {volume} {86}},\
  \bibinfo {pages} {578} (\bibinfo {year} {2026})},\ \Eprint
  {http://arxiv.org/abs/2603.08310} {arXiv:2603.08310 [gr-qc]} \BibitemShut
  {NoStop}%
\bibitem [{\citenamefont {{Ara{\'u}jo Filho}}\ \emph
  {et~al.}(2023)\citenamefont {{Ara{\'u}jo Filho}}, \citenamefont {Nascimento},
  \citenamefont {Petrov},\ and\ \citenamefont
  {Porf{\'i}rio}}]{AraujoFilho2023MetricAffine}%
  \BibitemOpen
  \bibfield  {author} {\bibinfo {author} {\bibfnamefont {A.~A.}\ \bibnamefont
  {{Ara{\'u}jo Filho}}}, \bibinfo {author} {\bibfnamefont {J.~R.}\ \bibnamefont
  {Nascimento}}, \bibinfo {author} {\bibfnamefont {A.~Yu.}\ \bibnamefont
  {Petrov}}, \ and\ \bibinfo {author} {\bibfnamefont {P.~J.}\ \bibnamefont
  {Porf{\'i}rio}},\ }\bibfield  {title} {\enquote {\bibinfo {title} {Vacuum
  solution within a metric-affine bumblebee gravity},}\ }\href {\doibase
  10.1103/PhysRevD.108.085010} {\bibfield  {journal} {\bibinfo  {journal}
  {Physical Review D}\ }\textbf {\bibinfo {volume} {108}},\ \bibinfo {pages}
  {085010} (\bibinfo {year} {2023})},\ \Eprint
  {http://arxiv.org/abs/2211.11821} {arXiv:2211.11821 [gr-qc]} \BibitemShut
  {NoStop}%
\bibitem [{\citenamefont {Ara{\'u}jo~Filho}\ \emph {et~al.}(2024)\citenamefont
  {Ara{\'u}jo~Filho}, \citenamefont {Nascimento}, \citenamefont {Petrov},\ and\
  \citenamefont {Porf{\'\i}rio}}]{AraujoFilho:2024ykw}%
  \BibitemOpen
  \bibfield  {author} {\bibinfo {author} {\bibfnamefont {A.~A.}\ \bibnamefont
  {Ara{\'u}jo~Filho}}, \bibinfo {author} {\bibfnamefont {J.~R.}\ \bibnamefont
  {Nascimento}}, \bibinfo {author} {\bibfnamefont {A.~Yu.}\ \bibnamefont
  {Petrov}}, \ and\ \bibinfo {author} {\bibfnamefont {P.~J.}\ \bibnamefont
  {Porf{\'\i}rio}},\ }\bibfield  {title} {\enquote {\bibinfo {title} {{An exact
  stationary axisymmetric vacuum solution within a metric-affine bumblebee
  gravity}},}\ }\href {\doibase 10.1088/1475-7516/2024/07/004} {\bibfield
  {journal} {\bibinfo  {journal} {JCAP}\ }\textbf {\bibinfo {volume} {07}},\
  \bibinfo {pages} {004} (\bibinfo {year} {2024})},\ \Eprint
  {http://arxiv.org/abs/2402.13014} {arXiv:2402.13014 [gr-qc]} \BibitemShut
  {NoStop}%
\bibitem [{\citenamefont {Ara{\'u}jo~Filho}\ \emph
  {et~al.}(2026{\natexlab{b}})\citenamefont {Ara{\'u}jo~Filho}, \citenamefont
  {Heidari}, \citenamefont {Bezerra},\ and\ \citenamefont
  {Lobo}}]{AraujoFilho:2026ktz}%
  \BibitemOpen
  \bibfield  {author} {\bibinfo {author} {\bibfnamefont {A.~A.}\ \bibnamefont
  {Ara{\'u}jo~Filho}}, \bibinfo {author} {\bibfnamefont {N.}~\bibnamefont
  {Heidari}}, \bibinfo {author} {\bibfnamefont {V.~B.}\ \bibnamefont
  {Bezerra}}, \ and\ \bibinfo {author} {\bibfnamefont {Francisco S.~N.}\
  \bibnamefont {Lobo}},\ }\bibfield  {title} {\enquote {\bibinfo {title} {{A
  general static and spherically symmetric black hole in traceless
  metric-affine bumblebee gravity}},}\ }\href@noop {} {\  (\bibinfo {year}
  {2026}{\natexlab{b}})},\ \Eprint {http://arxiv.org/abs/2609.29675}
  {arXiv:2609.29675 [gr-qc]} \BibitemShut {NoStop}%
\bibitem [{\citenamefont {Simpson}\ and\ \citenamefont
  {Visser}(2019)}]{SimpsonVisser2019}%
  \BibitemOpen
  \bibfield  {author} {\bibinfo {author} {\bibfnamefont {Alex}\ \bibnamefont
  {Simpson}}\ and\ \bibinfo {author} {\bibfnamefont {Matt}\ \bibnamefont
  {Visser}},\ }\bibfield  {title} {\enquote {\bibinfo {title} {Black-bounce to
  traversable wormhole},}\ }\href {\doibase 10.1088/1475-7516/2019/02/042}
  {\bibfield  {journal} {\bibinfo  {journal} {Journal of Cosmology and
  Astroparticle Physics}\ }\textbf {\bibinfo {volume} {2019}},\ \bibinfo
  {pages} {042} (\bibinfo {year} {2019})},\ \Eprint
  {http://arxiv.org/abs/1812.07114} {arXiv:1812.07114 [gr-qc]} \BibitemShut
  {NoStop}%
\bibitem [{\citenamefont {Lobo}\ \emph
  {et~al.}(2021{\natexlab{a}})\citenamefont {Lobo}, \citenamefont {Rodrigues},
  \citenamefont {Silva}, \citenamefont {Simpson},\ and\ \citenamefont
  {Visser}}]{LoboEtAl2021}%
  \BibitemOpen
  \bibfield  {author} {\bibinfo {author} {\bibfnamefont {Francisco S.~N.}\
  \bibnamefont {Lobo}}, \bibinfo {author} {\bibfnamefont {Manuel~E.}\
  \bibnamefont {Rodrigues}}, \bibinfo {author} {\bibfnamefont {Marcos V.
  de~S.}\ \bibnamefont {Silva}}, \bibinfo {author} {\bibfnamefont {Alex}\
  \bibnamefont {Simpson}}, \ and\ \bibinfo {author} {\bibfnamefont {Matt}\
  \bibnamefont {Visser}},\ }\bibfield  {title} {\enquote {\bibinfo {title}
  {Novel black-bounce spacetimes: wormholes, regularity, energy conditions, and
  causal structure},}\ }\href {\doibase 10.1103/PhysRevD.103.084052} {\bibfield
   {journal} {\bibinfo  {journal} {Physical Review D}\ }\textbf {\bibinfo
  {volume} {103}},\ \bibinfo {pages} {084052} (\bibinfo {year}
  {2021}{\natexlab{a}})},\ \Eprint {http://arxiv.org/abs/2009.12057}
  {arXiv:2009.12057 [gr-qc]} \BibitemShut {NoStop}%
\bibitem [{\citenamefont {Simpson}\ \emph {et~al.}(2019)\citenamefont
  {Simpson}, \citenamefont {Martin-Moruno},\ and\ \citenamefont
  {Visser}}]{SimpsonMartinMorunoVisser2019}%
  \BibitemOpen
  \bibfield  {author} {\bibinfo {author} {\bibfnamefont {Alex}\ \bibnamefont
  {Simpson}}, \bibinfo {author} {\bibfnamefont {Prado}\ \bibnamefont
  {Martin-Moruno}}, \ and\ \bibinfo {author} {\bibfnamefont {Matt}\
  \bibnamefont {Visser}},\ }\bibfield  {title} {\enquote {\bibinfo {title}
  {{Vaidya} spacetimes, black-bounces, and traversable wormholes},}\ }\href
  {\doibase 10.1088/1361-6382/ab28a5} {\bibfield  {journal} {\bibinfo
  {journal} {Classical and Quantum Gravity}\ }\textbf {\bibinfo {volume}
  {36}},\ \bibinfo {pages} {145007} (\bibinfo {year} {2019})},\ \Eprint
  {http://arxiv.org/abs/1902.04232} {arXiv:1902.04232 [gr-qc]} \BibitemShut
  {NoStop}%
\bibitem [{\citenamefont {Mazza}\ \emph {et~al.}(2021)\citenamefont {Mazza},
  \citenamefont {Franzin},\ and\ \citenamefont
  {Liberati}}]{MazzaFranzinLiberati2021}%
  \BibitemOpen
  \bibfield  {author} {\bibinfo {author} {\bibfnamefont {Jacopo}\ \bibnamefont
  {Mazza}}, \bibinfo {author} {\bibfnamefont {Edgardo}\ \bibnamefont
  {Franzin}}, \ and\ \bibinfo {author} {\bibfnamefont {Stefano}\ \bibnamefont
  {Liberati}},\ }\bibfield  {title} {\enquote {\bibinfo {title} {A novel family
  of rotating black hole mimickers},}\ }\href {\doibase
  10.1088/1475-7516/2021/04/082} {\bibfield  {journal} {\bibinfo  {journal}
  {Journal of Cosmology and Astroparticle Physics}\ }\textbf {\bibinfo {volume}
  {2021}},\ \bibinfo {pages} {082} (\bibinfo {year} {2021})},\ \Eprint
  {http://arxiv.org/abs/2102.01105} {arXiv:2102.01105 [gr-qc]} \BibitemShut
  {NoStop}%
\bibitem [{\citenamefont {Franzin}\ \emph {et~al.}(2021)\citenamefont
  {Franzin}, \citenamefont {Liberati}, \citenamefont {Mazza}, \citenamefont
  {Simpson},\ and\ \citenamefont {Visser}}]{FranzinEtAl2021Charged}%
  \BibitemOpen
  \bibfield  {author} {\bibinfo {author} {\bibfnamefont {Edgardo}\ \bibnamefont
  {Franzin}}, \bibinfo {author} {\bibfnamefont {Stefano}\ \bibnamefont
  {Liberati}}, \bibinfo {author} {\bibfnamefont {Jacopo}\ \bibnamefont
  {Mazza}}, \bibinfo {author} {\bibfnamefont {Alex}\ \bibnamefont {Simpson}}, \
  and\ \bibinfo {author} {\bibfnamefont {Matt}\ \bibnamefont {Visser}},\
  }\bibfield  {title} {\enquote {\bibinfo {title} {Charged black-bounce
  spacetimes},}\ }\href {\doibase 10.1088/1475-7516/2021/07/036} {\bibfield
  {journal} {\bibinfo  {journal} {Journal of Cosmology and Astroparticle
  Physics}\ }\textbf {\bibinfo {volume} {2021}},\ \bibinfo {pages} {036}
  (\bibinfo {year} {2021})},\ \Eprint {http://arxiv.org/abs/2104.11376}
  {arXiv:2104.11376 [gr-qc]} \BibitemShut {NoStop}%
\bibitem [{\citenamefont {Bronnikov}\ and\ \citenamefont
  {Walia}(2022)}]{BronnikovWalia2022}%
  \BibitemOpen
  \bibfield  {author} {\bibinfo {author} {\bibfnamefont {Kirill~A.}\
  \bibnamefont {Bronnikov}}\ and\ \bibinfo {author} {\bibfnamefont
  {Rahul~Kumar}\ \bibnamefont {Walia}},\ }\bibfield  {title} {\enquote
  {\bibinfo {title} {Field sources for {Simpson--Visser} space-times},}\ }\href
  {\doibase 10.1103/PhysRevD.105.044039} {\bibfield  {journal} {\bibinfo
  {journal} {Physical Review D}\ }\textbf {\bibinfo {volume} {105}},\ \bibinfo
  {pages} {044039} (\bibinfo {year} {2022})},\ \Eprint
  {http://arxiv.org/abs/2112.13198} {arXiv:2112.13198 [gr-qc]} \BibitemShut
  {NoStop}%
\bibitem [{\citenamefont {Rodrigues}\ and\ \citenamefont
  {Silva}(2023)}]{RodriguesSilva2023}%
  \BibitemOpen
  \bibfield  {author} {\bibinfo {author} {\bibfnamefont {Manuel~E.}\
  \bibnamefont {Rodrigues}}\ and\ \bibinfo {author} {\bibfnamefont {Marcos V.
  de~S.}\ \bibnamefont {Silva}},\ }\bibfield  {title} {\enquote {\bibinfo
  {title} {Source of black bounces in general relativity},}\ }\href {\doibase
  10.1103/PhysRevD.107.044064} {\bibfield  {journal} {\bibinfo  {journal}
  {Physical Review D}\ }\textbf {\bibinfo {volume} {107}},\ \bibinfo {pages}
  {044064} (\bibinfo {year} {2023})},\ \Eprint
  {http://arxiv.org/abs/2302.10772} {arXiv:2302.10772 [gr-qc]} \BibitemShut
  {NoStop}%
\bibitem [{\citenamefont {Alencar}\ \emph {et~al.}(2024)\citenamefont
  {Alencar}, \citenamefont {Bronnikov}, \citenamefont {Rodrigues},
  \citenamefont {S{\'a}ez-Chill{\'o}n~G{\'o}mez},\ and\ \citenamefont
  {Silva}}]{AlencarEtAl2024}%
  \BibitemOpen
  \bibfield  {author} {\bibinfo {author} {\bibfnamefont {G.}~\bibnamefont
  {Alencar}}, \bibinfo {author} {\bibfnamefont {Kirill~A.}\ \bibnamefont
  {Bronnikov}}, \bibinfo {author} {\bibfnamefont {Manuel~E.}\ \bibnamefont
  {Rodrigues}}, \bibinfo {author} {\bibfnamefont {Diego}\ \bibnamefont
  {S{\'a}ez-Chill{\'o}n~G{\'o}mez}}, \ and\ \bibinfo {author} {\bibfnamefont
  {Marcos V. de~S.}\ \bibnamefont {Silva}},\ }\bibfield  {title} {\enquote
  {\bibinfo {title} {On black bounce space-times in non-linear
  electrodynamics},}\ }\href {\doibase 10.1140/epjc/s10052-024-13119-4}
  {\bibfield  {journal} {\bibinfo  {journal} {European Physical Journal C}\
  }\textbf {\bibinfo {volume} {84}},\ \bibinfo {pages} {745} (\bibinfo {year}
  {2024})},\ \Eprint {http://arxiv.org/abs/2403.12897} {arXiv:2403.12897
  [gr-qc]} \BibitemShut {NoStop}%
\bibitem [{\citenamefont {Alencar}\ \emph {et~al.}(2025)\citenamefont
  {Alencar}, \citenamefont {Duran-Cabac{\'e}s}, \citenamefont
  {Rubiera-Garcia},\ and\ \citenamefont
  {S{\'a}ez-Chill{\'o}n~G{\'o}mez}}]{AlencarEtAl2025}%
  \BibitemOpen
  \bibfield  {author} {\bibinfo {author} {\bibfnamefont {G.}~\bibnamefont
  {Alencar}}, \bibinfo {author} {\bibfnamefont {Albert}\ \bibnamefont
  {Duran-Cabac{\'e}s}}, \bibinfo {author} {\bibfnamefont {Diego}\ \bibnamefont
  {Rubiera-Garcia}}, \ and\ \bibinfo {author} {\bibfnamefont {Diego}\
  \bibnamefont {S{\'a}ez-Chill{\'o}n~G{\'o}mez}},\ }\bibfield  {title}
  {\enquote {\bibinfo {title} {General spherically symmetric black bounces
  within nonlinear electrodynamics},}\ }\href {\doibase
  10.1103/PhysRevD.111.104020} {\bibfield  {journal} {\bibinfo  {journal}
  {Physical Review D}\ }\textbf {\bibinfo {volume} {111}},\ \bibinfo {pages}
  {104020} (\bibinfo {year} {2025})},\ \Eprint
  {http://arxiv.org/abs/2501.03909} {arXiv:2501.03909 [gr-qc]} \BibitemShut
  {NoStop}%
\bibitem [{\citenamefont {Ding}\ \emph {et~al.}(2025)\citenamefont {Ding},
  \citenamefont {Liu}, \citenamefont {Xiao},\ and\ \citenamefont
  {Chen}}]{Ding2025}%
  \BibitemOpen
  \bibfield  {author} {\bibinfo {author} {\bibfnamefont {Chikun}\ \bibnamefont
  {Ding}}, \bibinfo {author} {\bibfnamefont {Changqing}\ \bibnamefont {Liu}},
  \bibinfo {author} {\bibfnamefont {Yuehua}\ \bibnamefont {Xiao}}, \ and\
  \bibinfo {author} {\bibfnamefont {Jun}\ \bibnamefont {Chen}},\ }\bibfield
  {title} {\enquote {\bibinfo {title} {Phantom hairy black holes and wormholes
  in {Einstein}-bumblebee gravity},}\ }\href {\doibase
  10.1140/epjc/s10052-025-13815-9} {\bibfield  {journal} {\bibinfo  {journal}
  {European Physical Journal C}\ }\textbf {\bibinfo {volume} {85}},\ \bibinfo
  {pages} {73} (\bibinfo {year} {2025})},\ \Eprint
  {http://arxiv.org/abs/2407.16916} {arXiv:2407.16916 [gr-qc]} \BibitemShut
  {NoStop}%
\bibitem [{\citenamefont {Pereira}\ \emph {et~al.}(2025)\citenamefont
  {Pereira}, \citenamefont {Silva}, \citenamefont {Belich}, \citenamefont
  {Rodrigues}, \citenamefont {Fabris},\ and\ \citenamefont
  {Rodrigues}}]{Pereira2025}%
  \BibitemOpen
  \bibfield  {author} {\bibinfo {author} {\bibfnamefont {Carlos F.~S.}\
  \bibnamefont {Pereira}}, \bibinfo {author} {\bibfnamefont {Marcos V. de~S.}\
  \bibnamefont {Silva}}, \bibinfo {author} {\bibfnamefont {H.}~\bibnamefont
  {Belich}}, \bibinfo {author} {\bibfnamefont {Denis~C.}\ \bibnamefont
  {Rodrigues}}, \bibinfo {author} {\bibfnamefont {J{\'u}lio~C.}\ \bibnamefont
  {Fabris}}, \ and\ \bibinfo {author} {\bibfnamefont {Manuel~E.}\ \bibnamefont
  {Rodrigues}},\ }\bibfield  {title} {\enquote {\bibinfo {title} {Black-bounce
  solutions in a {$k$}-essence theory under the effects of bumblebee
  gravity},}\ }\href {\doibase 10.1103/m3gq-1t5l} {\bibfield  {journal}
  {\bibinfo  {journal} {Physical Review D}\ }\textbf {\bibinfo {volume}
  {111}},\ \bibinfo {pages} {124005} (\bibinfo {year} {2025})},\ \Eprint
  {http://arxiv.org/abs/2503.09920} {arXiv:2503.09920 [gr-qc]} \BibitemShut
  {NoStop}%
\bibitem [{\citenamefont {Li}\ and\ \citenamefont {L{\"u}}(2026)}]{LiLu2026}%
  \BibitemOpen
  \bibfield  {author} {\bibinfo {author} {\bibfnamefont {Zhi-Chao}\
  \bibnamefont {Li}}\ and\ \bibinfo {author} {\bibfnamefont {H.}~\bibnamefont
  {L{\"u}}},\ }\bibfield  {title} {\enquote {\bibinfo {title} {When bumblebee
  meets {NLED}: {Lorentz}-violating black holes and regular spacetimes},}\
  }\href {\doibase 10.1140/epjc/s10052-026-16219-5} {\bibfield  {journal}
  {\bibinfo  {journal} {European Physical Journal C}\ }\textbf {\bibinfo
  {volume} {86}},\ \bibinfo {pages} {959} (\bibinfo {year} {2026})},\ \Eprint
  {http://arxiv.org/abs/2605.13963} {arXiv:2605.13963 [gr-qc]} \BibitemShut
  {NoStop}%
\bibitem [{\citenamefont {Churilova}\ and\ \citenamefont
  {Stuchl{\'i}k}(2020)}]{ChurilovaStuchlik2020}%
  \BibitemOpen
  \bibfield  {author} {\bibinfo {author} {\bibfnamefont {M.~S.}\ \bibnamefont
  {Churilova}}\ and\ \bibinfo {author} {\bibfnamefont {Z.}~\bibnamefont
  {Stuchl{\'i}k}},\ }\bibfield  {title} {\enquote {\bibinfo {title} {Ringing of
  the regular black-hole/wormhole transition},}\ }\href {\doibase
  10.1088/1361-6382/ab7717} {\bibfield  {journal} {\bibinfo  {journal}
  {Classical and Quantum Gravity}\ }\textbf {\bibinfo {volume} {37}},\ \bibinfo
  {pages} {075014} (\bibinfo {year} {2020})},\ \Eprint
  {http://arxiv.org/abs/1911.11823} {arXiv:1911.11823 [gr-qc]} \BibitemShut
  {NoStop}%
\bibitem [{\citenamefont {Tsukamoto}(2021)}]{Tsukamoto2021}%
  \BibitemOpen
  \bibfield  {author} {\bibinfo {author} {\bibfnamefont {Naoki}\ \bibnamefont
  {Tsukamoto}},\ }\bibfield  {title} {\enquote {\bibinfo {title} {Gravitational
  lensing in the {Simpson--Visser} black-bounce spacetime in a strong
  deflection limit},}\ }\href {\doibase 10.1103/PhysRevD.103.024033} {\bibfield
   {journal} {\bibinfo  {journal} {Physical Review D}\ }\textbf {\bibinfo
  {volume} {103}},\ \bibinfo {pages} {024033} (\bibinfo {year} {2021})},\
  \Eprint {http://arxiv.org/abs/2011.03932} {arXiv:2011.03932 [gr-qc]}
  \BibitemShut {NoStop}%
\bibitem [{\citenamefont {Franzin}\ \emph {et~al.}(2022)\citenamefont
  {Franzin}, \citenamefont {Liberati}, \citenamefont {Mazza}, \citenamefont
  {Dey},\ and\ \citenamefont {Chakraborty}}]{FranzinEtAl2022}%
  \BibitemOpen
  \bibfield  {author} {\bibinfo {author} {\bibfnamefont {Edgardo}\ \bibnamefont
  {Franzin}}, \bibinfo {author} {\bibfnamefont {Stefano}\ \bibnamefont
  {Liberati}}, \bibinfo {author} {\bibfnamefont {Jacopo}\ \bibnamefont
  {Mazza}}, \bibinfo {author} {\bibfnamefont {Ramit}\ \bibnamefont {Dey}}, \
  and\ \bibinfo {author} {\bibfnamefont {Sumanta}\ \bibnamefont
  {Chakraborty}},\ }\bibfield  {title} {\enquote {\bibinfo {title} {Scalar
  perturbations around rotating regular black holes and wormholes: Quasinormal
  modes, ergoregion instability, and superradiance},}\ }\href {\doibase
  10.1103/PhysRevD.105.124051} {\bibfield  {journal} {\bibinfo  {journal}
  {Physical Review D}\ }\textbf {\bibinfo {volume} {105}},\ \bibinfo {pages}
  {124051} (\bibinfo {year} {2022})},\ \Eprint
  {http://arxiv.org/abs/2201.01650} {arXiv:2201.01650 [gr-qc]} \BibitemShut
  {NoStop}%
\bibitem [{\citenamefont {Duran-Cabac{\'e}s}\ \emph {et~al.}(2025)\citenamefont
  {Duran-Cabac{\'e}s}, \citenamefont {Rubiera-Garcia},\ and\ \citenamefont
  {S{\'a}ez-Chill{\'o}n~G{\'o}mez}}]{DuranCabacesEtAl2025}%
  \BibitemOpen
  \bibfield  {author} {\bibinfo {author} {\bibfnamefont {Albert}\ \bibnamefont
  {Duran-Cabac{\'e}s}}, \bibinfo {author} {\bibfnamefont {Diego}\ \bibnamefont
  {Rubiera-Garcia}}, \ and\ \bibinfo {author} {\bibfnamefont {Diego}\
  \bibnamefont {S{\'a}ez-Chill{\'o}n~G{\'o}mez}},\ }\bibfield  {title}
  {\enquote {\bibinfo {title} {Quasinormal modes and echoes of generalized
  black hole bounces and their correspondence with shadows},}\ }\href {\doibase
  10.1103/3v89-z8rf} {\bibfield  {journal} {\bibinfo  {journal} {Physical
  Review D}\ }\textbf {\bibinfo {volume} {112}},\ \bibinfo {pages} {044016}
  (\bibinfo {year} {2025})}\BibitemShut {NoStop}%
\bibitem [{\citenamefont {Liang}\ \emph {et~al.}(2022)\citenamefont {Liang},
  \citenamefont {Xu}, \citenamefont {Lu},\ and\ \citenamefont
  {Shao}}]{Liang2022Polarizations}%
  \BibitemOpen
  \bibfield  {author} {\bibinfo {author} {\bibfnamefont {Dicong}\ \bibnamefont
  {Liang}}, \bibinfo {author} {\bibfnamefont {Rui}\ \bibnamefont {Xu}},
  \bibinfo {author} {\bibfnamefont {Xuchen}\ \bibnamefont {Lu}}, \ and\
  \bibinfo {author} {\bibfnamefont {Lijing}\ \bibnamefont {Shao}},\ }\bibfield
  {title} {\enquote {\bibinfo {title} {Polarizations of gravitational waves in
  the bumblebee gravity model},}\ }\href {\doibase 10.1103/PhysRevD.106.124019}
  {\bibfield  {journal} {\bibinfo  {journal} {Physical Review D}\ }\textbf
  {\bibinfo {volume} {106}},\ \bibinfo {pages} {124019} (\bibinfo {year}
  {2022})},\ \Eprint {http://arxiv.org/abs/2207.14423} {arXiv:2207.14423
  [gr-qc]} \BibitemShut {NoStop}%
\bibitem [{\citenamefont {Amarilo}\ \emph {et~al.}(2024)\citenamefont
  {Amarilo}, \citenamefont {Filho}, \citenamefont {Filho},\ and\ \citenamefont
  {Reis}}]{Amarilo:2023wpn}%
  \BibitemOpen
  \bibfield  {author} {\bibinfo {author} {\bibfnamefont {K.~M.}\ \bibnamefont
  {Amarilo}}, \bibinfo {author} {\bibfnamefont {M.~B.~Ferreira}\ \bibnamefont
  {Filho}}, \bibinfo {author} {\bibfnamefont {A.~A.~Ara{\'u}jo}\ \bibnamefont
  {Filho}}, \ and\ \bibinfo {author} {\bibfnamefont {J.~A. A.~S.}\ \bibnamefont
  {Reis}},\ }\bibfield  {title} {\enquote {\bibinfo {title} {{Gravitational
  waves effects in a Lorentz{\textendash}violating scenario}},}\ }\href
  {\doibase 10.1016/j.physletb.2024.138785} {\bibfield  {journal} {\bibinfo
  {journal} {Phys. Lett. B}\ }\textbf {\bibinfo {volume} {855}},\ \bibinfo
  {pages} {138785} (\bibinfo {year} {2024})},\ \Eprint
  {http://arxiv.org/abs/2307.10937} {arXiv:2307.10937 [gr-qc]} \BibitemShut
  {NoStop}%
\bibitem [{\citenamefont {Mai}\ \emph {et~al.}(2024)\citenamefont {Mai},
  \citenamefont {Xu}, \citenamefont {Liang},\ and\ \citenamefont
  {Shao}}]{Mai2024}%
  \BibitemOpen
  \bibfield  {author} {\bibinfo {author} {\bibfnamefont {Zhan-Feng}\
  \bibnamefont {Mai}}, \bibinfo {author} {\bibfnamefont {Rui}\ \bibnamefont
  {Xu}}, \bibinfo {author} {\bibfnamefont {Dicong}\ \bibnamefont {Liang}}, \
  and\ \bibinfo {author} {\bibfnamefont {Lijing}\ \bibnamefont {Shao}},\
  }\bibfield  {title} {\enquote {\bibinfo {title} {Dynamic instability analysis
  for bumblebee black holes: The odd parity},}\ }\href {\doibase
  10.1103/PhysRevD.109.084076} {\bibfield  {journal} {\bibinfo  {journal}
  {Physical Review D}\ }\textbf {\bibinfo {volume} {109}},\ \bibinfo {pages}
  {084076} (\bibinfo {year} {2024})},\ \Eprint
  {http://arxiv.org/abs/2401.07757} {arXiv:2401.07757 [gr-qc]} \BibitemShut
  {NoStop}%
\bibitem [{\citenamefont {Li}\ \emph {et~al.}(2026)\citenamefont {Li},
  \citenamefont {Liu}, \citenamefont {Guo},\ and\ \citenamefont
  {Liu}}]{Li2026ChargedQNMs}%
  \BibitemOpen
  \bibfield  {author} {\bibinfo {author} {\bibfnamefont {Bo-Rui}\ \bibnamefont
  {Li}}, \bibinfo {author} {\bibfnamefont {Jia-Zhou}\ \bibnamefont {Liu}},
  \bibinfo {author} {\bibfnamefont {Wen-Di}\ \bibnamefont {Guo}}, \ and\
  \bibinfo {author} {\bibfnamefont {Yu-Xiao}\ \bibnamefont {Liu}},\ }\bibfield
  {title} {\enquote {\bibinfo {title} {Quasinormal modes of a charged
  spherically symmetric black hole in bumblebee gravity},}\ }\href {\doibase
  10.1140/epjc/s10052-026-16043-x} {\bibfield  {journal} {\bibinfo  {journal}
  {European Physical Journal C}\ }\textbf {\bibinfo {volume} {86}},\ \bibinfo
  {pages} {974} (\bibinfo {year} {2026})},\ \Eprint
  {http://arxiv.org/abs/2510.20503} {arXiv:2510.20503 [gr-qc]} \BibitemShut
  {NoStop}%
\bibitem [{\citenamefont {Berti}\ \emph {et~al.}(2009)\citenamefont {Berti},
  \citenamefont {Cardoso},\ and\ \citenamefont
  {Starinets}}]{BertiCardosoStarinets2009}%
  \BibitemOpen
  \bibfield  {author} {\bibinfo {author} {\bibfnamefont {Emanuele}\
  \bibnamefont {Berti}}, \bibinfo {author} {\bibfnamefont {Vitor}\ \bibnamefont
  {Cardoso}}, \ and\ \bibinfo {author} {\bibfnamefont {Andrei~O.}\ \bibnamefont
  {Starinets}},\ }\bibfield  {title} {\enquote {\bibinfo {title} {Quasinormal
  modes of black holes and black branes},}\ }\href {\doibase
  10.1088/0264-9381/26/16/163001} {\bibfield  {journal} {\bibinfo  {journal}
  {Classical and Quantum Gravity}\ }\textbf {\bibinfo {volume} {26}},\ \bibinfo
  {pages} {163001} (\bibinfo {year} {2009})},\ \Eprint
  {http://arxiv.org/abs/0905.2975} {arXiv:0905.2975 [gr-qc]} \BibitemShut
  {NoStop}%
\bibitem [{\citenamefont {Cardoso}\ \emph {et~al.}(2009)\citenamefont
  {Cardoso}, \citenamefont {Miranda}, \citenamefont {Berti}, \citenamefont
  {Witek},\ and\ \citenamefont {Zanchin}}]{Cardoso2009}%
  \BibitemOpen
  \bibfield  {author} {\bibinfo {author} {\bibfnamefont {Vitor}\ \bibnamefont
  {Cardoso}}, \bibinfo {author} {\bibfnamefont {Alex~S.}\ \bibnamefont
  {Miranda}}, \bibinfo {author} {\bibfnamefont {Emanuele}\ \bibnamefont
  {Berti}}, \bibinfo {author} {\bibfnamefont {Helvi}\ \bibnamefont {Witek}}, \
  and\ \bibinfo {author} {\bibfnamefont {Vilson~T.}\ \bibnamefont {Zanchin}},\
  }\bibfield  {title} {\enquote {\bibinfo {title} {Geodesic stability,
  {Lyapunov} exponents and quasinormal modes},}\ }\href {\doibase
  10.1103/PhysRevD.79.064016} {\bibfield  {journal} {\bibinfo  {journal}
  {Physical Review D}\ }\textbf {\bibinfo {volume} {79}},\ \bibinfo {pages}
  {064016} (\bibinfo {year} {2009})},\ \Eprint {http://arxiv.org/abs/0812.1806}
  {arXiv:0812.1806 [hep-th]} \BibitemShut {NoStop}%
\bibitem [{\citenamefont {Novello}\ \emph {et~al.}(2000)\citenamefont
  {Novello}, \citenamefont {De~Lorenci}, \citenamefont {Salim},\ and\
  \citenamefont {Klippert}}]{Novello2000}%
  \BibitemOpen
  \bibfield  {author} {\bibinfo {author} {\bibfnamefont {M.}~\bibnamefont
  {Novello}}, \bibinfo {author} {\bibfnamefont {V.~A.}\ \bibnamefont
  {De~Lorenci}}, \bibinfo {author} {\bibfnamefont {J.~M.}\ \bibnamefont
  {Salim}}, \ and\ \bibinfo {author} {\bibfnamefont {R.}~\bibnamefont
  {Klippert}},\ }\bibfield  {title} {\enquote {\bibinfo {title} {Geometrical
  aspects of light propagation in nonlinear electrodynamics},}\ }\href
  {\doibase 10.1103/PhysRevD.61.045001} {\bibfield  {journal} {\bibinfo
  {journal} {Physical Review D}\ }\textbf {\bibinfo {volume} {61}},\ \bibinfo
  {pages} {045001} (\bibinfo {year} {2000})},\ \Eprint
  {http://arxiv.org/abs/gr-qc/9911085} {arXiv:gr-qc/9911085 [gr-qc]}
  \BibitemShut {NoStop}%
\bibitem [{\citenamefont {Abalos}\ \emph {et~al.}(2015)\citenamefont {Abalos},
  \citenamefont {Carrasco}, \citenamefont {Goulart},\ and\ \citenamefont
  {Reula}}]{Abalos2015}%
  \BibitemOpen
  \bibfield  {author} {\bibinfo {author} {\bibfnamefont {Fernando}\
  \bibnamefont {Abalos}}, \bibinfo {author} {\bibfnamefont {Federico}\
  \bibnamefont {Carrasco}}, \bibinfo {author} {\bibfnamefont {{\'E}rico}\
  \bibnamefont {Goulart}}, \ and\ \bibinfo {author} {\bibfnamefont {Oscar}\
  \bibnamefont {Reula}},\ }\bibfield  {title} {\enquote {\bibinfo {title}
  {Nonlinear electrodynamics as a symmetric hyperbolic system},}\ }\href
  {\doibase 10.1103/PhysRevD.92.084024} {\bibfield  {journal} {\bibinfo
  {journal} {Physical Review D}\ }\textbf {\bibinfo {volume} {92}},\ \bibinfo
  {pages} {084024} (\bibinfo {year} {2015})},\ \Eprint
  {http://arxiv.org/abs/1507.02262} {arXiv:1507.02262 [gr-qc]} \BibitemShut
  {NoStop}%
\bibitem [{\citenamefont {Moreno}\ and\ \citenamefont
  {Sarbach}(2003)}]{MorenoSarbach2003}%
  \BibitemOpen
  \bibfield  {author} {\bibinfo {author} {\bibfnamefont {Claudia}\ \bibnamefont
  {Moreno}}\ and\ \bibinfo {author} {\bibfnamefont {Olivier}\ \bibnamefont
  {Sarbach}},\ }\bibfield  {title} {\enquote {\bibinfo {title} {Stability
  properties of black holes in self-gravitating nonlinear electrodynamics},}\
  }\href {\doibase 10.1103/PhysRevD.67.024028} {\bibfield  {journal} {\bibinfo
  {journal} {Physical Review D}\ }\textbf {\bibinfo {volume} {67}},\ \bibinfo
  {pages} {024028} (\bibinfo {year} {2003})},\ \Eprint
  {http://arxiv.org/abs/gr-qc/0208090} {arXiv:gr-qc/0208090 [gr-qc]}
  \BibitemShut {NoStop}%
\bibitem [{\citenamefont {Chaverra}\ \emph {et~al.}(2016)\citenamefont
  {Chaverra}, \citenamefont {Degollado}, \citenamefont {Moreno},\ and\
  \citenamefont {Sarbach}}]{Chaverra2016}%
  \BibitemOpen
  \bibfield  {author} {\bibinfo {author} {\bibfnamefont {Eliana}\ \bibnamefont
  {Chaverra}}, \bibinfo {author} {\bibfnamefont {Juan~Carlos}\ \bibnamefont
  {Degollado}}, \bibinfo {author} {\bibfnamefont {Claudia}\ \bibnamefont
  {Moreno}}, \ and\ \bibinfo {author} {\bibfnamefont {Olivier}\ \bibnamefont
  {Sarbach}},\ }\bibfield  {title} {\enquote {\bibinfo {title} {Black holes in
  nonlinear electrodynamics: Quasinormal spectra and parity splitting},}\
  }\href {\doibase 10.1103/PhysRevD.93.123013} {\bibfield  {journal} {\bibinfo
  {journal} {Physical Review D}\ }\textbf {\bibinfo {volume} {93}},\ \bibinfo
  {pages} {123013} (\bibinfo {year} {2016})},\ \Eprint
  {http://arxiv.org/abs/1605.04003} {arXiv:1605.04003 [gr-qc]} \BibitemShut
  {NoStop}%
\bibitem [{\citenamefont {Nomura}\ \emph {et~al.}(2020)\citenamefont {Nomura},
  \citenamefont {Yoshida},\ and\ \citenamefont {Soda}}]{Nomura2020}%
  \BibitemOpen
  \bibfield  {author} {\bibinfo {author} {\bibfnamefont {Kimihiro}\
  \bibnamefont {Nomura}}, \bibinfo {author} {\bibfnamefont {Daisuke}\
  \bibnamefont {Yoshida}}, \ and\ \bibinfo {author} {\bibfnamefont {Jiro}\
  \bibnamefont {Soda}},\ }\bibfield  {title} {\enquote {\bibinfo {title}
  {Stability of magnetic black holes in general nonlinear electrodynamics},}\
  }\href {\doibase 10.1103/PhysRevD.101.124026} {\bibfield  {journal} {\bibinfo
   {journal} {Physical Review D}\ }\textbf {\bibinfo {volume} {101}},\ \bibinfo
  {pages} {124026} (\bibinfo {year} {2020})},\ \Eprint
  {http://arxiv.org/abs/2004.07560} {arXiv:2004.07560 [gr-qc]} \BibitemShut
  {NoStop}%
\bibitem [{\citenamefont {Jansen}(2017)}]{Jansen2017}%
  \BibitemOpen
  \bibfield  {author} {\bibinfo {author} {\bibfnamefont {Aron}\ \bibnamefont
  {Jansen}},\ }\bibfield  {title} {\enquote {\bibinfo {title} {Overdamped modes
  in {Schwarzschild--de Sitter} and a {Mathematica} package for the numerical
  computation of quasinormal modes},}\ }\href {\doibase
  10.1140/epjp/i2017-11825-9} {\bibfield  {journal} {\bibinfo  {journal}
  {European Physical Journal Plus}\ }\textbf {\bibinfo {volume} {132}},\
  \bibinfo {pages} {546} (\bibinfo {year} {2017})},\ \Eprint
  {http://arxiv.org/abs/1709.09178} {arXiv:1709.09178 [gr-qc]} \BibitemShut
  {NoStop}%
\bibitem [{\citenamefont {Mart{\'i}n-Garc{\'i}a}(2021)}]{xTensor}%
  \BibitemOpen
  \bibfield  {author} {\bibinfo {author} {\bibfnamefont {Jos{\'e}~M.}\
  \bibnamefont {Mart{\'i}n-Garc{\'i}a}},\ }\href@noop {} {\enquote {\bibinfo
  {title} {{xTensor}: Fast abstract tensor computer algebra},}\ }\bibinfo
  {howpublished} {\url{https://xact.es/xTensor/}} (\bibinfo {year} {2021}),\
  \bibinfo {note} {version 1.2.0; part of the xAct system; accessed 10 October
  2026}\BibitemShut {NoStop}%
\bibitem [{\citenamefont {Yllanes}\ and\ \citenamefont
  {Mart{\'i}n-Garc{\'i}a}(2021)}]{xCoba}%
  \BibitemOpen
  \bibfield  {author} {\bibinfo {author} {\bibfnamefont {David}\ \bibnamefont
  {Yllanes}}\ and\ \bibinfo {author} {\bibfnamefont {Jos{\'e}~M.}\ \bibnamefont
  {Mart{\'i}n-Garc{\'i}a}},\ }\href@noop {} {\enquote {\bibinfo {title}
  {{xCoba}: General component tensor computer algebra},}\ }\bibinfo
  {howpublished} {\url{https://xact.es/xCoba/}} (\bibinfo {year} {2021}),\
  \bibinfo {note} {version 0.8.6; part of the xAct system; accessed 10 October
  2026}\BibitemShut {NoStop}%
\bibitem [{\citenamefont {Lobo}\ \emph
  {et~al.}(2021{\natexlab{b}})\citenamefont {Lobo}, \citenamefont {Rodrigues},
  \citenamefont {de~S.~Silva}, \citenamefont {Simpson},\ and\ \citenamefont
  {Visser}}]{Lobo2021}%
  \BibitemOpen
  \bibfield  {author} {\bibinfo {author} {\bibfnamefont {Francisco S.~N.}\
  \bibnamefont {Lobo}}, \bibinfo {author} {\bibfnamefont {Manuel~E.}\
  \bibnamefont {Rodrigues}}, \bibinfo {author} {\bibfnamefont {Marcos~V.}\
  \bibnamefont {de~S.~Silva}}, \bibinfo {author} {\bibfnamefont {Alex}\
  \bibnamefont {Simpson}}, \ and\ \bibinfo {author} {\bibfnamefont {Matt}\
  \bibnamefont {Visser}},\ }\bibfield  {title} {\enquote {\bibinfo {title}
  {Novel black-bounce spacetimes: Wormholes, regularity, energy conditions, and
  causal structure},}\ }\href {\doibase 10.1103/PhysRevD.103.084052} {\bibfield
   {journal} {\bibinfo  {journal} {Phys. Rev. D}\ }\textbf {\bibinfo {volume}
  {103}},\ \bibinfo {pages} {084052} (\bibinfo {year} {2021}{\natexlab{b}})},\
  \Eprint {http://arxiv.org/abs/2009.12057} {arXiv:2009.12057 [gr-qc]}
  \BibitemShut {NoStop}%
\bibitem [{\citenamefont {Claudel}\ \emph {et~al.}(2001)\citenamefont
  {Claudel}, \citenamefont {Virbhadra},\ and\ \citenamefont
  {Ellis}}]{Claudel2001}%
  \BibitemOpen
  \bibfield  {author} {\bibinfo {author} {\bibfnamefont {Clarissa-Marie}\
  \bibnamefont {Claudel}}, \bibinfo {author} {\bibfnamefont {K.~S.}\
  \bibnamefont {Virbhadra}}, \ and\ \bibinfo {author} {\bibfnamefont
  {G.~F.~R.}\ \bibnamefont {Ellis}},\ }\bibfield  {title} {\enquote {\bibinfo
  {title} {The geometry of photon surfaces},}\ }\href {\doibase
  10.1063/1.1308507} {\bibfield  {journal} {\bibinfo  {journal} {J. Math.
  Phys.}\ }\textbf {\bibinfo {volume} {42}},\ \bibinfo {pages} {818--838}
  (\bibinfo {year} {2001})},\ \Eprint {http://arxiv.org/abs/gr-qc/0005050}
  {arXiv:gr-qc/0005050} \BibitemShut {NoStop}%
\bibitem [{\citenamefont {Perlick}(2004)}]{Perlick2004}%
  \BibitemOpen
  \bibfield  {author} {\bibinfo {author} {\bibfnamefont {Volker}\ \bibnamefont
  {Perlick}},\ }\bibfield  {title} {\enquote {\bibinfo {title} {Exact
  gravitational lens equation in spherically symmetric and static
  spacetimes},}\ }\href {\doibase 10.1103/PhysRevD.69.064017} {\bibfield
  {journal} {\bibinfo  {journal} {Phys. Rev. D}\ }\textbf {\bibinfo {volume}
  {69}},\ \bibinfo {pages} {064017} (\bibinfo {year} {2004})},\ \Eprint
  {http://arxiv.org/abs/gr-qc/0307072} {arXiv:gr-qc/0307072} \BibitemShut
  {NoStop}%
\bibitem [{\citenamefont {Ishihara}\ \emph {et~al.}(2016)\citenamefont
  {Ishihara}, \citenamefont {Suzuki}, \citenamefont {Ono}, \citenamefont
  {Kitamura},\ and\ \citenamefont {Asada}}]{Ishihara2016}%
  \BibitemOpen
  \bibfield  {author} {\bibinfo {author} {\bibfnamefont {Asahi}\ \bibnamefont
  {Ishihara}}, \bibinfo {author} {\bibfnamefont {Yusuke}\ \bibnamefont
  {Suzuki}}, \bibinfo {author} {\bibfnamefont {Toshiaki}\ \bibnamefont {Ono}},
  \bibinfo {author} {\bibfnamefont {Takao}\ \bibnamefont {Kitamura}}, \ and\
  \bibinfo {author} {\bibfnamefont {Hideki}\ \bibnamefont {Asada}},\ }\bibfield
   {title} {\enquote {\bibinfo {title} {Gravitational bending angle of light
  for finite distance and the {Gauss--Bonnet} theorem},}\ }\href {\doibase
  10.1103/PhysRevD.94.084015} {\bibfield  {journal} {\bibinfo  {journal} {Phys.
  Rev. D}\ }\textbf {\bibinfo {volume} {94}},\ \bibinfo {pages} {084015}
  (\bibinfo {year} {2016})},\ \Eprint {http://arxiv.org/abs/1604.08308}
  {arXiv:1604.08308 [gr-qc]} \BibitemShut {NoStop}%
\bibitem [{\citenamefont {Iyer}\ and\ \citenamefont
  {Petters}(2007)}]{IyerPetters2007}%
  \BibitemOpen
  \bibfield  {author} {\bibinfo {author} {\bibfnamefont {Savitri~V.}\
  \bibnamefont {Iyer}}\ and\ \bibinfo {author} {\bibfnamefont {Arlie~O.}\
  \bibnamefont {Petters}},\ }\bibfield  {title} {\enquote {\bibinfo {title}
  {Light's bending angle due to black holes: From the photon sphere to
  infinity},}\ }\href {\doibase 10.1007/s10714-007-0481-8} {\bibfield
  {journal} {\bibinfo  {journal} {Gen. Relativ. Gravit.}\ }\textbf {\bibinfo
  {volume} {39}},\ \bibinfo {pages} {1563--1582} (\bibinfo {year} {2007})},\
  \Eprint {http://arxiv.org/abs/gr-qc/0611086} {arXiv:gr-qc/0611086}
  \BibitemShut {NoStop}%
\bibitem [{\citenamefont {Bozza}(2002)}]{Bozza2002}%
  \BibitemOpen
  \bibfield  {author} {\bibinfo {author} {\bibfnamefont {Valerio}\ \bibnamefont
  {Bozza}},\ }\bibfield  {title} {\enquote {\bibinfo {title} {Gravitational
  lensing in the strong field limit},}\ }\href {\doibase
  10.1103/PhysRevD.66.103001} {\bibfield  {journal} {\bibinfo  {journal} {Phys.
  Rev. D}\ }\textbf {\bibinfo {volume} {66}},\ \bibinfo {pages} {103001}
  (\bibinfo {year} {2002})},\ \Eprint {http://arxiv.org/abs/gr-qc/0208075}
  {arXiv:gr-qc/0208075} \BibitemShut {NoStop}%
\bibitem [{\citenamefont {Tsukamoto}(2017)}]{Tsukamoto2017}%
  \BibitemOpen
  \bibfield  {author} {\bibinfo {author} {\bibfnamefont {Naoki}\ \bibnamefont
  {Tsukamoto}},\ }\bibfield  {title} {\enquote {\bibinfo {title} {Deflection
  angle in the strong deflection limit in a general asymptotically flat,
  static, spherically symmetric spacetime},}\ }\href {\doibase
  10.1103/PhysRevD.95.064035} {\bibfield  {journal} {\bibinfo  {journal} {Phys.
  Rev. D}\ }\textbf {\bibinfo {volume} {95}},\ \bibinfo {pages} {064035}
  (\bibinfo {year} {2017})},\ \Eprint {http://arxiv.org/abs/1612.08251}
  {arXiv:1612.08251 [gr-qc]} \BibitemShut {NoStop}%
\bibitem [{\citenamefont {Darwin}(1959)}]{Darwin1959}%
  \BibitemOpen
  \bibfield  {author} {\bibinfo {author} {\bibfnamefont {C.~G.}\ \bibnamefont
  {Darwin}},\ }\bibfield  {title} {\enquote {\bibinfo {title} {The gravity
  field of a particle},}\ }\href {\doibase 10.1098/rspa.1959.0015} {\bibfield
  {journal} {\bibinfo  {journal} {Proc. R. Soc. Lond. A}\ }\textbf {\bibinfo
  {volume} {249}},\ \bibinfo {pages} {180--194} (\bibinfo {year}
  {1959})}\BibitemShut {NoStop}%
\bibitem [{\citenamefont {Biesiada}(2003)}]{Biesiada2003}%
  \BibitemOpen
  \bibfield  {author} {\bibinfo {author} {\bibfnamefont {Marek}\ \bibnamefont
  {Biesiada}},\ }\bibfield  {title} {\enquote {\bibinfo {title} {Epicyclic
  orbital oscillations in {Newton}'s and {Einstein}'s gravity from the geodesic
  deviation equation},}\ }\href {\doibase 10.1023/A:1024599021730} {\bibfield
  {journal} {\bibinfo  {journal} {Gen. Relativ. Gravit.}\ }\textbf {\bibinfo
  {volume} {35}},\ \bibinfo {pages} {1503--1510} (\bibinfo {year} {2003})},\
  \Eprint {http://arxiv.org/abs/gr-qc/0303083} {arXiv:gr-qc/0303083}
  \BibitemShut {NoStop}%
\bibitem [{\citenamefont {Cutler}\ \emph {et~al.}(1994)\citenamefont {Cutler},
  \citenamefont {Kennefick},\ and\ \citenamefont {Poisson}}]{Cutler1994}%
  \BibitemOpen
  \bibfield  {author} {\bibinfo {author} {\bibfnamefont {Curt}\ \bibnamefont
  {Cutler}}, \bibinfo {author} {\bibfnamefont {Daniel}\ \bibnamefont
  {Kennefick}}, \ and\ \bibinfo {author} {\bibfnamefont {Eric}\ \bibnamefont
  {Poisson}},\ }\bibfield  {title} {\enquote {\bibinfo {title} {Gravitational
  radiation reaction for bound motion around a {Schwarzschild} black hole},}\
  }\href {\doibase 10.1103/PhysRevD.50.3816} {\bibfield  {journal} {\bibinfo
  {journal} {Phys. Rev. D}\ }\textbf {\bibinfo {volume} {50}},\ \bibinfo
  {pages} {3816--3835} (\bibinfo {year} {1994})}\BibitemShut {NoStop}%
\bibitem [{\citenamefont {Hopper}\ \emph {et~al.}(2016)\citenamefont {Hopper},
  \citenamefont {Kavanagh},\ and\ \citenamefont {Ottewill}}]{Hopper2016}%
  \BibitemOpen
  \bibfield  {author} {\bibinfo {author} {\bibfnamefont {Seth}\ \bibnamefont
  {Hopper}}, \bibinfo {author} {\bibfnamefont {Chris}\ \bibnamefont
  {Kavanagh}}, \ and\ \bibinfo {author} {\bibfnamefont {Adrian~C.}\
  \bibnamefont {Ottewill}},\ }\bibfield  {title} {\enquote {\bibinfo {title}
  {Analytic self-force calculations in the post-{Newtonian} regime: Eccentric
  orbits on a {Schwarzschild} background},}\ }\href {\doibase
  10.1103/PhysRevD.93.044010} {\bibfield  {journal} {\bibinfo  {journal} {Phys.
  Rev. D}\ }\textbf {\bibinfo {volume} {93}},\ \bibinfo {pages} {044010}
  (\bibinfo {year} {2016})},\ \Eprint {http://arxiv.org/abs/1512.01556}
  {arXiv:1512.01556 [gr-qc]} \BibitemShut {NoStop}%
\bibitem [{\citenamefont {Levin}\ and\ \citenamefont
  {Perez-Giz}(2008)}]{LevinPerezGiz2008}%
  \BibitemOpen
  \bibfield  {author} {\bibinfo {author} {\bibfnamefont {Janna}\ \bibnamefont
  {Levin}}\ and\ \bibinfo {author} {\bibfnamefont {Gabe}\ \bibnamefont
  {Perez-Giz}},\ }\bibfield  {title} {\enquote {\bibinfo {title} {A periodic
  table for black hole orbits},}\ }\href {\doibase 10.1103/PhysRevD.77.103005}
  {\bibfield  {journal} {\bibinfo  {journal} {Phys. Rev. D}\ }\textbf {\bibinfo
  {volume} {77}},\ \bibinfo {pages} {103005} (\bibinfo {year} {2008})},\
  \Eprint {http://arxiv.org/abs/0802.0459} {arXiv:0802.0459 [gr-qc]}
  \BibitemShut {NoStop}%
\bibitem [{\citenamefont {Perlick}\ and\ \citenamefont
  {Tsupko}(2022)}]{PerlickTsupko2022}%
  \BibitemOpen
  \bibfield  {author} {\bibinfo {author} {\bibfnamefont {Volker}\ \bibnamefont
  {Perlick}}\ and\ \bibinfo {author} {\bibfnamefont {Oleg~Yu.}\ \bibnamefont
  {Tsupko}},\ }\bibfield  {title} {\enquote {\bibinfo {title} {Calculating
  black hole shadows: Review of analytical studies},}\ }\href {\doibase
  10.1016/j.physrep.2021.10.004} {\bibfield  {journal} {\bibinfo  {journal}
  {Phys. Rept.}\ }\textbf {\bibinfo {volume} {947}},\ \bibinfo {pages} {1--39}
  (\bibinfo {year} {2022})},\ \Eprint {http://arxiv.org/abs/2105.07101}
  {arXiv:2105.07101 [gr-qc]} \BibitemShut {NoStop}%
\bibitem [{\citenamefont {{Event Horizon Telescope Collaboration}}\ \emph
  {et~al.}(2022{\natexlab{a}})\citenamefont {{Event Horizon Telescope
  Collaboration}}, \citenamefont {Akiyama} \emph {et~al.}}]{EHTSgrA2022VI}%
  \BibitemOpen
  \bibfield  {author} {\bibinfo {author} {\bibnamefont {{Event Horizon
  Telescope Collaboration}}}, \bibinfo {author} {\bibfnamefont {Kazunori}\
  \bibnamefont {Akiyama}},  \emph {et~al.},\ }\bibfield  {title} {\enquote
  {\bibinfo {title} {First {Sagittarius A*} {Event Horizon Telescope} results.
  {VI}. testing the black hole metric},}\ }\href {\doibase
  10.3847/2041-8213/ac6756} {\bibfield  {journal} {\bibinfo  {journal}
  {Astrophys. J. Lett.}\ }\textbf {\bibinfo {volume} {930}},\ \bibinfo {pages}
  {L17} (\bibinfo {year} {2022}{\natexlab{a}})},\ \Eprint
  {http://arxiv.org/abs/2311.09484} {arXiv:2311.09484 [astro-ph.HE]}
  \BibitemShut {NoStop}%
\bibitem [{\citenamefont {{Event Horizon Telescope Collaboration}}\ \emph
  {et~al.}(2019)\citenamefont {{Event Horizon Telescope Collaboration}},
  \citenamefont {Akiyama} \emph {et~al.}}]{EHTM872019VI}%
  \BibitemOpen
  \bibfield  {author} {\bibinfo {author} {\bibnamefont {{Event Horizon
  Telescope Collaboration}}}, \bibinfo {author} {\bibfnamefont {Kazunori}\
  \bibnamefont {Akiyama}},  \emph {et~al.},\ }\bibfield  {title} {\enquote
  {\bibinfo {title} {First {M87} {Event Horizon Telescope} results. {VI}. the
  shadow and mass of the central black hole},}\ }\href {\doibase
  10.3847/2041-8213/ab1141} {\bibfield  {journal} {\bibinfo  {journal}
  {Astrophys. J. Lett.}\ }\textbf {\bibinfo {volume} {875}},\ \bibinfo {pages}
  {L6} (\bibinfo {year} {2019})},\ \Eprint {http://arxiv.org/abs/1906.11243}
  {arXiv:1906.11243 [astro-ph.GA]} \BibitemShut {NoStop}%
\bibitem [{\citenamefont {{Event Horizon Telescope Collaboration}}\ \emph
  {et~al.}(2022{\natexlab{b}})\citenamefont {{Event Horizon Telescope
  Collaboration}}, \citenamefont {Akiyama} \emph {et~al.}}]{EHTSgrA2022IV}%
  \BibitemOpen
  \bibfield  {author} {\bibinfo {author} {\bibnamefont {{Event Horizon
  Telescope Collaboration}}}, \bibinfo {author} {\bibfnamefont {Kazunori}\
  \bibnamefont {Akiyama}},  \emph {et~al.},\ }\bibfield  {title} {\enquote
  {\bibinfo {title} {First {Sagittarius A*} {Event Horizon Telescope} results.
  {IV}. variability, morphology, and black hole mass},}\ }\href {\doibase
  10.3847/2041-8213/ac6736} {\bibfield  {journal} {\bibinfo  {journal}
  {Astrophys. J. Lett.}\ }\textbf {\bibinfo {volume} {930}},\ \bibinfo {pages}
  {L15} (\bibinfo {year} {2022}{\natexlab{b}})},\ \Eprint
  {http://arxiv.org/abs/2311.08697} {arXiv:2311.08697 [astro-ph.HE]}
  \BibitemShut {NoStop}%
\bibitem [{\citenamefont {{Event Horizon Telescope Collaboration}}\ \emph
  {et~al.}(2025)\citenamefont {{Event Horizon Telescope Collaboration}},
  \citenamefont {Akiyama} \emph {et~al.}}]{EHTM872025Variability}%
  \BibitemOpen
  \bibfield  {author} {\bibinfo {author} {\bibnamefont {{Event Horizon
  Telescope Collaboration}}}, \bibinfo {author} {\bibfnamefont {Kazunori}\
  \bibnamefont {Akiyama}},  \emph {et~al.},\ }\bibfield  {title} {\enquote
  {\bibinfo {title} {Horizon-scale variability of {M87*} from 2017--2021 {EHT}
  observations},}\ }\href {\doibase 10.1051/0004-6361/202555855} {\bibfield
  {journal} {\bibinfo  {journal} {Astron. Astrophys.}\ }\textbf {\bibinfo
  {volume} {704}},\ \bibinfo {pages} {A91} (\bibinfo {year} {2025})},\ \Eprint
  {http://arxiv.org/abs/2509.24593} {arXiv:2509.24593 [astro-ph.HE]}
  \BibitemShut {NoStop}%
\bibitem [{\citenamefont
  {Ara{\'u}jo~Filho}(2025{\natexlab{a}})}]{AraujoFilho:2024xhm}%
  \BibitemOpen
  \bibfield  {author} {\bibinfo {author} {\bibfnamefont {A.~A.}\ \bibnamefont
  {Ara{\'u}jo~Filho}},\ }\bibfield  {title} {\enquote {\bibinfo {title}
  {{Analysis of a nonlinear electromagnetic generalization of the
  Reissner{\textendash}Nordstr{\"o}m black hole}},}\ }\href {\doibase
  10.1140/epjc/s10052-025-14182-1} {\bibfield  {journal} {\bibinfo  {journal}
  {Eur. Phys. J. C}\ }\textbf {\bibinfo {volume} {85}},\ \bibinfo {pages} {454}
  (\bibinfo {year} {2025}{\natexlab{a}})},\ \Eprint
  {http://arxiv.org/abs/2410.12060} {arXiv:2410.12060 [gr-qc]} \BibitemShut
  {NoStop}%
\bibitem [{\citenamefont
  {Ara{\'u}jo~Filho}(2025{\natexlab{b}})}]{AraujoFilho:2024lsi}%
  \BibitemOpen
  \bibfield  {author} {\bibinfo {author} {\bibfnamefont {A.~A.}\ \bibnamefont
  {Ara{\'u}jo~Filho}},\ }\bibfield  {title} {\enquote {\bibinfo {title}
  {{Remarks on a nonlinear electromagnetic extension in AdS
  Reissner-Nordstr{\"o}m spacetime}},}\ }\href {\doibase
  10.1088/1475-7516/2025/01/072} {\bibfield  {journal} {\bibinfo  {journal}
  {JCAP}\ }\textbf {\bibinfo {volume} {01}},\ \bibinfo {pages} {072} (\bibinfo
  {year} {2025}{\natexlab{b}})},\ \Eprint {http://arxiv.org/abs/2410.23165}
  {arXiv:2410.23165 [gr-qc]} \BibitemShut {NoStop}%
\bibitem [{\citenamefont {Filho}(2024)}]{Filho:2023qxu}%
  \BibitemOpen
  \bibfield  {author} {\bibinfo {author} {\bibfnamefont {A.~A.~Ara{\'u}jo}\
  \bibnamefont {Filho}},\ }\bibfield  {title} {\enquote {\bibinfo {title}
  {{Implications of a Simpson{\textendash}Visser solution in
  Verlinde{\textquoteright}s framework}},}\ }\href {\doibase
  10.1140/epjc/s10052-023-12328-7} {\bibfield  {journal} {\bibinfo  {journal}
  {Eur. Phys. J. C}\ }\textbf {\bibinfo {volume} {84}},\ \bibinfo {pages} {73}
  (\bibinfo {year} {2024})},\ \Eprint {http://arxiv.org/abs/2308.04939}
  {arXiv:2308.04939 [gr-qc]} \BibitemShut {NoStop}%
\bibitem [{\citenamefont {Liu}\ \emph {et~al.}(2026{\natexlab{b}})\citenamefont
  {Liu}, \citenamefont {Liu}, \citenamefont {Liu}, \citenamefont {Su},\ and\
  \citenamefont {Zeng}}]{Liu:2026cxs}%
  \BibitemOpen
  \bibfield  {author} {\bibinfo {author} {\bibfnamefont {Hui-Fa}\ \bibnamefont
  {Liu}}, \bibinfo {author} {\bibfnamefont {Wentao}\ \bibnamefont {Liu}},
  \bibinfo {author} {\bibfnamefont {Yu-Xiao}\ \bibnamefont {Liu}}, \bibinfo
  {author} {\bibfnamefont {Qi}~\bibnamefont {Su}}, \ and\ \bibinfo {author}
  {\bibfnamefont {Ding-fang}\ \bibnamefont {Zeng}},\ }\bibfield  {title}
  {\enquote {\bibinfo {title} {{Gravitational-bumblebee perturbations: exact
  decoupling and isospectrality}},}\ }\href {\doibase 10.1007/JHEP08(2026)027}
  {\bibfield  {journal} {\bibinfo  {journal} {JHEP}\ }\textbf {\bibinfo
  {volume} {08}},\ \bibinfo {pages} {027} (\bibinfo {year}
  {2026}{\natexlab{b}})},\ \Eprint {http://arxiv.org/abs/2605.02820}
  {arXiv:2605.02820 [gr-qc]} \BibitemShut {NoStop}%
\bibitem [{\citenamefont {Santos}\ \emph {et~al.}(2026)\citenamefont {Santos},
  \citenamefont {Lessa}, \citenamefont {Maluf},\ and\ \citenamefont
  {Olmo}}]{Santos:2025xbk}%
  \BibitemOpen
  \bibfield  {author} {\bibinfo {author} {\bibfnamefont {Alana C.~L.}\
  \bibnamefont {Santos}}, \bibinfo {author} {\bibfnamefont {Leandro~A.}\
  \bibnamefont {Lessa}}, \bibinfo {author} {\bibfnamefont {Roberto~V.}\
  \bibnamefont {Maluf}}, \ and\ \bibinfo {author} {\bibfnamefont {Gonzalo~J.}\
  \bibnamefont {Olmo}},\ }\bibfield  {title} {\enquote {\bibinfo {title}
  {{Echoes and quasinormal modes of asymmetric black bounces}},}\ }\href
  {\doibase 10.1088/1475-7516/2026/04/010} {\bibfield  {journal} {\bibinfo
  {journal} {JCAP}\ }\textbf {\bibinfo {volume} {04}},\ \bibinfo {pages} {010}
  (\bibinfo {year} {2026})},\ \Eprint {http://arxiv.org/abs/2508.11096}
  {arXiv:2508.11096 [gr-qc]} \BibitemShut {NoStop}%
\bibitem [{\citenamefont {Iyer}\ and\ \citenamefont
  {Will}(1987)}]{IyerWill1987}%
  \BibitemOpen
  \bibfield  {author} {\bibinfo {author} {\bibfnamefont {Sai}\ \bibnamefont
  {Iyer}}\ and\ \bibinfo {author} {\bibfnamefont {Clifford~M.}\ \bibnamefont
  {Will}},\ }\bibfield  {title} {\enquote {\bibinfo {title} {Black-hole normal
  modes: A {WKB} approach. {I}. foundations and application of a higher-order
  {WKB} analysis of potential-barrier scattering},}\ }\href {\doibase
  10.1103/PhysRevD.35.3621} {\bibfield  {journal} {\bibinfo  {journal}
  {Physical Review D}\ }\textbf {\bibinfo {volume} {35}},\ \bibinfo {pages}
  {3621--3631} (\bibinfo {year} {1987})}\BibitemShut {NoStop}%
\bibitem [{\citenamefont {Konoplya}\ \emph {et~al.}(2019)\citenamefont
  {Konoplya}, \citenamefont {Zhidenko},\ and\ \citenamefont
  {Zinhailo}}]{KonoplyaZhidenkoZinhailo2019}%
  \BibitemOpen
  \bibfield  {author} {\bibinfo {author} {\bibfnamefont {R.~A.}\ \bibnamefont
  {Konoplya}}, \bibinfo {author} {\bibfnamefont {A.}~\bibnamefont {Zhidenko}},
  \ and\ \bibinfo {author} {\bibfnamefont {A.~F.}\ \bibnamefont {Zinhailo}},\
  }\bibfield  {title} {\enquote {\bibinfo {title} {Higher order {WKB} formula
  for quasinormal modes and grey-body factors: recipes for quick and accurate
  calculations},}\ }\href {\doibase 10.1088/1361-6382/ab2e25} {\bibfield
  {journal} {\bibinfo  {journal} {Classical and Quantum Gravity}\ }\textbf
  {\bibinfo {volume} {36}},\ \bibinfo {pages} {155002} (\bibinfo {year}
  {2019})},\ \Eprint {http://arxiv.org/abs/1904.10333} {arXiv:1904.10333
  [gr-qc]} \BibitemShut {NoStop}%
\end{thebibliography}%

\end{document}